\documentclass[preprint,12pt]{elsarticle}

\usepackage{color,amsmath,amssymb,graphicx,latexsym,subfigure}
\usepackage{threeparttable}

\newcommand{\sv}{\ensuremath{\langle\sigma v\rangle}}

\begin{document}

\begin{frontmatter}
\title{Implications of the AMS-02 positron fraction in cosmic rays}

\author{Qiang Yuan$^a$, Xiao-Jun Bi$^a$\footnote{The corresponding 
author (email: bixj@ihep.ac.cn)}, Guo-Ming Chen$^a$,
Yi-Qing Guo$^a$, Su-Jie Lin$^a$, Xinmin Zhang$^b$}

\address[a]{Key Laboratory of Particle Astrophysics,
Institute of High Energy Physics, Chinese Academy of Science,
Beijing 100049, P.R.China}
\address[b]{Theoretical Physics Division, Institute of High Energy
Physics, Chinese Academy of Sciences, Beijing 10049, P.R. China
}

\begin{abstract}

The AMS-02 collaboration has just released its first result of the cosmic 
positron fraction $e^+/(e^-+e^+)$ with high precision up to $\sim 350$
GeV. The AMS-02 result shows the same trend with the previous PAMELA result,
which requires extra electron/positron sources on top of the conventional
cosmic ray background, either from astrophysical sources or from dark matter
annihilation/decay. In this paper we try to figure out the nature of
the extra sources by fitting to the AMS-02 $e^+/(e^-+e^+)$ data, as well 
as the electron and proton spectra by PAMELA and the $(e^-+e^+)$
spectrum by Fermi and HESS. We adopt the GALPROP package to calculate 
the propagation of the Galactic cosmic rays and the Markov Chain Monte 
Carlo sampler to do the fit. We find that under the conventional 
assumptions about the background and the extra source of the $e^-+e^+$, 
we cannot fit the AMS-02 and Fermi/HESS data well simultaneously. 
The AMS-02 data require less electrons/positrons from the extra sources 
than that required by Fermi/HESS. It may indicate that the model needs 
to be refined or the data between these experiments have systematic 
uncertainties. The pulsar scenario generally fits the data better 
than the DM scenario. Furthermore, the constraints from $\gamma$-rays 
also disfavor the DM scenario to explain the cosmic ray lepton data.

\end{abstract}
\end{frontmatter}

\section{Introduction}

The Alpha Magnetic Spectrometer (AMS-02) was launched in May 2011.
After nearly two years operation and analysis, the AMS-02 collaboration
has released its first physical result, i.e. the positron fraction 
$e^+/(e^-+e^+)$ in cosmic rays (CRs) \cite{2013PhRvL.110n1102A}. 
The data show very high precision with the energy range from $\sim 0.5$ 
GeV to $\sim 350 $ GeV. The fraction rises above $\sim 8$ GeV up to the 
energy end at $\sim 350 $ GeV, which is consistent with the previous 
PAMELA result of the cosmic positron fraction \cite{2009Natur.458..607A,
2010APh....34....1A}. The result is NOT consistent with the conventional 
CR expectation \cite{2012PhRvD..85d3507L}. A great number of works
have dedicated to explaining the PAMELA result, either by astrophysical
sources \cite{2009PhRvL.103e1101Y,2009JCAP...01..025H,2012CEJPh..10....1P,
2009PhRvD..80f3005M,2009ApJ...700L.170H,2009PhRvL.103e1104B} or by
dark matter (DM) \cite{2008PhRvD..78j3520B,2009PhLB..672..141B,
2009NuPhB.813....1C,2009PhRvD..79b3512Y,2009PhRvD..80b3007Z}.

To determine the parameters of the CR background and the nature of
the extra sources we have to consider all the relevant results.
The available results at present include the pure electron spectrum
measured by PAMELA \cite{2011PhRvL.106t1101A} and the antiproton flux 
and ratio $\bar{p}/p$ \cite{2009PhRvL.102e1101A,2010PhRvL.105l1101A} by
PAMELA. Another important result from PAMELA is the precise measurement
of the proton spectrum \cite{2011Sci...332...69A}, which determines
the secondary positron spectrum by collision with the interstellar
medium (ISM) when propagating in the Galaxy. There are also precise
measurements of the total electron and positron $(e^-+e^+)$ spectrum,
by the Fermi-LAT collaboration \cite{2009PhRvL.102r1101A,
2010PhRvD..82i2004A} and ATIC collaboration \cite{2008Natur.456..362C}.
The ground-based atmospheric Cerenkov telescopes HESS also gives
measurement of the total $e^{\pm}$ spectrum up to higher energies
\cite{2008PhRvL.101z1104A,2009A&A...508..561A}.

Fitting to the PAMELA positron fraction data \cite{2009Natur.458..607A},
PAMELA electron spectrum and Fermi/HESS total $e^{\pm}$ spectrum shows
that both the astrophysical source, such as pulsars, and the DM scenarios
can give a good explanation to the data \cite{2012PhRvD..85d3507L}.
It is hard to discriminate the two scenarios with the CR spectra
mentioned above. As the AMS-02 data show much higher precision and wider
energy extension, especially it shows softer behavior than the PAMELA 2008
result, it is necessary to re-examine the previous conclusion with all of
the newly available data. In this work we have done such a global fitting
to all the relevant data, including AMS-02 $e^+/(e^-+e^+)$, Fermi and HESS
$(e^-+e^+)$ total spectrum and PAMELA proton and electron spectra.

We adopt the CosRayMC code, which embeds the CR propagation code
in the Markov Chain Monte Carlo (MCMC) sampler and enables efficient
survey of the high-dimensional parameter space \cite{2010PhRvD..81b3516L,
2012PhRvD..85d3507L}. The CR propagation is treated by the GALPROP package
\cite{1998ApJ...509..212S}. The CR transportation process in the Galaxy 
is characterized by the secondary particles. Therefore the 
secondary-to-primary ratio, such as B/C, (Sc+Ti+V)/Fe, and the 
unstable-to-stable ratio of secondary particles, such as $^{10}$Be/$^9$Be, 
$^{26}$Al/$^{27}$Al are often used to determine the propagation parameters
\cite{1990cup..book.....G,1998ApJ...509..212S,2001ApJ...555..585M,
2010APh....34..274D}. In this work we fix the CR propagation parameters
to the values which give the best fitting to the currently available 
B/C and $^{10}$Be/$^9$Be data with the MCMC method. The fitting process 
will be reported separately in \cite{mcmc:prop}.

We then fit the parameters of the electrons and positrons to the
relevant data, both from the CR background and the extra sources.
We emphasize that the global fitting is important because when
both components contribute to the observations neither one should
be determined seperately. In the work we have considered the continuously 
distributed pulsars\footnote{In the following part of this paper, 
``pulsar'' actually means the pulsar-like astrophysical sources which 
can produce $e^{\pm}$ pairs.} and DM annihilation/decay as two 
typical scenarios of the extra positron/electron sources. 
Note that this assumption may be over-simplified, because for 
energies up to hundreds of GeV the variance comes from discrete 
distribution of the sources can be very important. The discreteness
will make the problem more complicated and uncertain. The current data
may not be able to discriminate the continuous scenario from discrete
scenario. However, the phenomenological consequences of both scenarios,
i.e., the locally spectra of $e^{\pm}$ should be similar in order to
match the data.
The injection parameters of primary electrons are free parameters to be 
fitted. The cosmic positrons include secondaries from CR interaction with 
the ISM and the primary ones from the extra sources. The secondary positron 
spectrum is determined by the spectrum of cosmic protons (including Helium 
and a few heavier nuclei) and the primary positron spectrum is determined
by the nature of the extra sources. In principle the injection parameters
of the protons can be fitted independently and then be employed to
calculate the secondary positrons. However, since both the proton and
electron spectra are measured by PAMELA at almost the same time, they
are modulated by the solar activity with a similar magnitude. Therefore,
in this work we have adopted two ways to fit the cosmic proton injection
parameters: either by fitting the proton spectrum independently or by
fitting the proton and electron spectra simultaneously. 

This paper is organized as follows. We give a brief introduce of
the propagation of Galactic CRs in Sec. \ref{crprop}. The experimental
data and fitting method are described in Sec. \ref{cosraymc}. The results
are presented in Sec. \ref{result}. In Sec. \ref{discuss} we give
discussion about the fitting results, and finally a brief summary is
given in Sec. \ref{summary}.

\section{Cosmic ray propagation}
\label{crprop}

The propagation of charged CRs in the Galaxy is described by the
diffusive equation \cite{2007ARNPS..57..285S}
\begin{eqnarray}
\frac{\partial \psi}{\partial t}
&=& Q({\bf x},p)+\nabla\cdot(D_{xx}\nabla \psi-{\bf
V_c}\psi)+\frac{\partial}{\partial p}p^2D_{pp}\frac{\partial}
{\partial p}\frac{1}{p^2}\psi \nonumber \\
&-& \frac{\partial}{\partial p}
\left[\dot{p}\psi-\frac{p}{3}(\nabla\cdot{\bf V_c}\psi)\right]-
\frac{\psi}{\tau_f}-\frac{\psi}{\tau_r}, \label{prop}
\end{eqnarray}
where $\psi$ is the density of CR particles per unit momentum
interval, $Q({\bf x},p)$ is the source term, $D_{xx}$ is the spatial
diffusion coefficient, ${\bf V_c}$ is the convection velocity,
$D_{pp}$ is the diffusion coefficient in momentum space used to
describe the reacceleration process, $\dot{p}\equiv{\rm d}p/{\rm d}t$
is the momentum loss rate, $\tau_f$ and $\tau_r$ are time scales for
fragmentation and radioactive decay respectively. $D_{xx}$ is usually 
assumed to be only rigidity dependent and has a power-law form
$D_{xx}=D_0\beta(R/R_0)^{\delta}$, with $\delta$ reflecting the
property of the interstellar medium (ISM) turbulence.
The reacceleration is described by the diffusion in momentum space.
The momentum diffusion coefficient $D_{pp}$ relates with the spatial
diffusion coefficient $D_{xx}$ as \cite{1994ApJ...431..705S}
\begin{equation}
D_{pp}D_{xx}=\frac{4p^2v_A^2}{3\delta(4-\delta^2)(4-\delta)w},
\end{equation}
where $v_A$ is the Alfven speed, $w$ is the ratio of
magnetohydrodynamic wave energy density to the magnetic field energy
density, which characterizes the level of turbulence. In the usual way
we take $w$ to be $1$ and use the Alfven speed $v_A$ to
describe the reacceleration \cite{1994ApJ...431..705S}. The CRs
propagate in an extended halo with characteristic height $z_h$,
beyond which free escape of CRs is assumed. Thus the major propagation
parameters include $D_0$, $\delta$, $v_A$, $V_c$ and $z_h$.

There are publicly available numerical codes to compute the
CRs propagation in the Galaxy, such as 
GALPROP\footnote{http://galprop.stanford.edu/}
\cite{1998ApJ...509..212S} and
DRAGON\footnote{http://www.desy.de/\~maccione/DRAGON/}
\cite{2008JCAP...10..018E}. We have embedded the GALPROP package with 
the MCMC sampler and fitted to the data to constrain the propagation 
parameters \cite{mcmc:prop}.

\begin{table}[!htb]
\centering
\caption {Propagation parameters taken in the work}
\begin{tabular}{lc}
\hline \hline
$D_0$($10^{28}$ cm$^2$s$^{-1}$)\footnotemark[1] & $5.94$ \\
$\delta$ & $0.377$ \\
$z_h$(kpc) & $4.04$ \\
$v_A$(km s$^{-1}$) & $36.4$ \\
\hline
\hline
\end{tabular}\vspace{3mm}\\
\footnotemark[1]{At $R_0=4$ GV.}
\label{table:prop}
\end{table}

Recently several groups employed the MCMC technique to fit the CR
propagation parameters
\cite{2009A&A...497..991P,2010A&A...516A..66P,2011ApJ...729..106T}.
Using the currently available data of B/C and $^{10}$Be/$^9$Be, we did
an independent MCMC fit to the propagation parameters \cite{mcmc:prop}. 
We find the reacceleration model gives quite good description to the 
present data, while the convection model gives worse fitting. Therefore 
we adopt the reacceleration (i.e., $V_c=0$) scenario as the starting 
point of the present study. The values of the propagation parameters 
we adopted in the work are listed in Table \ref{table:prop}, as given 
in \cite{mcmc:prop}. We have tested that varying the propagation 
parameters within extreme ranges allowed by the B/C data 
\cite{2012ApJ...761...91A}, the qualitative results of this work do
not change.

\section{Fitting description}
\label{cosraymc}

\subsection{Model}

The basic framework of the CR model is as follows. The primary
sources of CRs, such as the supernova remnants (SNRs), accelerate
CR nuclei and electrons and inject them into the Galaxy. These
particles then propagate in the Galaxy, experiencing diffusion,
reacceleration, interactions and radiation during the propagation
process. The interactions with ISM produce secondary particles
including the secondary nuclei (such as Li, Be, B, and Sc, Ti, V), 
positrons, antiprotons and diffuse $\gamma$-rays. Such a scenario 
gives consistent description of most of the CR data as well as the
all-sky diffuse $\gamma$-ray emission \cite{2007ARNPS..57..285S}.
To distinguish from the extra sources of the $e^{\pm}$ as will be
described below, we call this component as the background. The 
model to fit the AMS-02 data includes this CR background (primary 
nuclei and electrons, and the secondary positrons) and the 
extra positron/electron sources.

The injection spectra of the primary protons (heavier nuclei are less
relevant) and electrons are assumed to be broken power-law functions 
with respect to the momentum $p$
\begin{equation}
q(p)\propto\left(\frac{p}{p_{\rm br}^{p,e}}\right)^{-\nu_1/\nu_2},
\end{equation}
where $\nu_1$ and $\nu_2$ are the spectral indices below and above 
the break momentum $p_{\rm br}$. In the following we use $\nu_1$ and 
$\nu_2$ to represent the proton spectrum and use $\gamma_1$ and 
$\gamma_2$ to represent the primary electron spectrum. The propagated 
fluxes of protons and electrons are then normalized to factors $A_p$ 
and $A_e$ to get the absolute fluxes.

The spatial distribution of the primary CR particles is adopted
to be the supernova remnants (SNR) like distribution
\begin{equation}
f(R,z)\propto\left(\frac{R}{R_{\odot}}\right)^a\exp\left[-\frac{b(R-
R_{\odot})}{R_{\odot}}\right]\exp\left(-\frac{|z|}{z_s}\right),
\label{source}
\end{equation}
where $R_{\odot}=8.5$ kpc is the distance of solar system from the
Galactic center, $z_s\approx 0.2$ kpc is the characteristic height
of the Galactic disk, $a$ and $b$ are the shape parameters which 
can be fitted according to the survey data of SNRs or other kinds of
assumed CR sources. We employ $a=1.25$ and $b=3.56$ following
\cite{2011ApJ...729..106T}.

For the secondary electrons/positrons we adopt the same GALPROP model
to calculate their propagation. The production spectra of the secondary
electrons/positrons are calculated using the parameterization given in 
\cite{2006ApJ...647..692K} based on the propagated proton spectrum. 
A free factor $c_{e^+}$ to normalize the secondary positron/electron 
flux is included in the fitting, which represents the possible uncertainties 
from the hadronic interactions, propagation models, the ISM density 
distributions, and the nuclear enhancement factor from heavy elements.
A caveat is that the nuclear collision may not be simply scaled
from the $pp$ collision, i.e., the nuclear enhancement factor is
energy dependent \cite{2009APh....31..341M}. Since the energy dependence 
is weak, we keep the constant factor as an approximation here.

About the extra electron/positron sources, we study two popular kinds
of sources: the astrophysical sources such as pulsars and the DM 
annihilation scenario. The injection spectrum of the astrophysical sources
is parameterized as a power-law function with an exponential cutoff
\begin{equation}
q(p)=A_{\rm psr}p^{-\alpha}\exp(-p/p_c),
\end{equation}
where $A_{\rm psr}$ is the normalization factor, $\alpha$ is the
spectral index and $p_c$ is the cutoff momentum. The source population
is taken to be a continuous form with distribution function (\ref{source}), 
but with different parameters $a=2.35$ and $b=5.56$, given in
\cite{2004IAUS..218..105L}. The effect of nearby isolate sources
(e.g., \cite{2009JCAP...01..025H,2009PhRvL.103e1101Y,2012CEJPh..10....1P},
see also the discussion in Sec. V) is not covered in the present study.
Note there were proposals that the hadronic interactions around
the CR sources and the subsequent acceleration of the secondary $e^{\pm}$
could be responsible for the $e^{\pm}$ excesses \cite{2009PhRvL.103e1104B,
2009PhRvD..80f3003F}. Such a scenario is not in conflict with the
pulsar-like scenario assumed here (with slight difference of the spatial
distribution which has little effect on the charged CR propagation).
But we should keep in mind the simultaneously produced antiprotons
and secondary nuclei may constrain this model \cite{2009PhRvL.103h1103B,
2009PhRvL.103h1104M,2009PhRvD..80l3017A}.

For the DM annihilation scenario, the density profile of the Milky Way halo
is adopted to be the Navarro-Frenk-White (NFW) distribution
\cite{1997ApJ...490..493N}
\begin{equation}
\rho(r)=\frac{\rho_s}{(r/r_s)(1+r/r_s)^2},
\end{equation}
with $r_s=20$ kpc and $\rho_s=0.26$ GeV cm$^{-3}$. Such a value of
$\rho_s$ will correspond to a local density of $0.3$ GeV cm$^{-3}$. 
For higher values of the local density as revealed by several recent 
studies \cite{2010JCAP...08..004C,2010A&A...523A..83S,2010PhRvD..82b3531P},
the annihilation cross section will be different by a constant factor. 
Since the measurement of CR antiprotons by PAMELA
\cite{2009PhRvL.102e1101A,2010PhRvL.105l1101A}
constrain the hadronic annihilation channels strongly
\cite{2009NuPhB.813....1C,2009PhRvD..79b3512Y,2009PhRvL.102g1301D,
2011JCAP...09..007C}, we will focus on the leptonic annihilation 
channels here. The positron spectrum from the DM annihilation products 
is calculated using the PYTHIA simulation package \cite{2006JHEP...05..026S}.

The CRs at low energy (typically with rigidity below $\sim30$
GV) are affected by the solar environment when entering the solar 
system, known as solar modulation effect. The force field approximation 
is often employed to describe the solar modulation effect 
\cite{1968ApJ...154.1011G}, which has only one single parameter --- 
the modulation potential $\phi$. However, the low energy data about 
the positron fraction measured by PAMELA and AMS-02 may imply that the 
simple force field approximation is not enough to explain the data, 
and the charge-sign dependent modulation effect is necessary 
\cite{1996ApJ...464..507C,2009NJPh...11j5021B,2012AdSpR..49.1587D}.
Therefore, in order to avoid possible inconsistency of the low energy 
behavior we do not include the AMS-02 data below $5$ GeV in our fit.
The solar modulation effect for the positron fraction above $5$ GeV 
is negligible \cite{2012AdSpR..49.1587D,2013PhRvL.110h1101M}.

In summary the full parameter space investigated in this work is
\begin{eqnarray}
\mathcal{P}=\begin{cases}
\{A_p,\nu_1,\nu_2,p_{\rm br}^p\},&\mbox{bkg protons,} \\
\{A_e,\gamma_1,\gamma_2,p_{\rm br}^e\},&\mbox{bkg electrons,}\\
\{A_{\rm psr},\alpha,p_c\}\ \mbox{or}\ \{m_{\chi},\sv,\mbox{ch}\}, &\mbox{exotic sources},\\
\{c_{e^+},\phi\},&\mbox{others},
\end{cases}
\end{eqnarray}
where ``ch'' represents the channel of DM annihilation, which is set to be
one of $\{\mu^+\mu^-,\,\tau^+\tau^-,\,W^+W^-,\,b\bar{b}\}$ and does not
enter in the fitting. Such a model works well for the PAMELA and Fermi data 
\cite{2012PhRvD..85d3507L}.

\subsection{Data}

\begin{table*}[!htb]
\centering
\caption {Definition of fitting}
\begin{tabular}{c|l}
\hline \hline
I-a & AMS $e^+/e^{\pm}$ + PAMELA $e^-$ + Fermi/HESS $e^{\pm}$ \\
II-a & AMS $e^+/e^{\pm}$ + PAMELA $e^-$ \\
\hline
I-b & AMS $e^+/e^{\pm}$ + PAMELA $e^-$ + Fermi/HESS $e^{\pm}$ +  PAMELA $p$\\
II-b & AMS $e^+/e^{\pm}$ + PAMELA $e^-$ + PAMELA $p$\\
\hline
\hline
\end{tabular}
\label{table:def}
\end{table*}

In this study the data to be fitted include the latest positron fraction 
by AMS-02\cite{2013PhRvL.110n1102A}, the electron spectrum by PAMELA
\cite{2011PhRvL.106t1101A}, the total electron and positron spectra
by Fermi \cite{2009PhRvL.102r1101A,2010PhRvD..82i2004A} and
HESS \cite{2008PhRvL.101z1104A,2009A&A...508..561A}, and the proton
spectrum by PAMELA \cite{2011Sci...332...69A}. We choose two ways 
to deal with the proton spectrum as described in Sec. I: a) to fit it
separately, and b) to include it in the global fitting.

Note that the CR spectral hardening at $\sim200$ GV reported by 
ATIC \cite{2007BRASP..71..494P}, CREAM \cite{2010ApJ...714L..89A} 
and PAMELA \cite{2011Sci...332...69A} implies that single power-law 
can not fully describe the high energy spectra of the CR nuclei above 
$\sim10$ GV\footnote{Note, however, the preliminary data about the
proton and Helium spectra show no hardening below $\sim$TeV 
\cite{2013ICRC-AMS02}. Combining with the CREAM data, it may still
show a hardening of these spectra at higher energies.}. 
Therefore we only take the PAMELA proton data below $150$ GeV in the 
fit. To better fit the high energy part (and the CREAM data), 
we need a further break or a curved injection spectrum of the protons
\cite{2011PhRvD..84d3002Y}. 

Since the Fermi and HESS data may have larger systematic uncertainties
we also investigate the case without the Fermi/HESS data. The definition 
of the fittings are compiled in Table \ref{table:def}.

\section{Results}
\label{result}

\subsection{Fixing the proton spectrum}

\begin{figure}[!htb]
\centering
\includegraphics[width=0.7\textwidth]{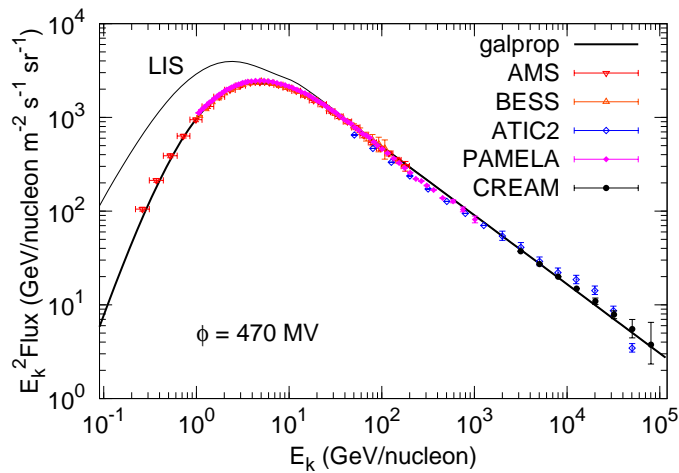}
\caption{Proton spectrum derived through fitting PAMELA data.
References of the proton data: AMS \cite{2000PhLB..490...27A}, 
BESS \cite{2000ApJ...545.1135S}, ATIC2 \cite{2007BRASP..71..494P}, 
PAMELA \cite{2011Sci...332...69A} and CREAM 
\cite{2010ApJ...714L..89A}.
\label{fig:proton_fix}}
\end{figure}

We first fit the proton spectrum independently with the PAMELA 
(and CREAM) data. Here we add the high energy CREAM data 
\cite{2010ApJ...714L..89A} in the fitting to give a description 
of the proton behavior in a wider energy range. The best fitting 
injection parameters of protons are: $\nu_1=1.79$, $\nu_2=2.36$, 
$p_{\rm br}^p=11.7$ GeV and the solar modulation potential 
$\phi=470$ MV. The propagated proton spectrum for the best fitting 
parameters is shown in Fig. \ref{fig:proton_fix}. 

\begin{figure*}[!htb]
\centering
\includegraphics[width=0.48\textwidth]{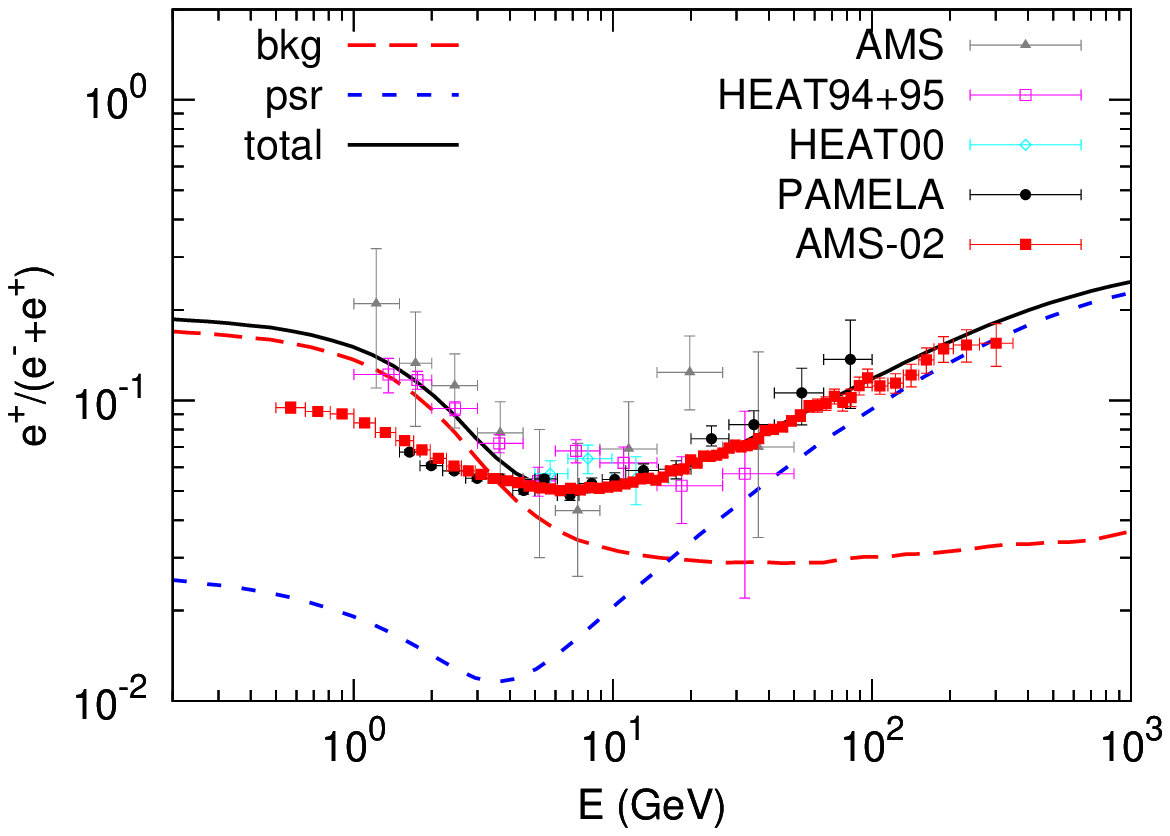}
\includegraphics[width=0.48\textwidth]{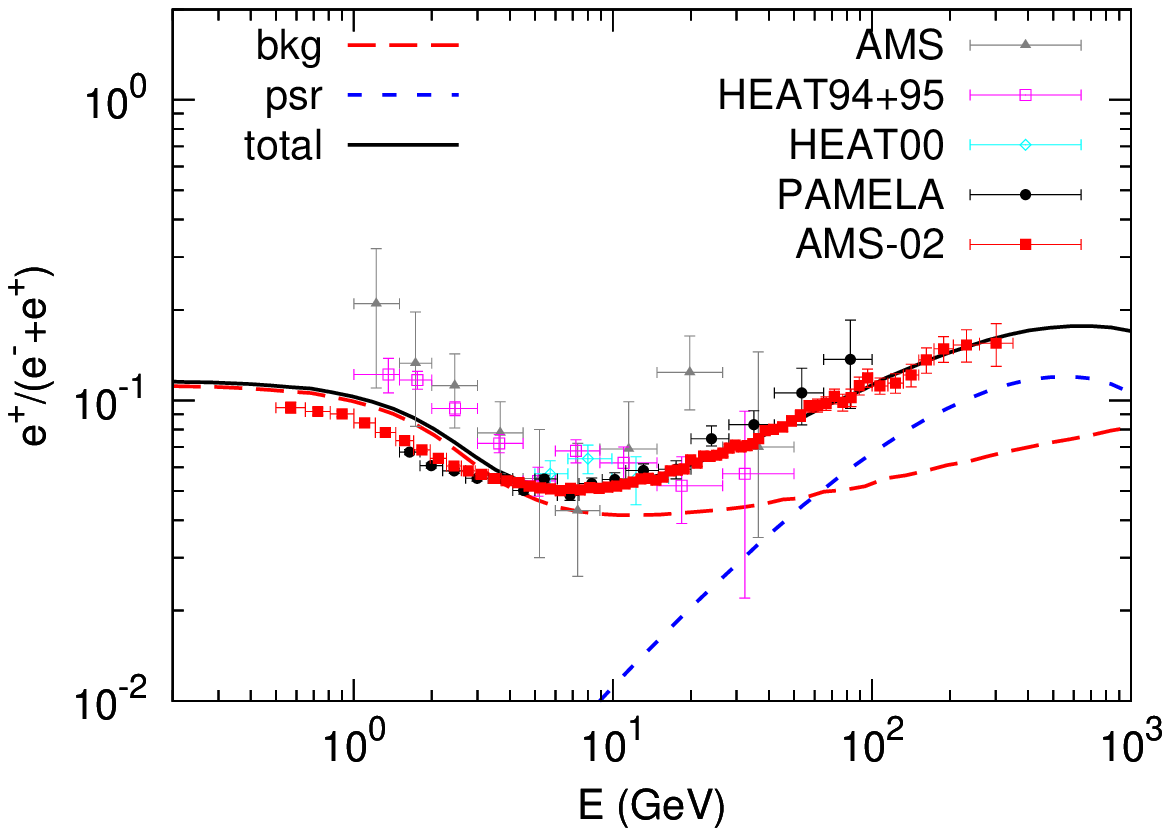}
\includegraphics[width=0.48\textwidth]{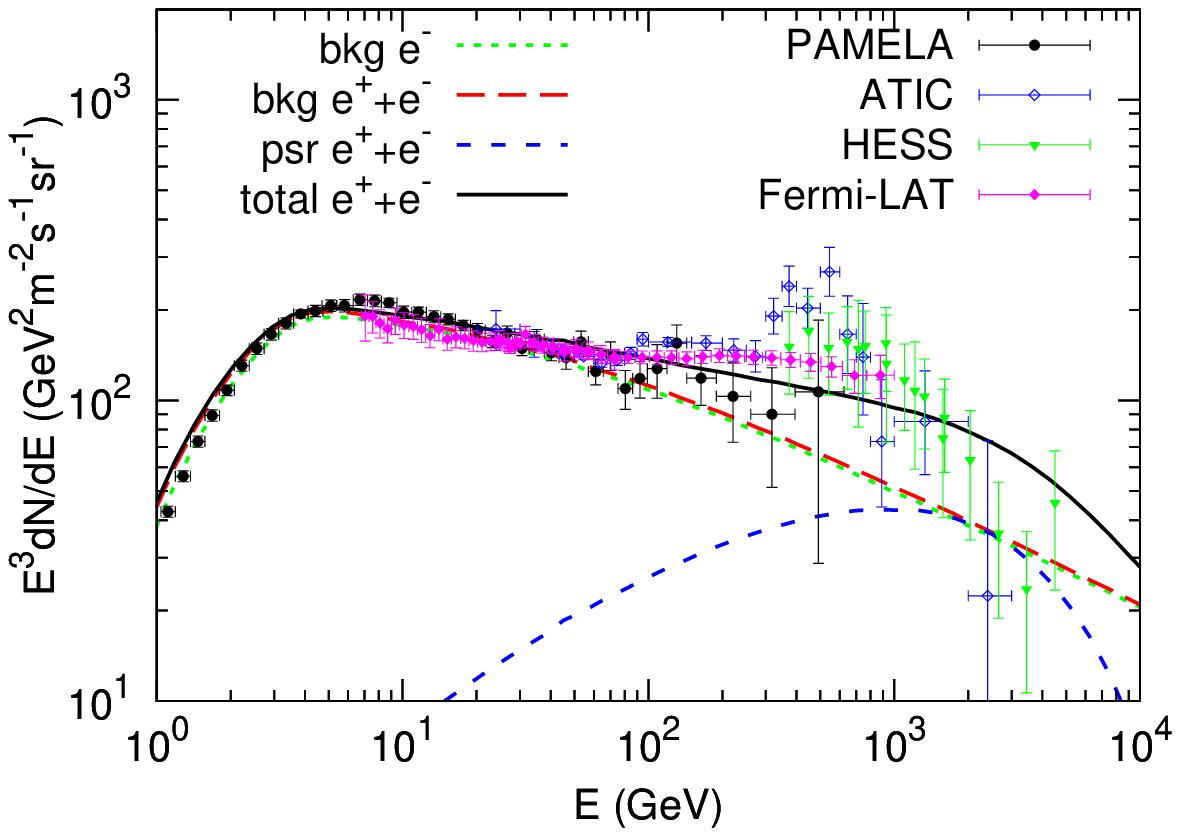}
\includegraphics[width=0.48\textwidth]{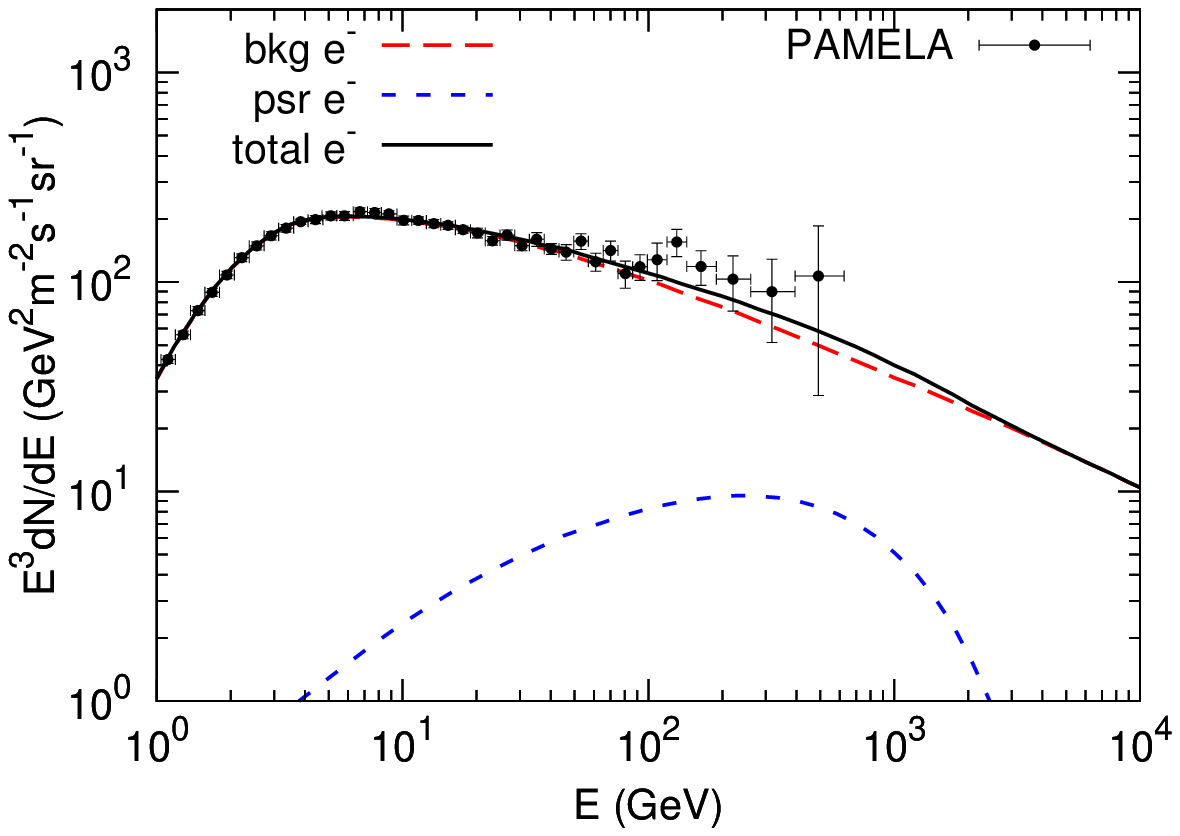}
\caption{The positron fraction (upper) and electron spectra (lower)
for the background together with a pulsar component of the exotic 
$e^{\pm}$. The panels from left to right are for fits I-a
and II-a respectively. References of the data: positron fraction ---
AMS \cite{2007PhLB..646..145A}, HEAT94+95 \cite{1997ApJ...482L.191B},
HEAT00 \cite{2001ICRC....5.1687C}, PAMELA \cite{2009Natur.458..607A},
AMS-02\cite{2013PhRvL.110n1102A} ; electron --- 
PAMELA \cite{2011PhRvL.106t1101A}, ATIC \cite{2008Natur.456..362C}, 
HESS \cite{2008PhRvL.101z1104A,2009A&A...508..561A}, 
Fermi-LAT \cite{2010PhRvD..82i2004A}.
\label{fig:psr_fixp}}
\end{figure*}

\begin{figure*}[!htb]
\centering
\includegraphics[width=0.48\textwidth]{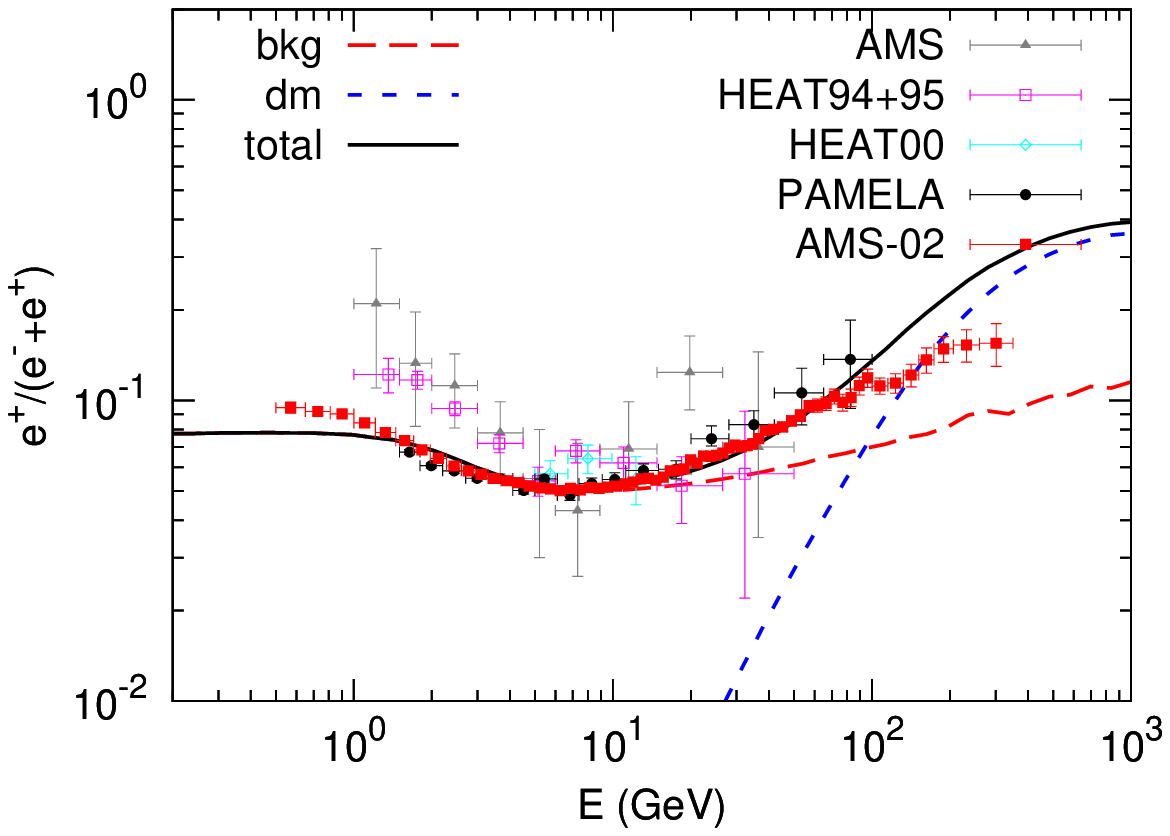}
\includegraphics[width=0.48\textwidth]{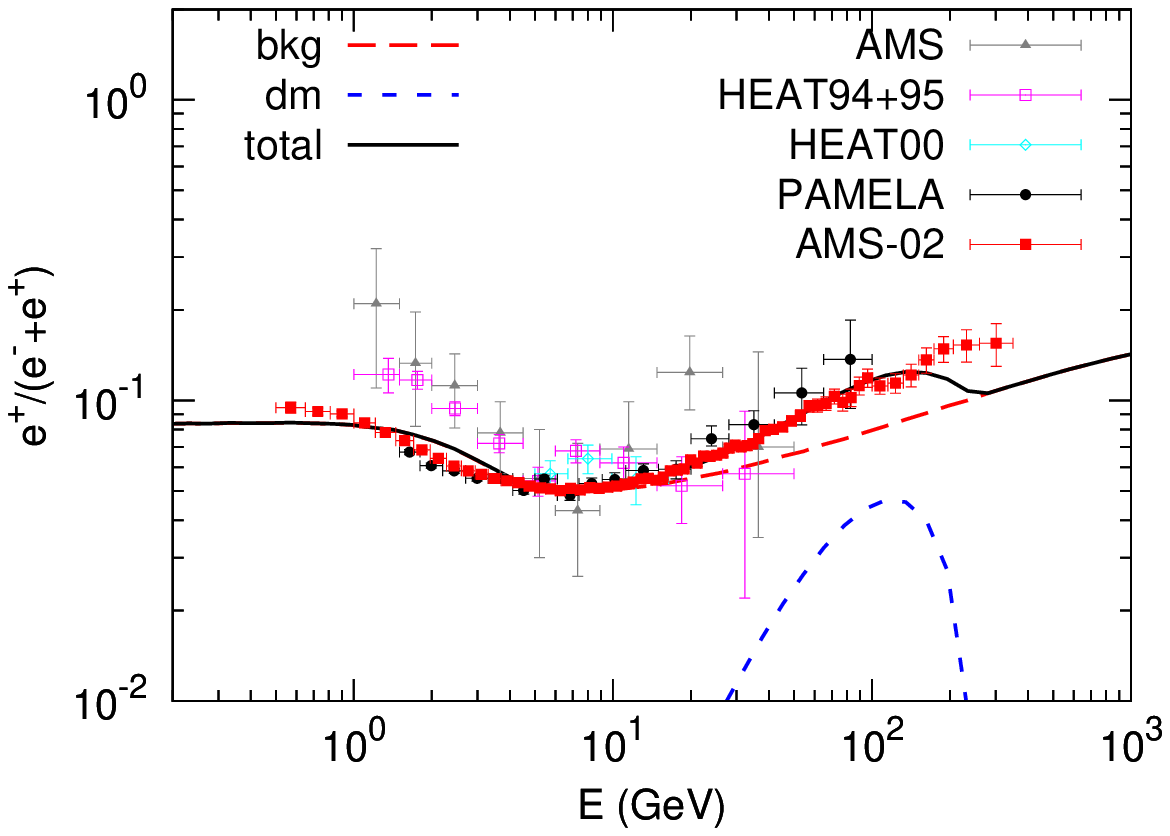}
\includegraphics[width=0.48\textwidth]{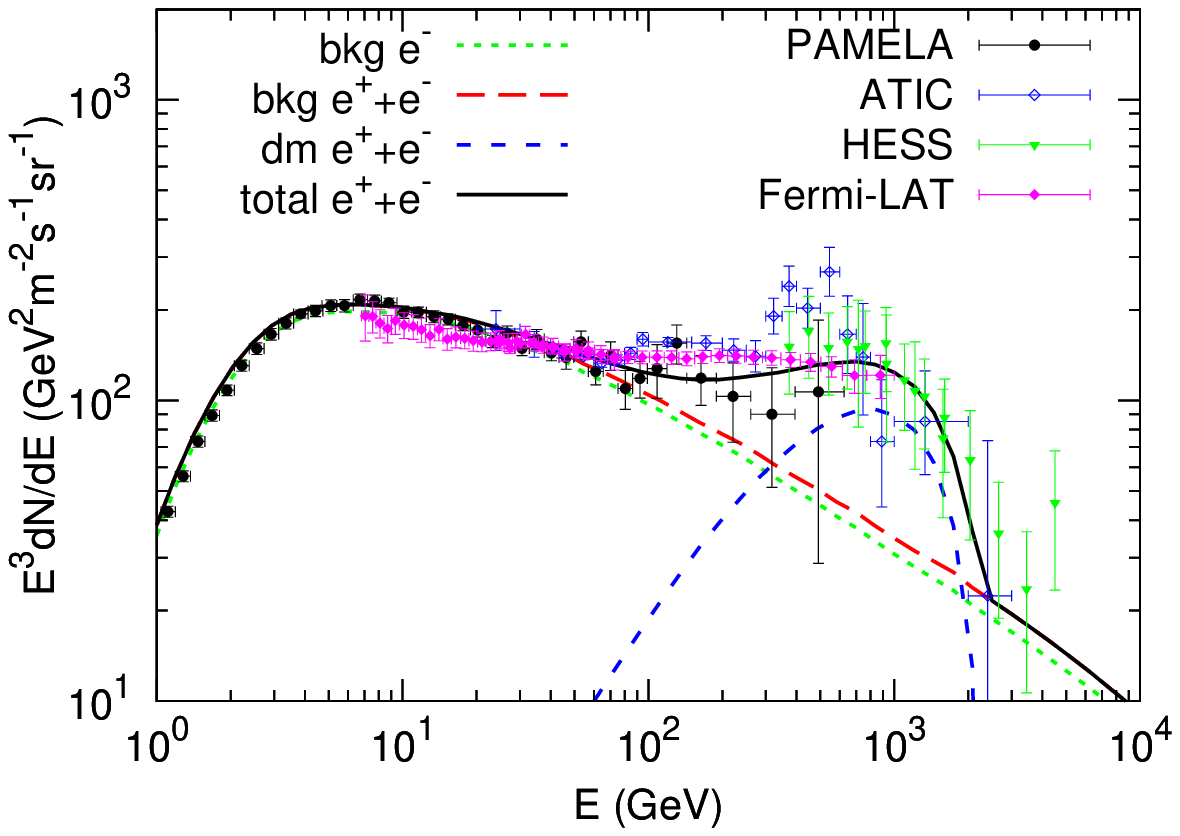}
\includegraphics[width=0.48\textwidth]{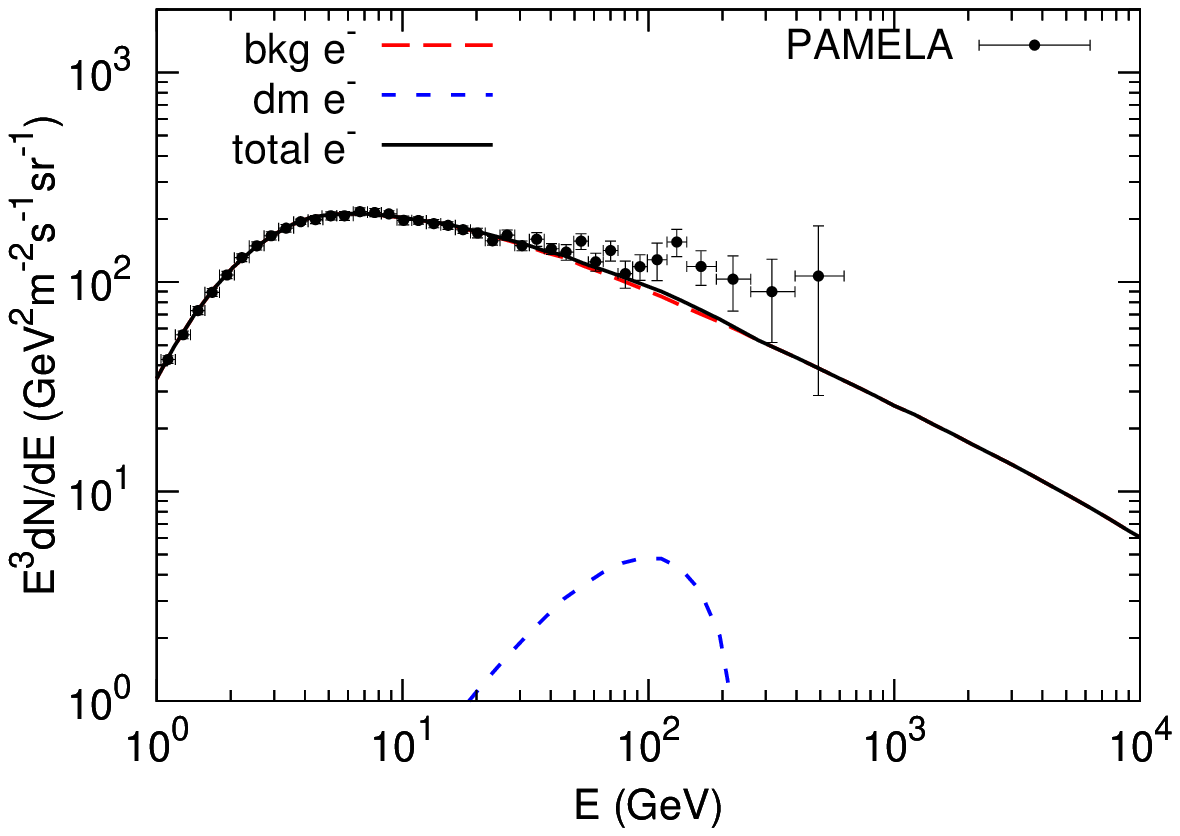}
\caption{Same as Fig. \ref{fig:psr_fixp} but the exotic $e^{\pm}$ 
are assumed to be from DM annihilation. The annihilation channel is
$\mu^+\mu^-$.
\label{fig:mu_fixp}}
\end{figure*}

\begin{figure*}[!htb]
\centering
\includegraphics[width=0.48\textwidth]{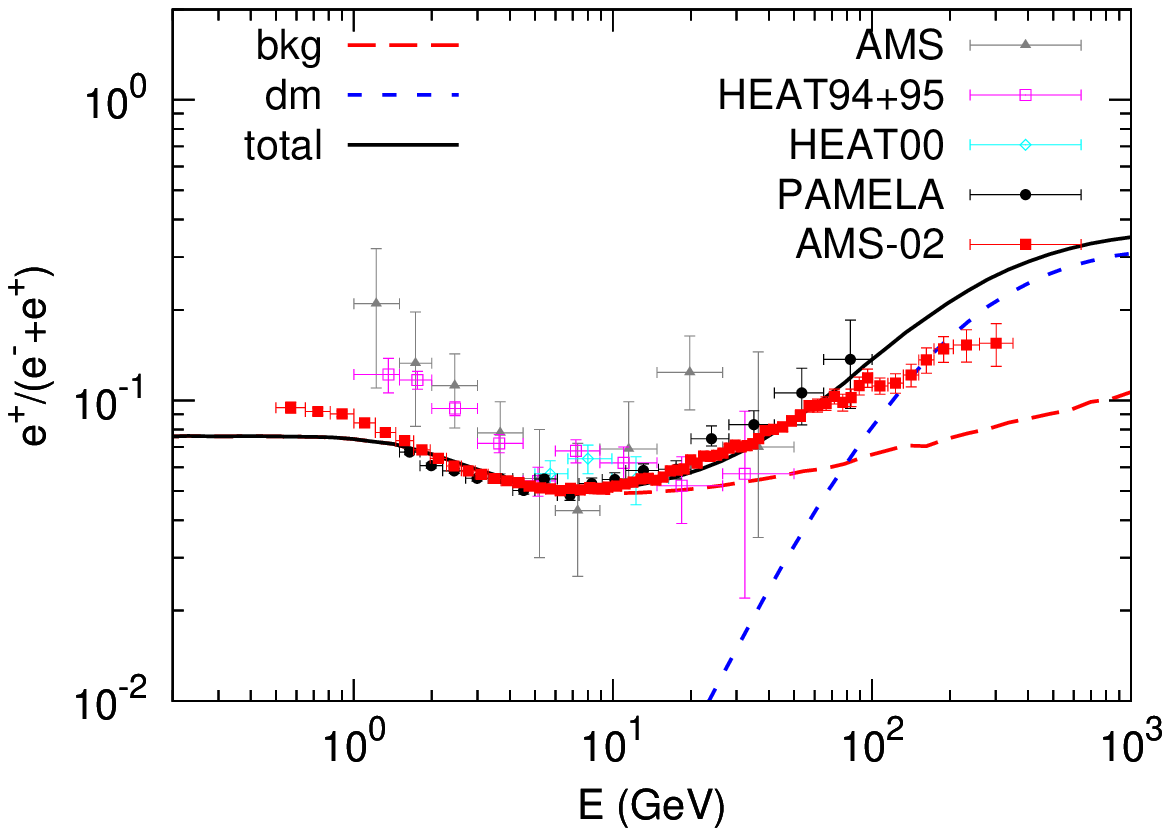}
\includegraphics[width=0.48\textwidth]{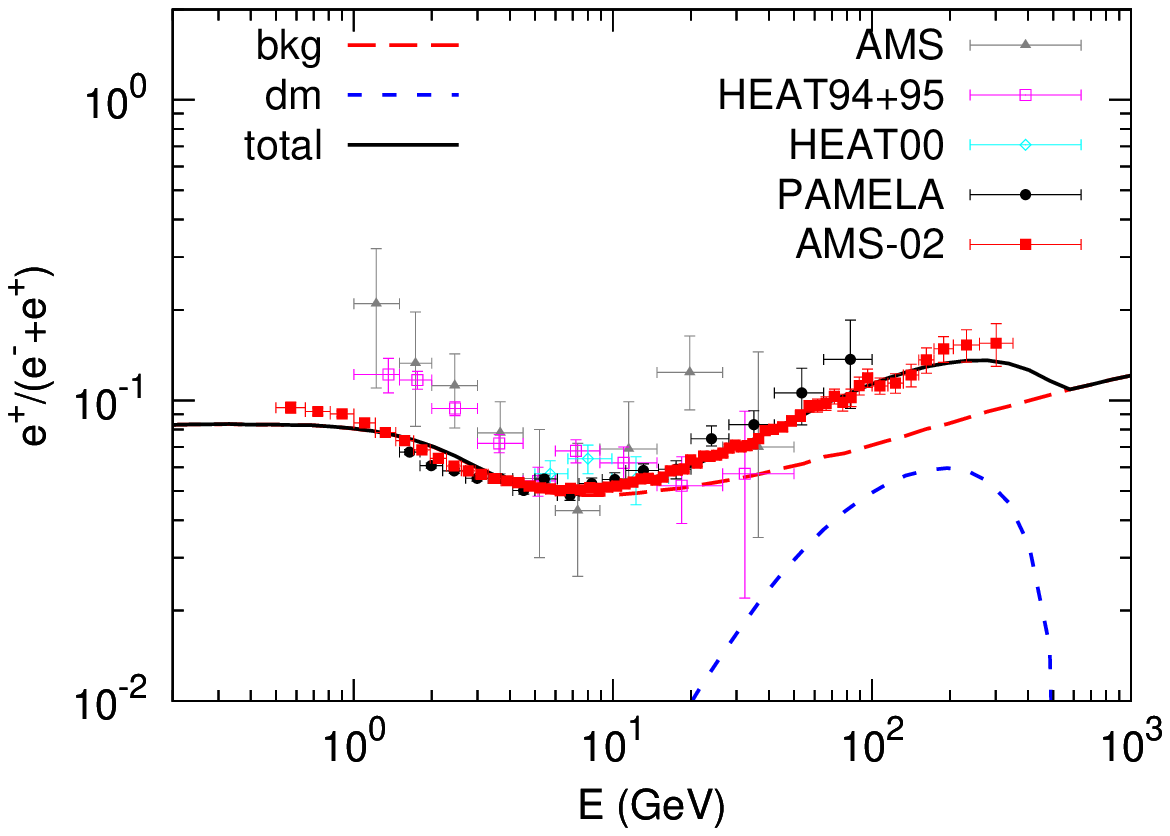}
\includegraphics[width=0.48\textwidth]{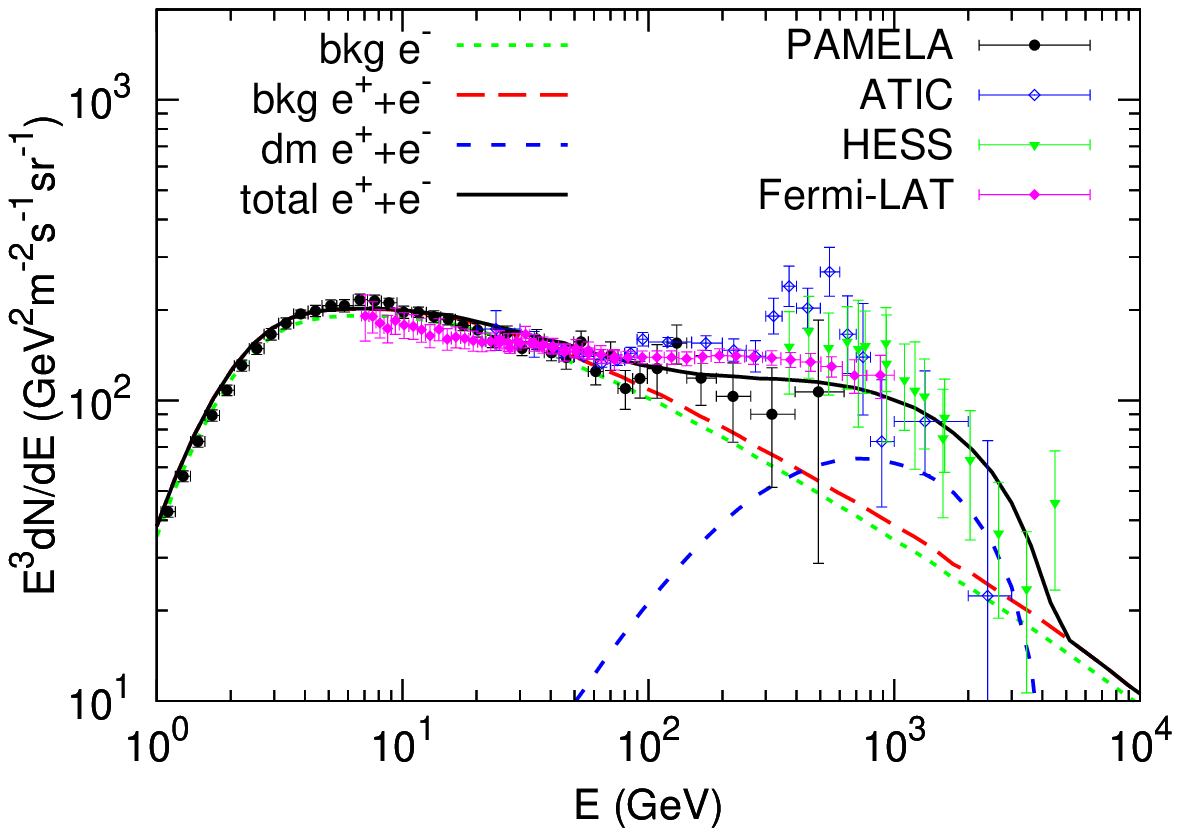}
\includegraphics[width=0.48\textwidth]{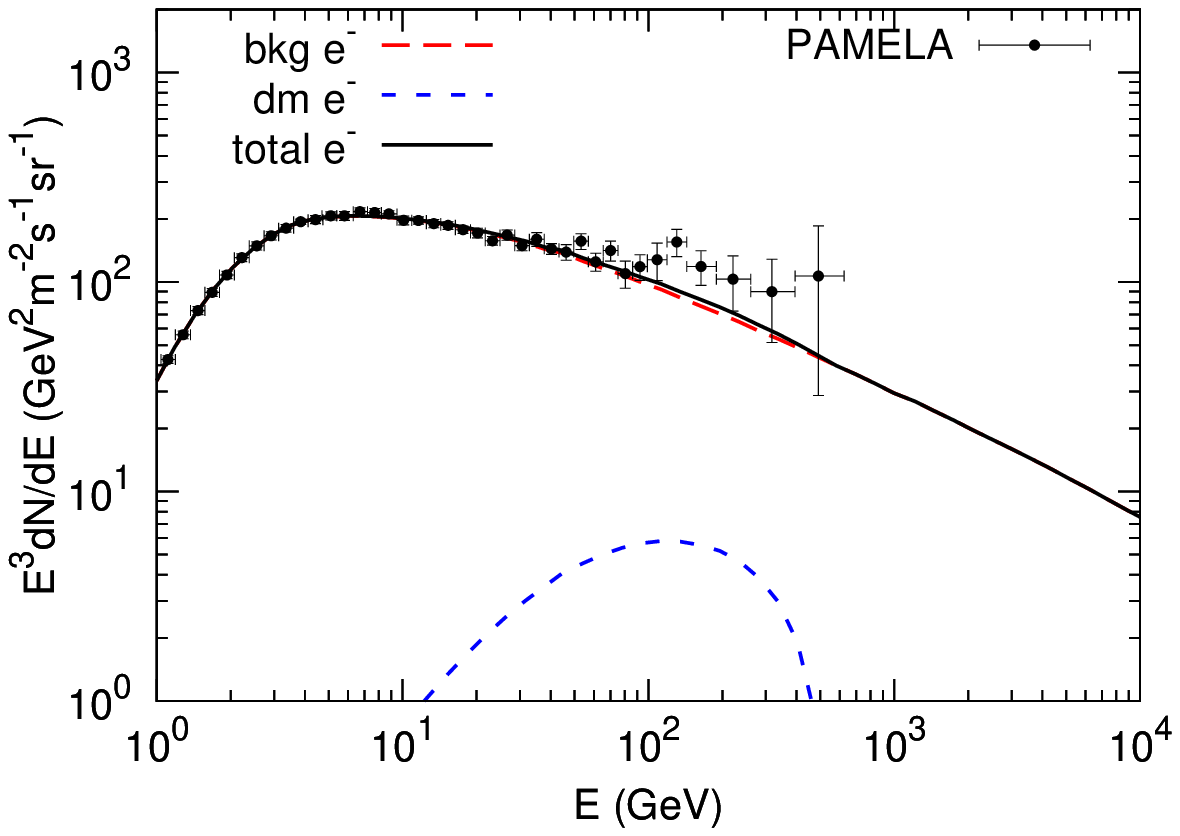}
\caption{Same as Fig. \ref{fig:psr_fixp} but the exotic $e^{\pm}$ are
assumed to be from DM annihilation. The annihilation channel is
$\tau^+\tau^-$.
\label{fig:tau_fixp}}
\end{figure*}

We then run the fits I-a, II-a to derive the parameters
of CR electrons/positrons with the best fitting proton spectrum.
The resulting positron fraction and electron spectrum for the
best fitting parameters of each fit, for both the pulsar 
and DM scenarios, are shown in Figs. \ref{fig:psr_fixp} - 
\ref{fig:tau_fixp}. For each figure, the panels from left to 
right correspond to the fits I-a and II-a  respectively.
The best fitting parameters, mean values and their $1\sigma$ 
uncertainties are compiled in Tables \ref{table:psr_fixp} - 
\ref{table:tau_fixp}, and the best fitting $\chi^2$ over the number 
of degree of freedom (dof) are given in Table \ref{table:chi2}.

\begin{table*}[!htb]
\centering
\caption {Fitting results of pulsar model with proton spectrum fixed}
\begin{tabular}{c|rr|rr}
\hline \hline
 & \multicolumn{2}{c|}{I-a} & \multicolumn{2}{c}{II-a}  \\
\hline
 & best & mean & best & mean  \\
\hline
$\log(A_e\footnotemark[1])$ & $-8.978$ & $-8.974\pm0.005$ &  $-8.925$ & $-8.921\pm0.011$ \\
$\gamma_1$  & $1.504$ & $1.512\pm0.010$ &  $1.708$ & $1.704\pm0.084$ \\
$\gamma_2$  & $2.645$ & $2.652\pm0.010$ &  $2.794$ & $2.796\pm0.028$ \\
$\log(p_{\rm br}^e/{\rm MeV})$ & $3.599$ & $3.587\pm0.022$ &  $3.597$ & $3.600\pm0.046$ \\
$\log(A_{\rm psr}\footnotemark[2])$ & $-24.867$ & $-24.918\pm0.146$ &  $-25.257$ & $-25.226\pm0.562$ \\
$\alpha$  & $1.912$ & $1.903\pm0.029$ &  $1.856$ & $1.863\pm0.116$ \\
$\log(p_c/{\rm MeV})$ & $6.640$ & $6.632\pm0.111$ &  $5.927$ & $6.097\pm0.412$ \\
$c_{e^+}$ & $1.272$ & $1.327\pm0.075$ &  $2.206$ & $2.222\pm0.242$ \\
$\phi/{\rm MV}$ & $500$ & $527\pm30$ & $818$ & $830\pm72$ \\
\hline
\hline
\end{tabular}\vspace{3mm}\\
\footnotemark[1]{Normalization at 25 GeV in unit of
cm$^{-2}$s$^{-1}$sr$^{-1}$MeV$^{-1}$.}\\
\footnotemark[2]{Normalization at 1 MeV in unit of
cm$^{-3}$s$^{-1}$MeV$^{-1}$.}
\label{table:psr_fixp}
\end{table*}

From Fig. \ref{fig:psr_fixp} we can see that adding a pulsar component 
can roughly reproduce the AMS-02 positron fraction data and the Fermi/HESS
total electron spectra. However, the fitting seems not good enough. 
The model prediction overproduces the positron fraction compared with 
AMS-02 data but underproduces total $e^+e^-$ spectra compared with 
Fermi/HESS data. This is clearly seen from the fits II-a, in which the 
constraints of Fermi/HESS data are removed. We see that the fit II-a 
gives good fit to the AMS-02 data, but deviate from the Fermi data 
more obviously. It may indicate that the model described in Sec. 3.1
needs to be refined, or there is a {\em tension} between the AMS-02 
positron fraction and the Fermi/HESS electron spectra\footnote{The recently 
reported $e^{\pm}$ spectra measured by AMS-02 do show the discrepancy 
with the Fermi data \cite{2013ICRC-AMS02}, which is however, most
aparent at low energies. We test the fit I-a with only the Fermi data
above 70 GeV, the $\chi^2$/dof is about $147/114$, which means the
tension still exists and the basic conclusion does not change even we
drop the low energy Fermi data. Better determination of the high energy
behavior of the $e^{\pm}$ spectra by AMS-02 is necessary to finally
solve this problem.}. Quantitatively, the best 
fitting $\chi^2$ is about $279$ for fit I-a. For $151$ dof such a $\chi^2$ 
means $\sim6.1\sigma$ deviation from what expected. Similar conclusion
has also been derived in \cite{2009NuPhB.813....1C,2013JCAP...11..026J,
2013PhRvD..88b3013C}.

Compared with the fitting results with PAMELA positron fraction data 
\cite{2012PhRvD..85d3507L}, the spectrum of the pulsar component 
becomes much softer (with power-law index $\sim 1.9$), which may be more 
reasonable according to the pulsar modeling \cite{2012CEJPh..10....1P}.
The contribution of $e^{\pm}$ from the pulsars is also smaller
than previous estimated according to the PAMELA data. Our fit shows
that up to TeV the positron fraction is only $\sim 20\%$, while it is 
more than $30\%$ or even reaching $40\%$ according to the fitting to 
the PAMELA data. 

\begin{table*}[!htb]
\centering
\caption {Fitting results of DM annihilation to $\mu^+\mu^-$ with 
proton spectrum fixed}
\begin{tabular}{c|rr|rr}
\hline \hline
 & \multicolumn{2}{c|}{I-a} & \multicolumn{2}{c}{II-a}  \\
\hline
 & best & mean & best & mean  \\
\hline
$\log(A_e\footnotemark[1])$ & $-8.916$ & $-8.916\pm0.003$ &  $-8.915$ & $-8.918\pm0.006$ \\
$\gamma_1$  & $1.894$ & $1.870\pm0.036$ &  $1.896$ & $1.868\pm0.036$ \\
$\gamma_2$  & $2.839$ & $2.839\pm0.006$ &  $2.904$ & $2.906\pm0.014$ \\
$\log(p_{\rm br}^e/{\rm MeV})$ & $3.608$ & $3.592\pm0.035$ &  $3.692$ & $3.674\pm0.037$ \\
$\log(m_\chi/{\rm GeV})$  & $3.371$ & $3.368\pm0.039$ &  $2.415$ & $2.423\pm0.039$ \\
$\log(\sv/{\rm cm^3s^{-1}})$ & $-22.307$ & $-22.313\pm0.067$ &  $-24.169$ & $-24.166\pm0.069$ \\
$c_{e^+}$ & $2.881$ & $2.881\pm0.030$ &  $3.052$ & $3.038\pm0.047$ \\
$\phi/{\rm MV}$ & $999$ & $996\pm4$ &  $999$ & $991\pm8$ \\
\hline
\hline
\end{tabular}\vspace{3mm}\\
\footnotemark[1]{Normalization at 25 GeV in unit of
cm$^{-2}$s$^{-1}$sr$^{-1}$MeV$^{-1}$.}\\
\label{table:mu_fixp}
\end{table*}

\begin{table*}[!htb]
\centering
\caption {Fitting results of DM annihilation to $\tau^+\tau^-$ with 
proton spectrum fixed}
\begin{tabular}{c|rr|rr}
\hline \hline
 & \multicolumn{2}{c|}{I-a} & \multicolumn{2}{c}{II-a}  \\
\hline
 & best & mean & best & mean \\
\hline
$\log(A_e\footnotemark[1])$ & $-8.915$ & $-8.916\pm0.003$ &  $-8.907$ & $-8.909\pm0.006$ \\
$\gamma_1$  & $1.879$ & $1.878\pm0.037$ &  $1.869$ & $1.817\pm0.070$ \\
$\gamma_2$  & $2.813$ & $2.813\pm0.007$ &  $2.863$ & $2.856\pm0.015$ \\
$\log(p_{\rm br}^e/{\rm MeV})$ & $3.570$ & $3.571\pm0.037$ &  $3.637$ & $3.608\pm0.058$ \\
$\log(m_\chi/{\rm GeV})$  & $3.667$ & $3.665\pm0.045$ & $2.765$ & $2.747\pm0.046$ \\
$\log(\sv/{\rm cm^3s^{-1}})$ & $-21.699$ & $-21.703\pm0.073$ & $-23.261$ & $-23.269\pm0.064$ \\
$c_{e^+}$ & $2.773$ & $2.769\pm0.035$ &  $2.900$ & $2.844\pm0.068$ \\
$\phi/{\rm MV}$ & $999$ & $994\pm5$ &  $998$ & $974\pm20$ \\
\hline
\hline
\end{tabular}\vspace{3mm}\\
\footnotemark[1]{Normalization at 25 GeV in unit of
cm$^{-2}$s$^{-1}$sr$^{-1}$MeV$^{-1}$.}\\
\label{table:tau_fixp}
\end{table*}

Figs. \ref{fig:mu_fixp} and \ref{fig:tau_fixp} give the results for the 
DM annihilation scenario. The fitting results are even worse than the
pulsar scenario. For DM annihilation into $\mu^+\mu^-$, the reduced 
$\chi^2$ for fit I-a is as high as $3.3$. The reason of the poor fit is 
that the positron spectrum from DM annihilation to a pair of $\mu^+\mu^-$ 
is too hard. This can be seen from the top-left panel of Fig. 
\ref{fig:mu_fixp}. The DM component will over-produce high energy 
positrons ($>100$ GeV) but under-produce positrons at tens of GeV. 
For fits I-a, heavy DM with mass $\sim2-3$ TeV is required due to the 
constraints of Fermi/HESS data. If we throw away the Fermi/HESS $e^{\pm}$ 
data (fit II-a), we find the AMS-02 data tend to favor lighter DM 
particles. In this case the $10-50$ GeV AMS-02 data can be better 
reproduced, however, the data above $100$ GeV are still difficult to 
be explained. Due to the lack of constraints from high energy
data, this fit may under-estimate the contribution to the $e^{\pm}$
excess from DM annihilation.

The situation for $\tau^+\tau^-$ final state is better. Similar with the 
$\mu^+\mu^-$ channel, the positron spectrum from tauon decay is still 
too hard. Relaxing the constraints from Fermi/HESS data we find a lighter 
DM mass, $\sim 550-750$ GeV is favored. In this case the AMS-02 data can 
be fitted relatively well. It is interesting to note that even only
the AMS-02 data and PAMELA electron data are considered (fit II-a), 
the mass of DM particles can be constrained in a small region. Such a
strong constraint comes from the very high precise AMS-02 data at lower
energy (a few tens GeV). However, since Fermi and HESS do observe
plenty of electrons/positrons up to TeV energies, and no hint of significant
drop of the total $e^+e^-$ spectra is shown below TeV. Therefore, we 
should not take these values too seriously.

\begin{table}[!htb]
\centering
\caption {Summary of fitting $\chi^2$/dof. Note that the AMS-02 data 
above 5 GeV and PAMELA proton data below $150$ GeV are used to 
calculate the $\chi^2$.}
\begin{tabular}{cccccc}
\hline \hline
 & pulsar & DM ($\mu^+\mu^-$) & DM ($\tau^+\tau^-$)\\
\hline
I-a   & 278.7/151 & 506.7/152 & 496.5/152 \\
II-a & 51.5/80  & 83.1/81 & 56.7/81 \\
\hline
I-b   & 288.0/205 & 615.3/206 & 584.6/206 \\
II-b & 83.0/134  & 238.7/135 & 164.3/135 \\
\hline
\hline
\end{tabular}
\label{table:chi2}
\end{table}

\begin{figure*}[!htb]
\centering
\includegraphics[width=0.48\textwidth]{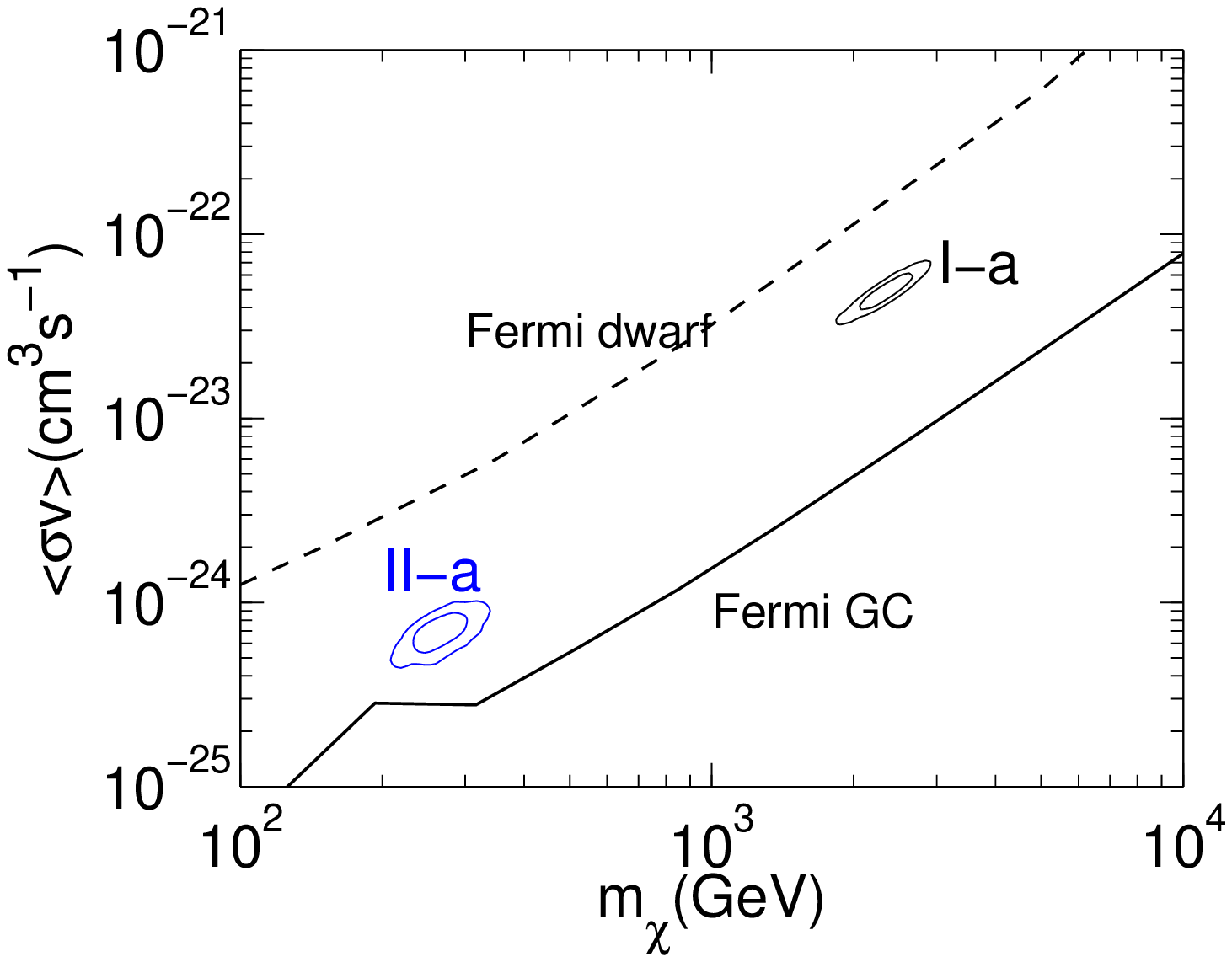}
\includegraphics[width=0.48\textwidth]{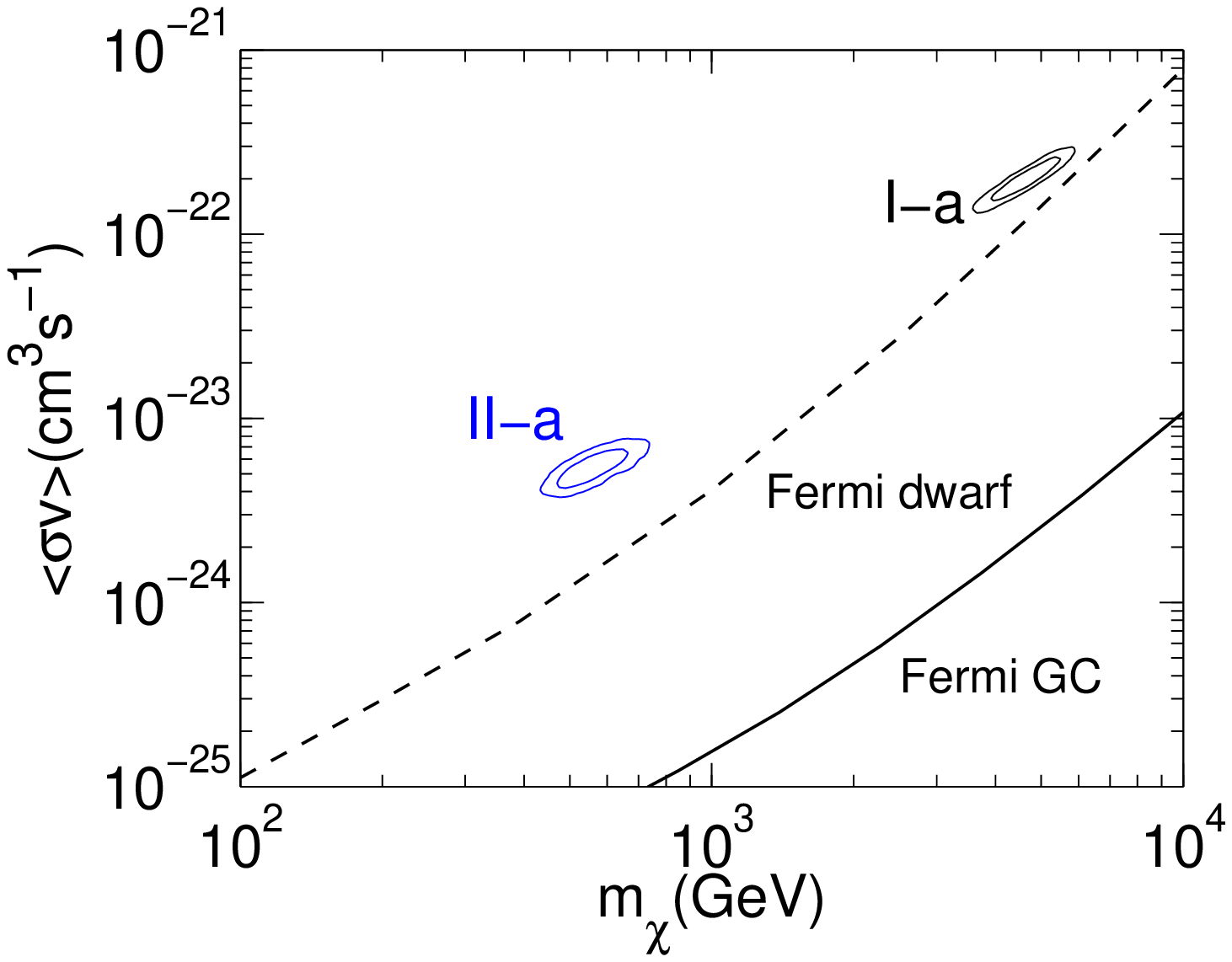}
\caption{$1\sigma$ and $2\sigma$ confidence regions on the DM mass and
cross section plane, for the fits I-a and II-a respectively. The left panel
is for $\mu^+\mu^-$ channel, and the right panel is for $\tau^+\tau^-$
channel. The solid lines show the $95\%$ upper limit of Fermi $\gamma$-ray
observations of the Galactic center (with normalization of the local
density corrected) \cite{2012JCAP...11..048H} and dwarf galaxies 
\cite{Drlica-Wagner2012}.
\label{fig:msv_lep_fixp}}
\end{figure*}

Fig. \ref{fig:msv_lep_fixp} gives the $1\sigma$ and $2\sigma$ 
contour for the DM  mass and annihilation cross section.
But we should keep in mind that such results should not be considered 
statistically meaningful as the fits are quite bad. The solid lines 
shown in Fig. \ref{fig:msv_lep_fixp} are the exclusion limits derived 
by the Fermi $\gamma$-ray observations of the Galactic center 
\cite{2012JCAP...11..048H} and dwarf galaxies \cite{Drlica-Wagner2012}.
We can see that $\gamma$-rays tend to give strong constraints on the DM 
scenario, especially for the $\tau^+\tau^-$ final state. Note, 
however, the Galactic center results may suffer from uncertainties
from the density profile. When calculating the $\gamma$-ray constraints 
the inverse Compton scattering component from the muon/tauon decaying 
electrons/positrons is not included, therefore these constraints should 
be somehow conservative.

It should be pointed out that fits II without Fermi/HESS data
may under-estimate the contribution to the $e^{\pm}$ fluxes from the
extra sources. The preliminary data of the electron spectrum by AMS-02 
extend to $\sim500$ GeV following the median values of PAMELA data 
without any features \cite{2013ICRC-AMS02} also favors the existance 
of $e^{\pm}$ excesses up to sub-TeV. Therefore we may be cautious
to use the fits II to interprete the data because the lack of 
constraints from high energies will lead to improper understanding
of the physics. The results of fits II may help understand the 
behaviors derived in fits I.

We further note that for the DM scenario, the parameter $\phi$ is very 
large. The solar modulation potential is assumed to vary between $300$ 
and $1000$ MV in these fits. From Tables \ref{table:psr_fixp} - 
\ref{table:tau_fixp} we see that almost in all cases the modulation 
potential tends to the upper end. This might be inconsistent with the 
fact that PAMELA and AMS-02 work approaching the solar minimum.

\subsection{Relaxing the proton spectrum}

\begin{figure*}[!htb]
\centering
\includegraphics[width=0.48\textwidth]{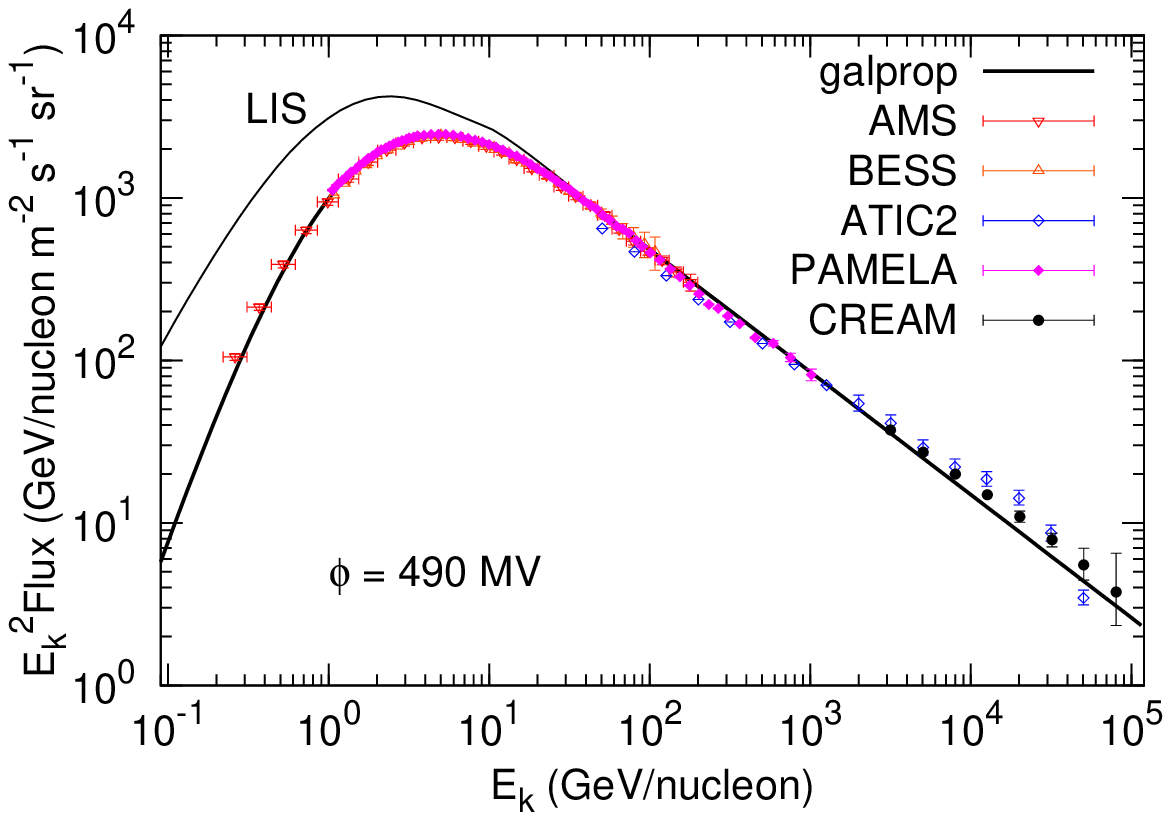}
\includegraphics[width=0.48\textwidth]{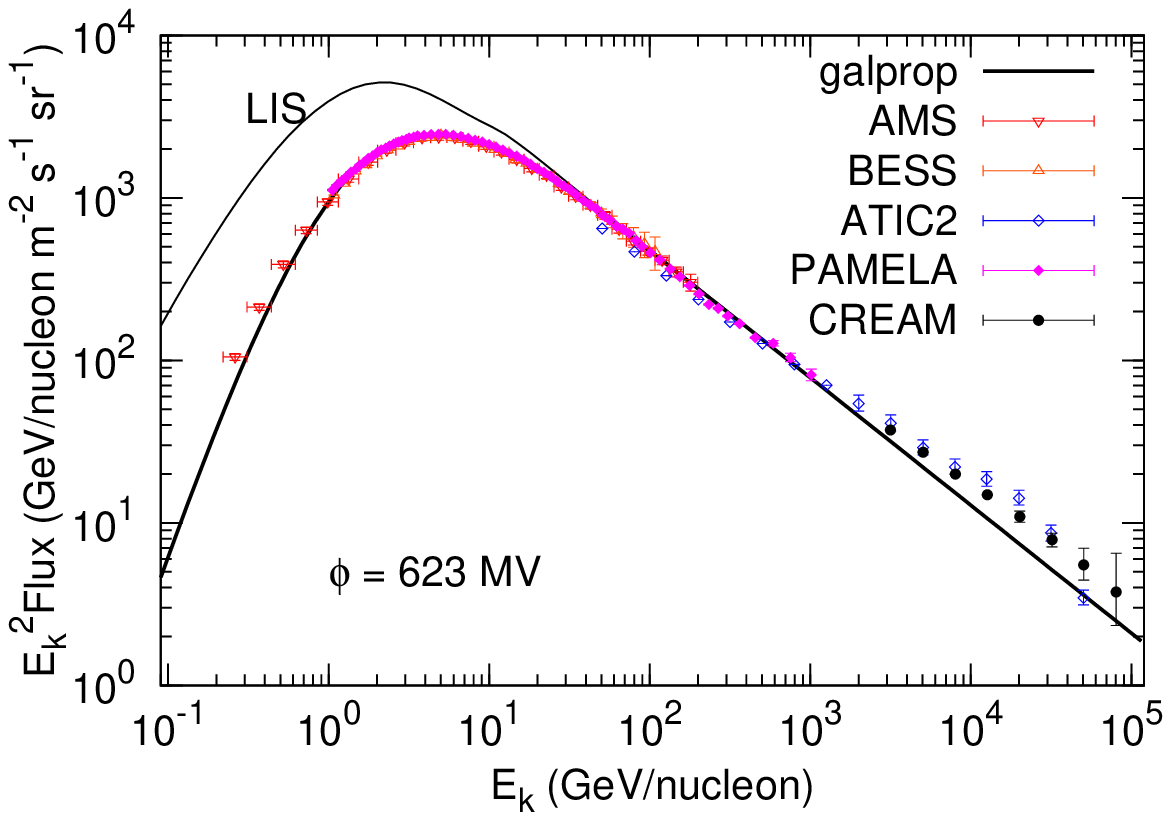}
\includegraphics[width=0.48\textwidth]{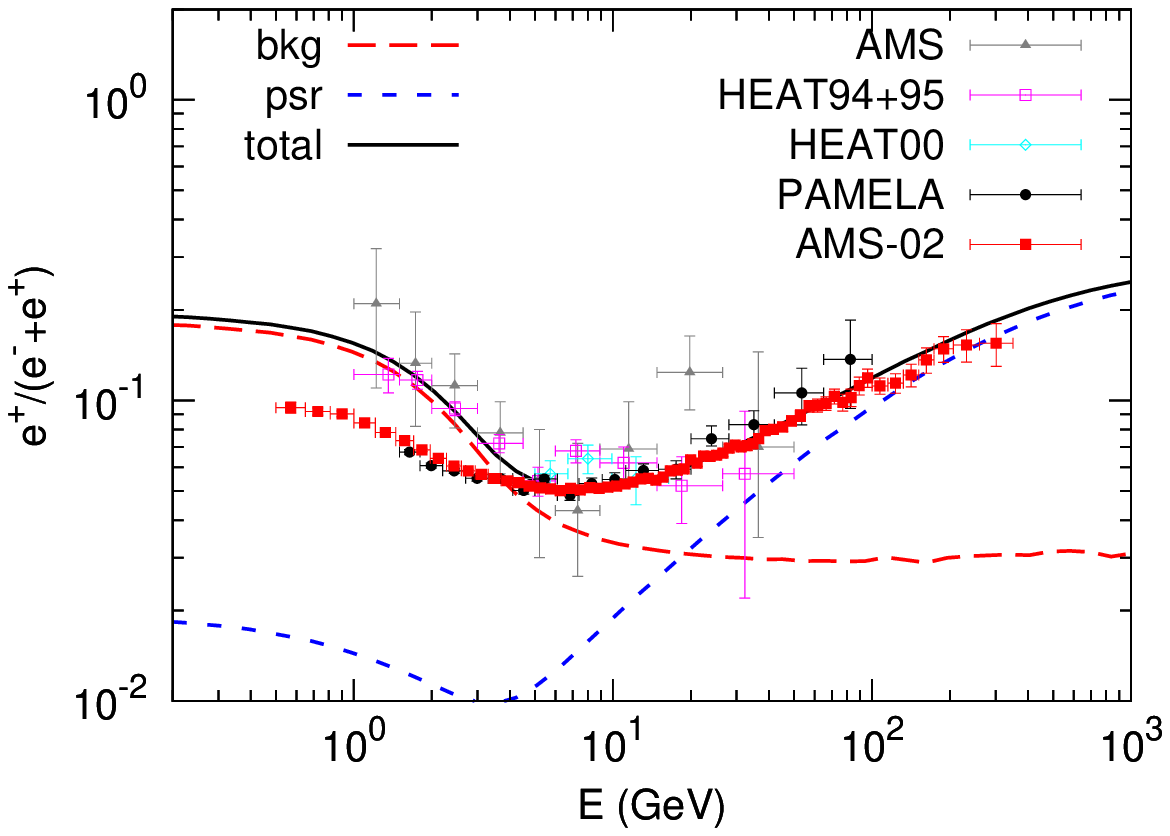}
\includegraphics[width=0.48\textwidth]{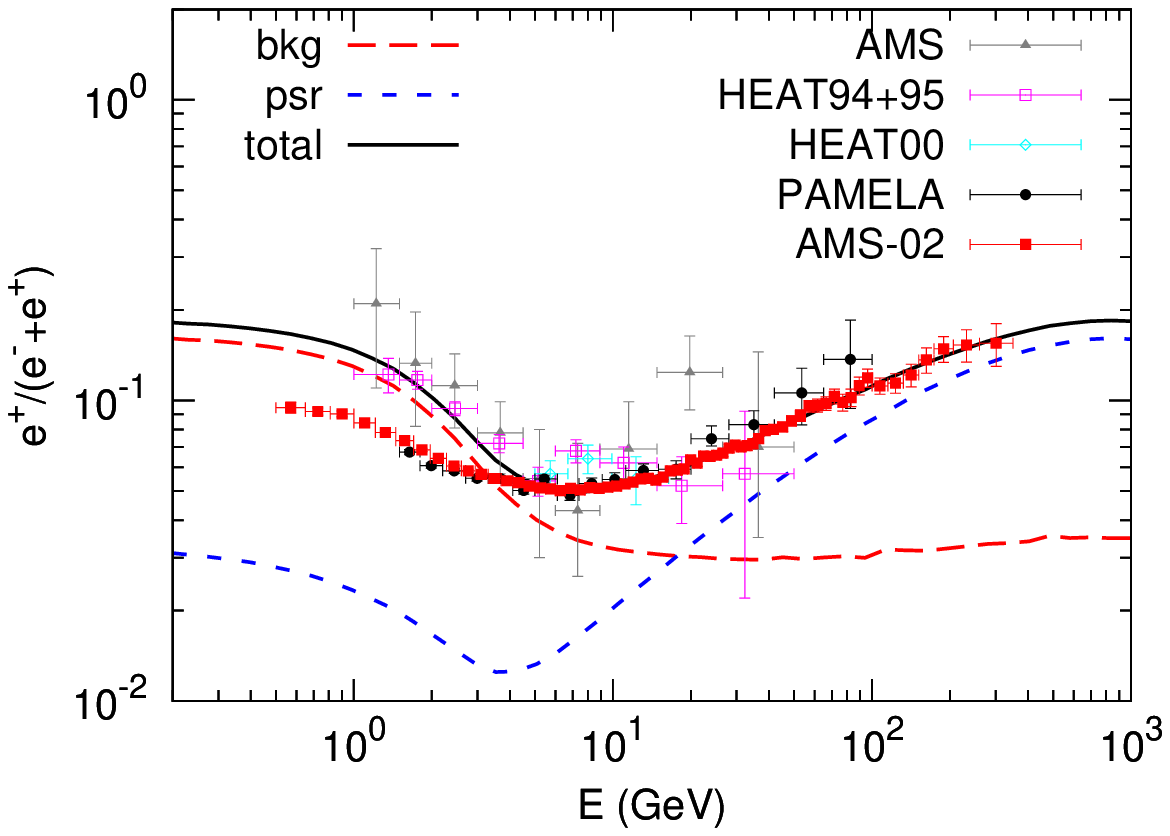}
\includegraphics[width=0.48\textwidth]{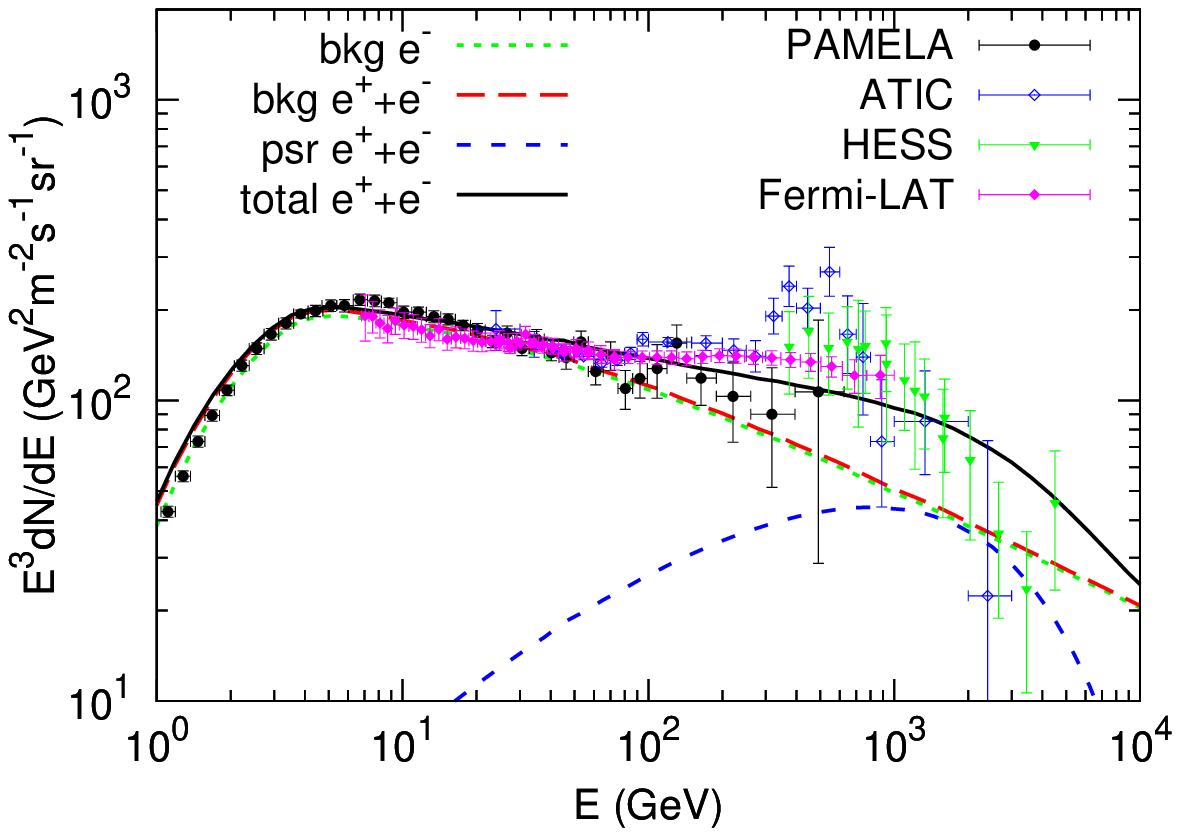}
\includegraphics[width=0.48\textwidth]{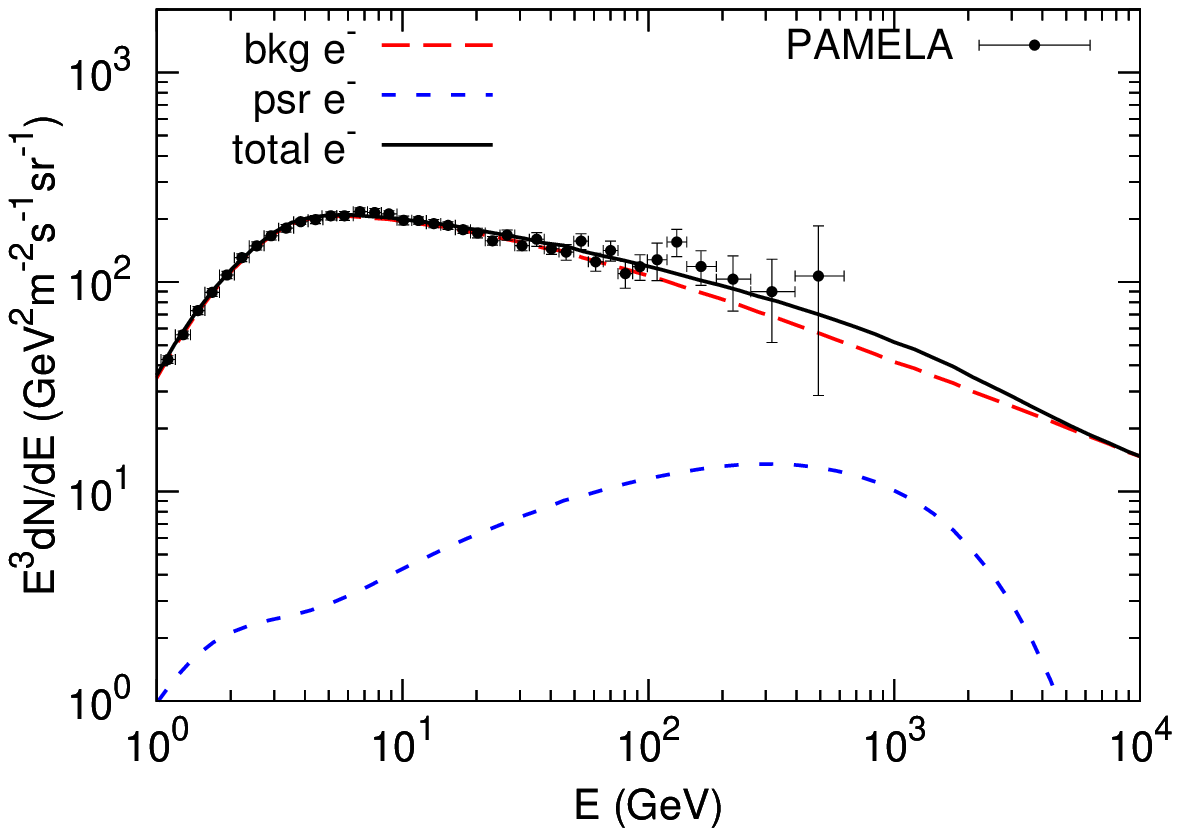}
\caption{From top to bottom: the proton, positron fraction and electron 
spectra for the background together with a pulsar component of
the exotic $e^{\pm}$. The left and right panels are for fits I-b and II-b 
respectively. 
\label{fig:psr}}
\end{figure*}

\begin{figure*}[!htb]
\centering
\includegraphics[width=0.48\textwidth]{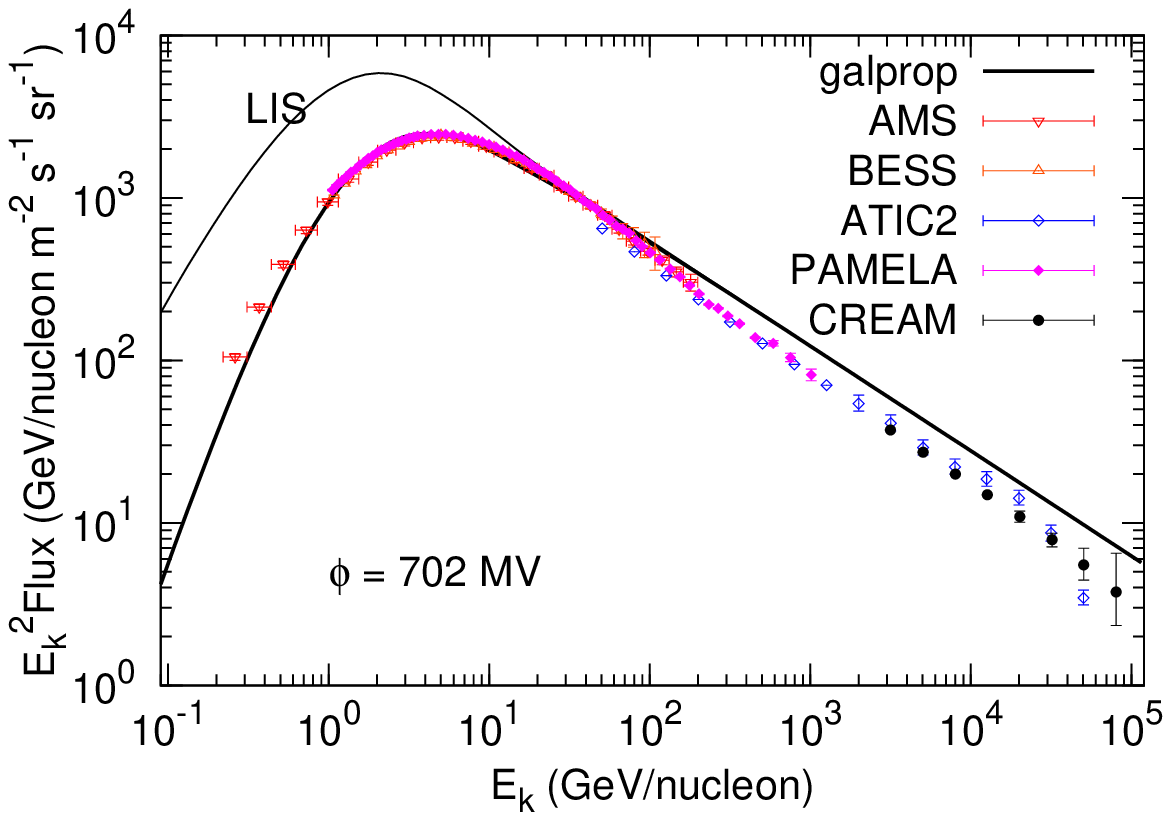}
\includegraphics[width=0.48\textwidth]{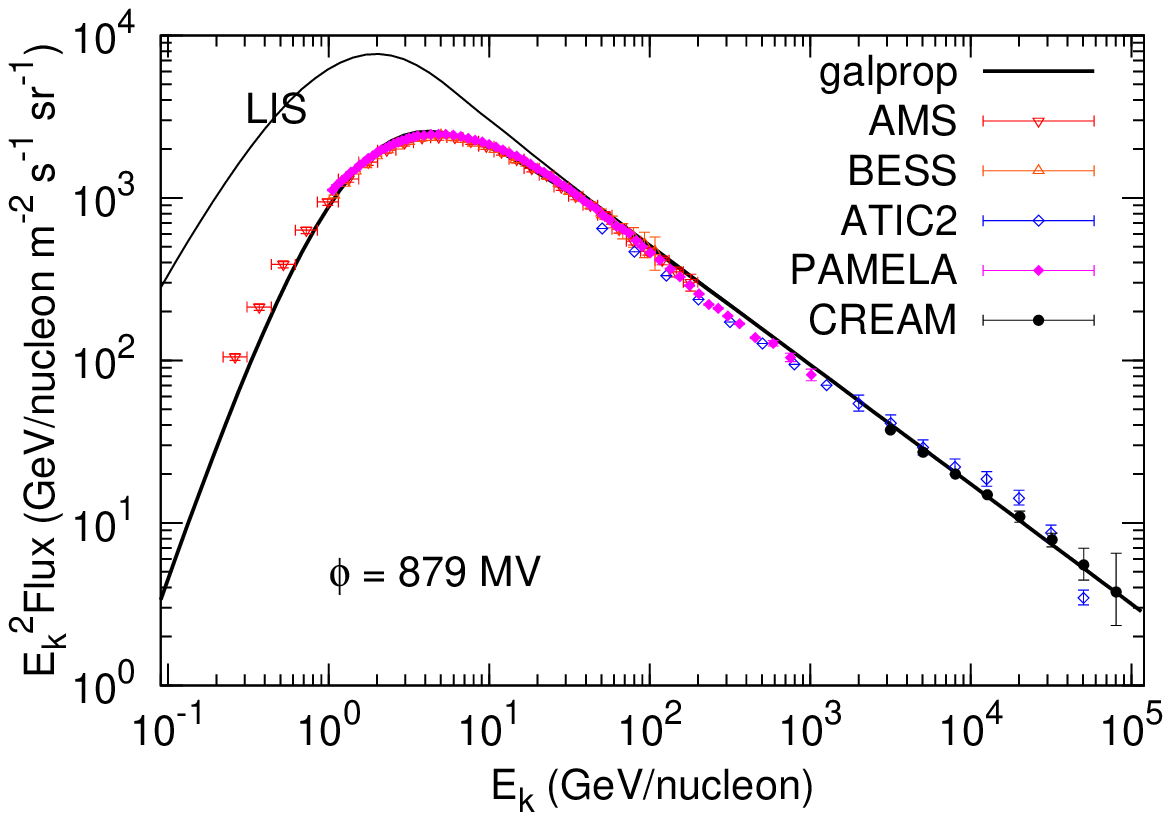}
\includegraphics[width=0.48\textwidth]{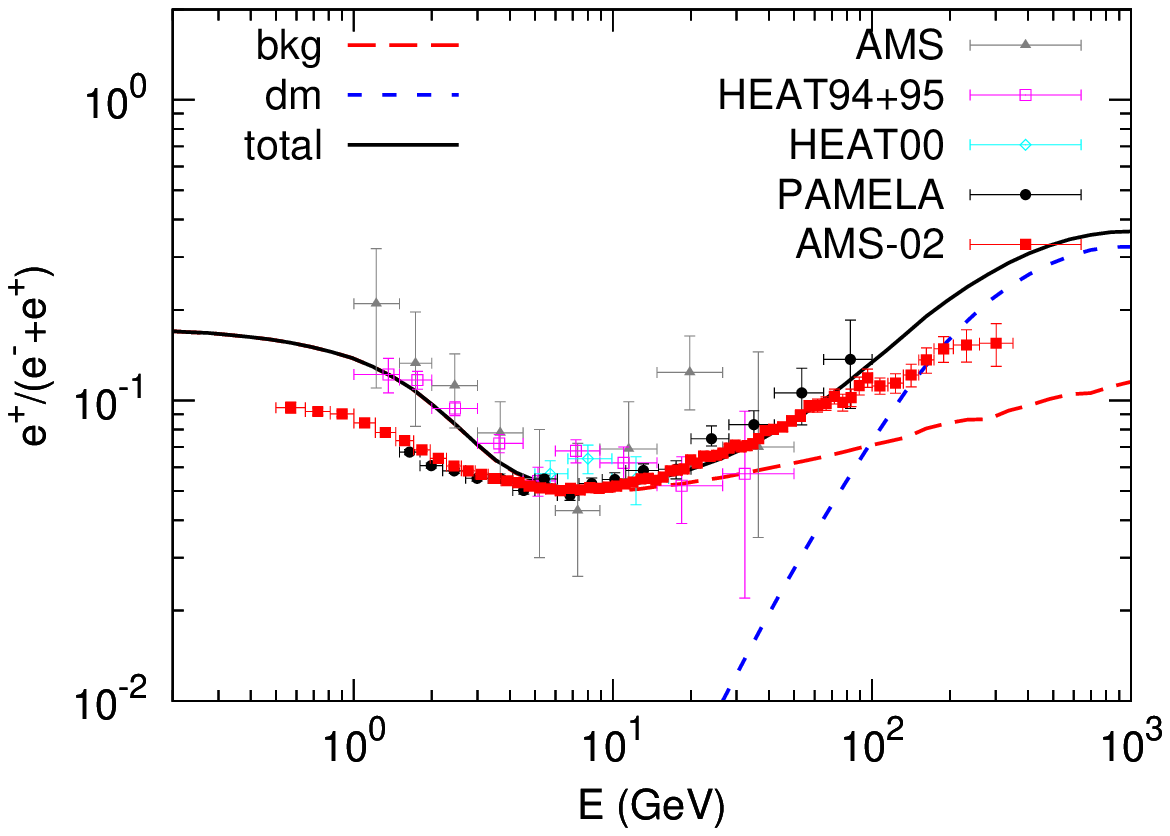}
\includegraphics[width=0.48\textwidth]{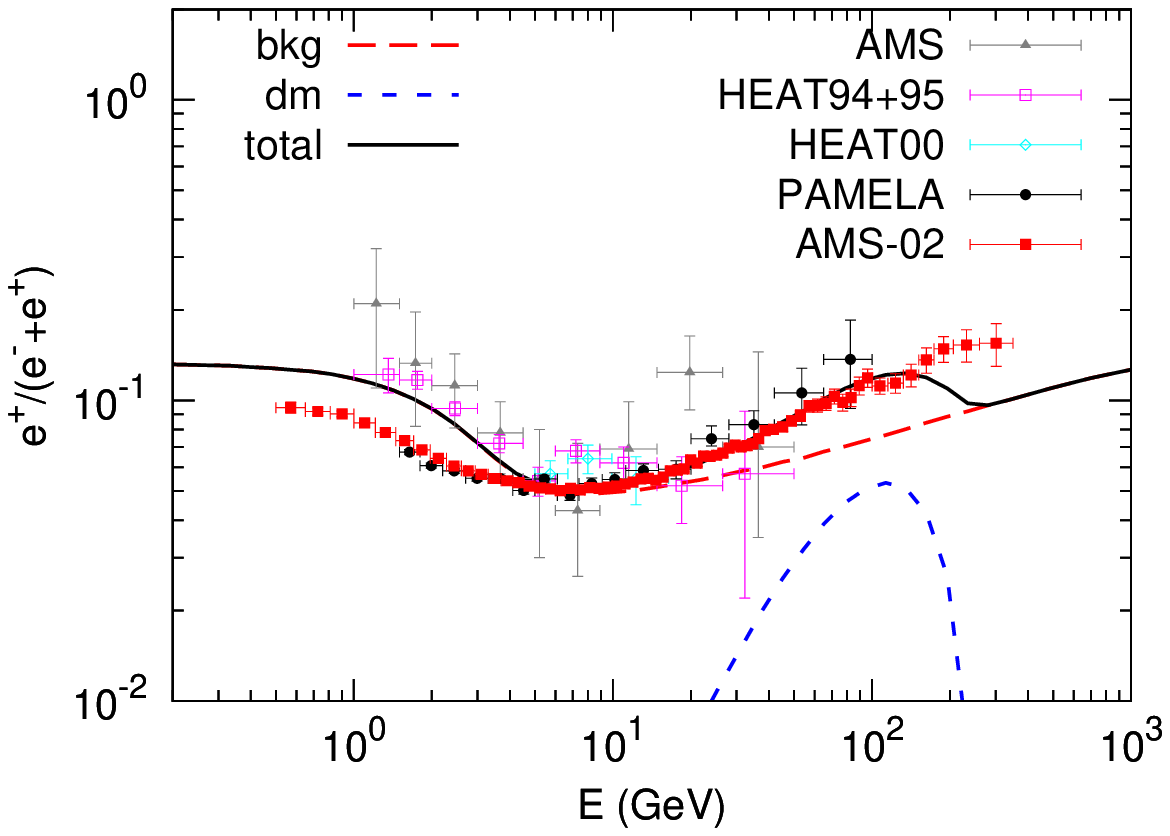}
\includegraphics[width=0.48\textwidth]{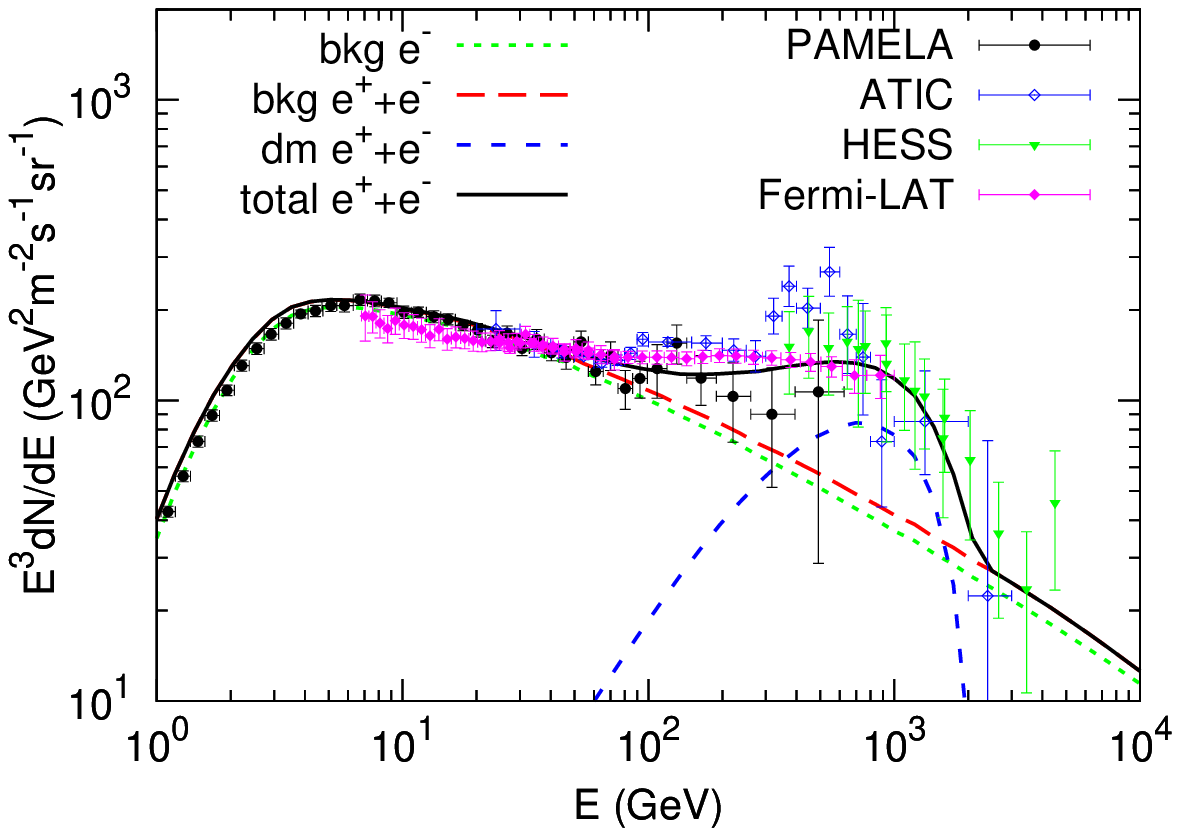}
\includegraphics[width=0.48\textwidth]{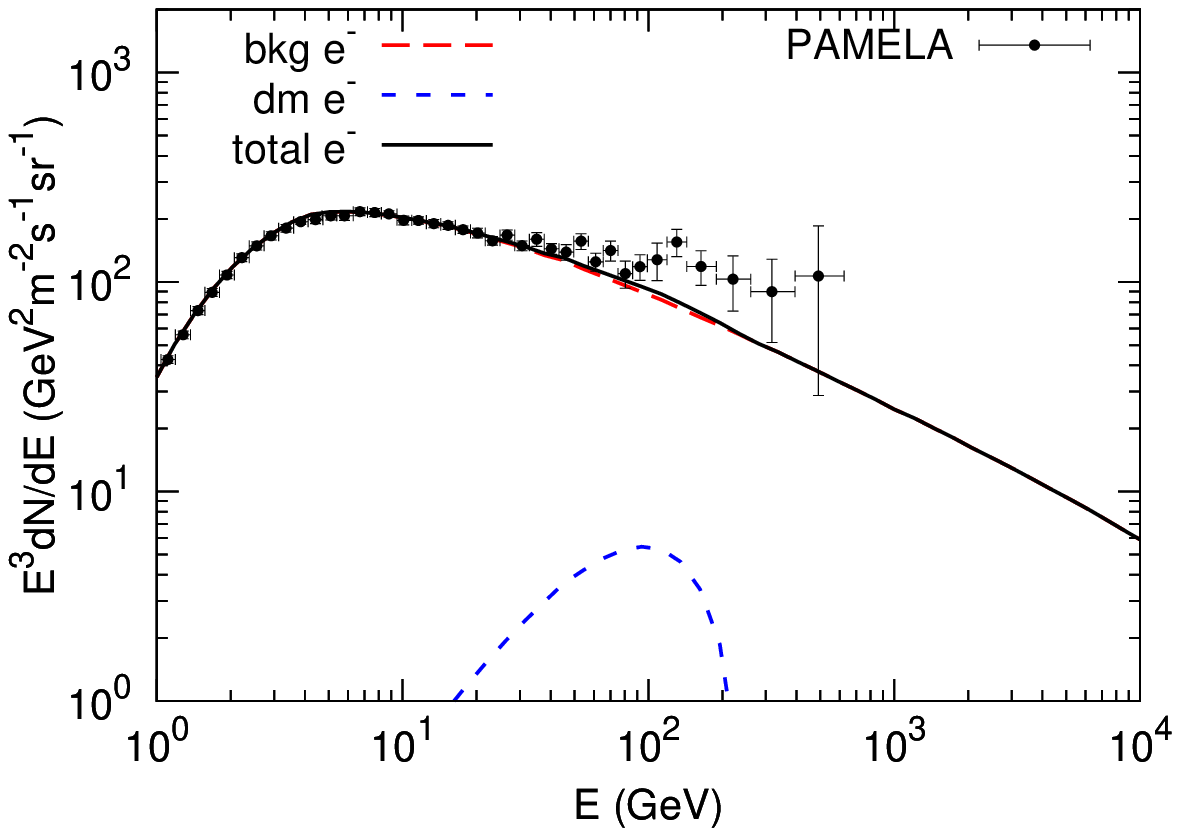}
\caption{Same as Fig. \ref{fig:psr} but the exotic $e^{\pm}$ are
assumed to be from DM annihilation. The annihilation channel is 
$\mu^+\mu^-$.
\label{fig:mu}}
\end{figure*}

\begin{figure*}[!htb]
\centering
\includegraphics[width=0.48\textwidth]{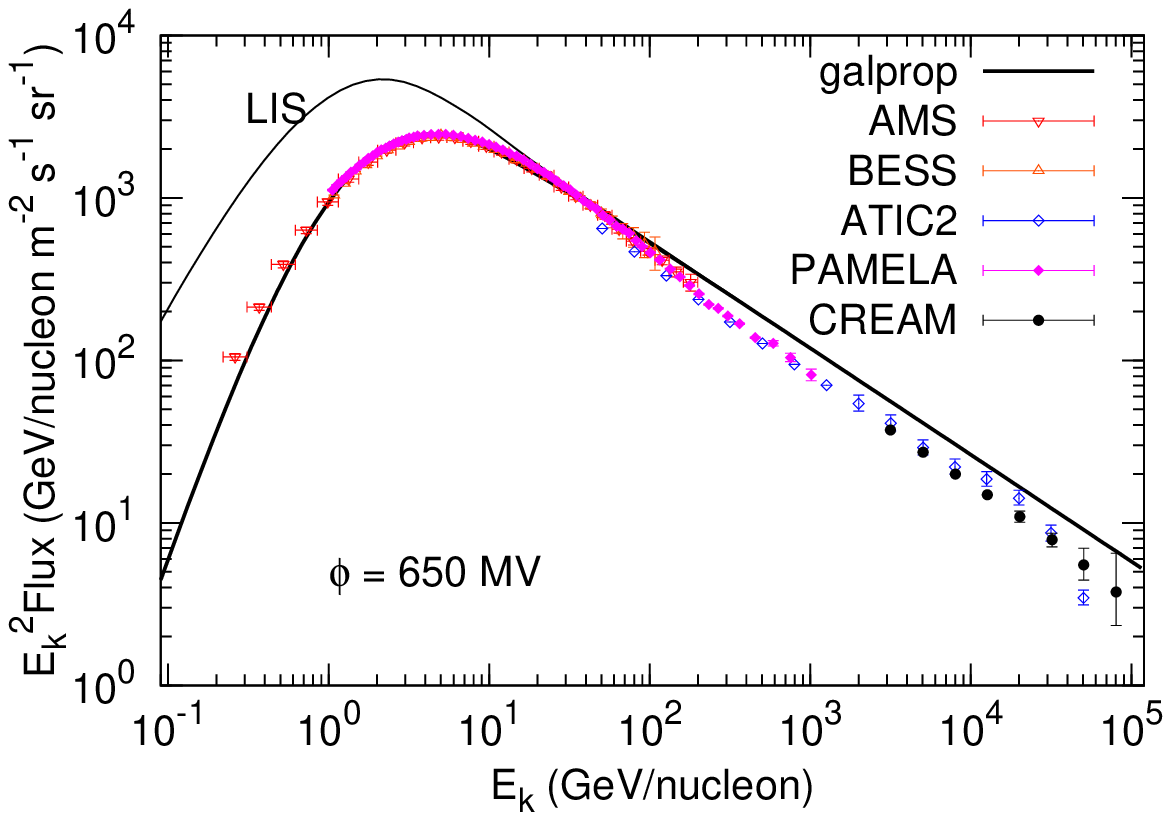}
\includegraphics[width=0.48\textwidth]{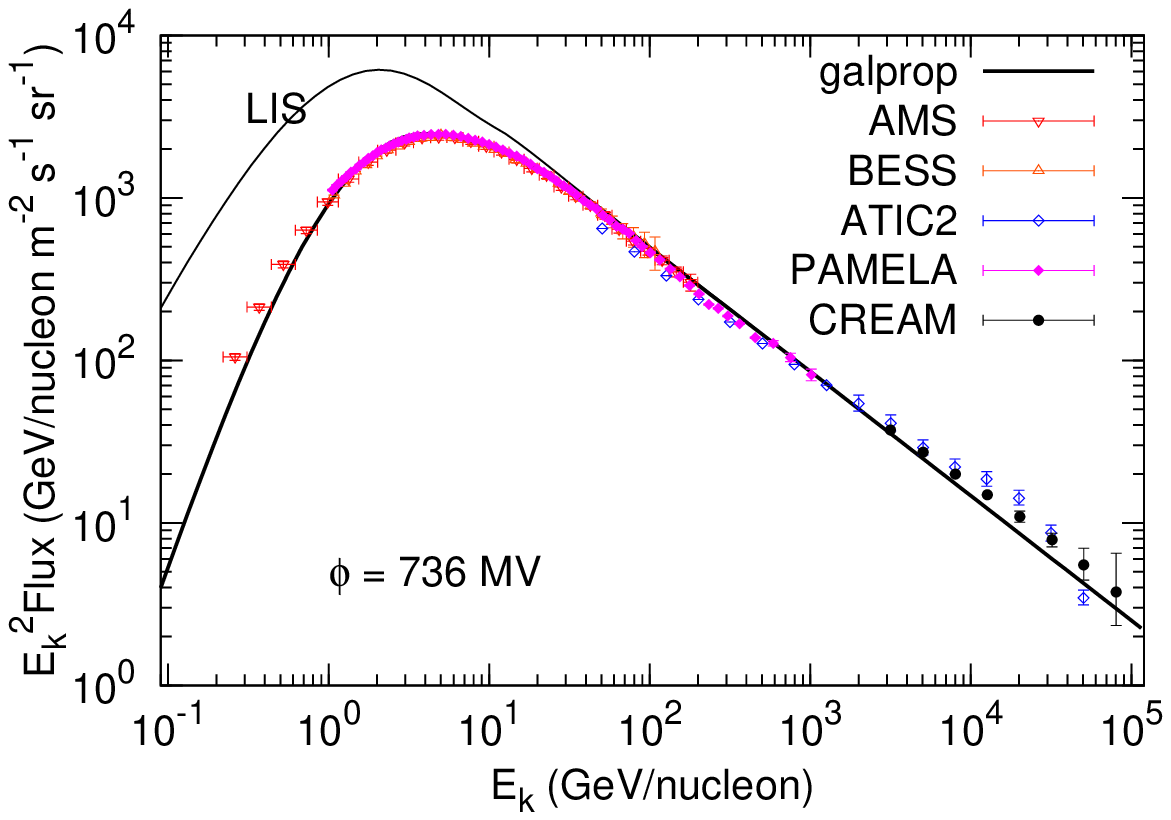}
\includegraphics[width=0.48\textwidth]{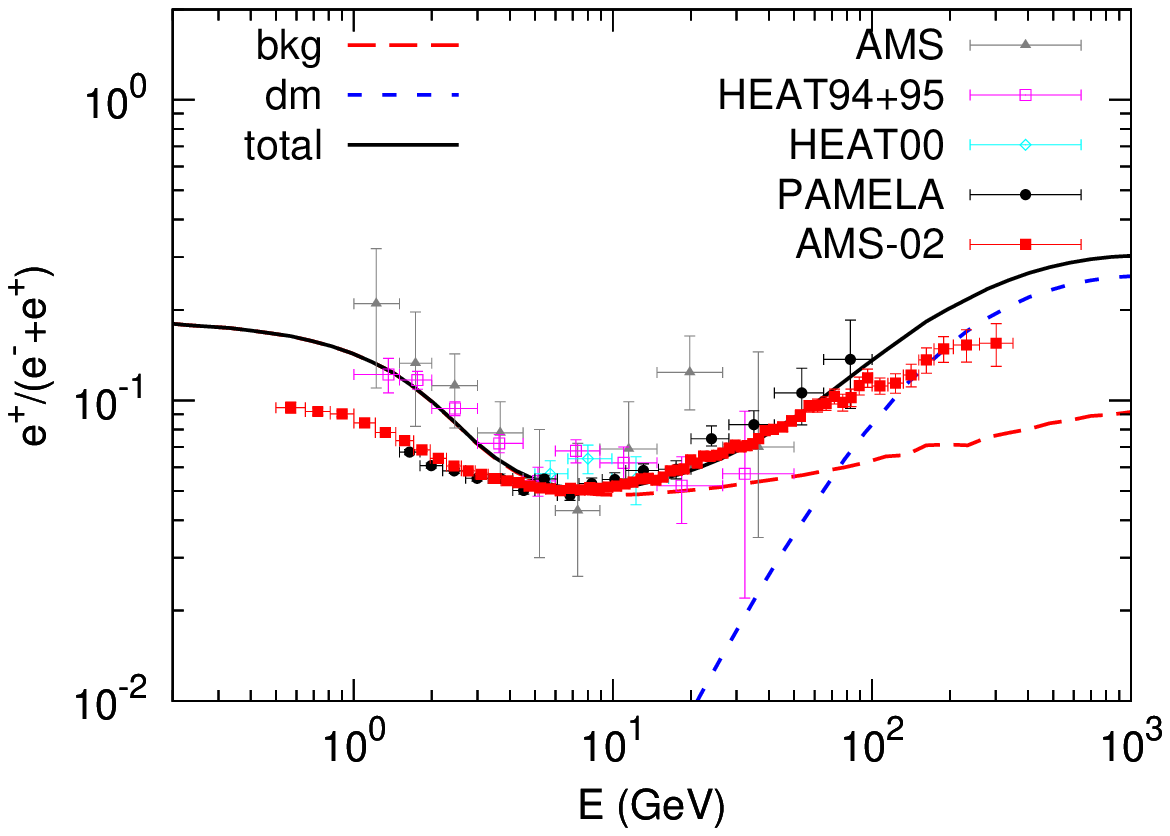}
\includegraphics[width=0.48\textwidth]{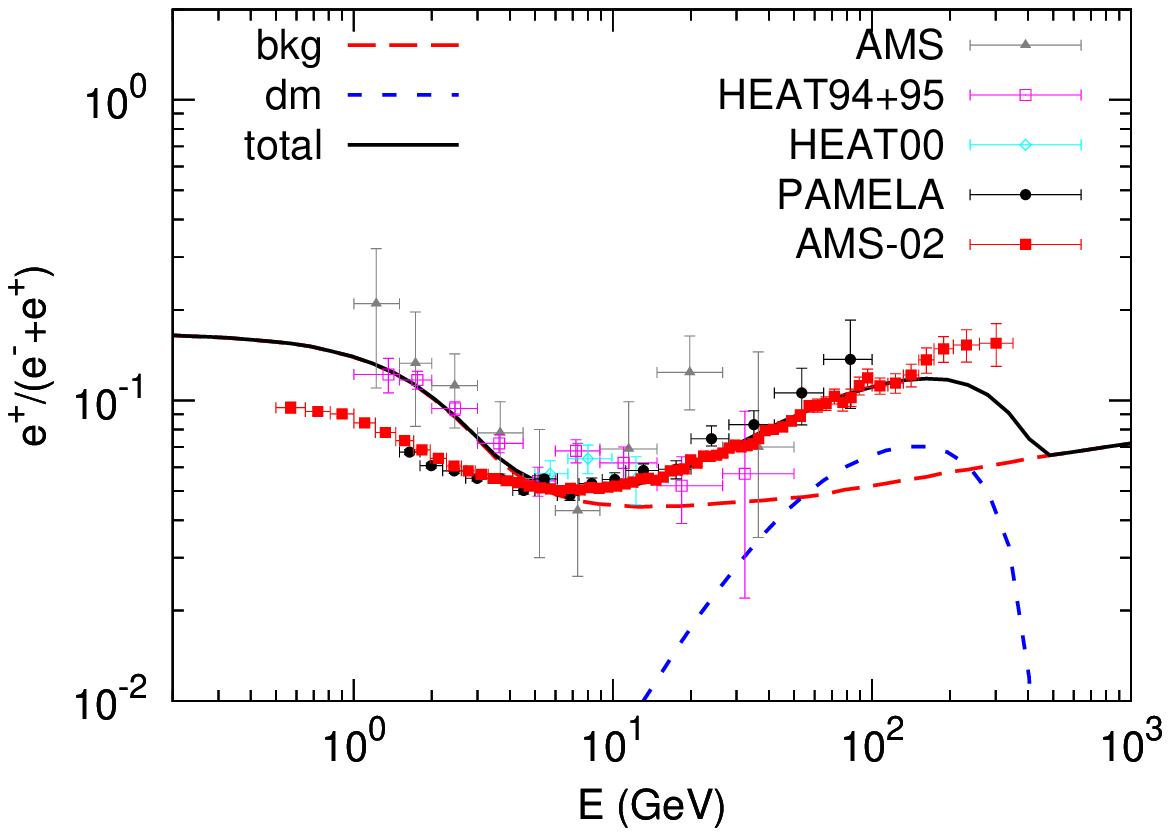}
\includegraphics[width=0.48\textwidth]{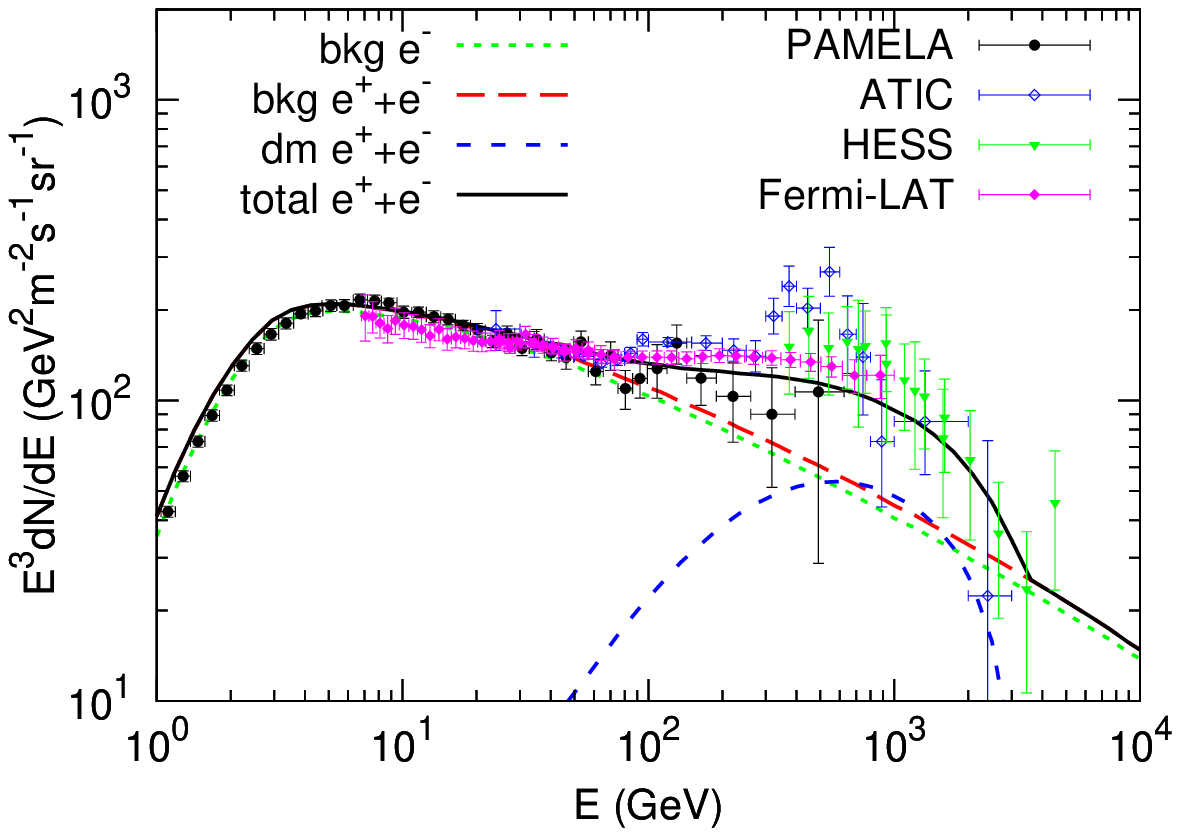}
\includegraphics[width=0.48\textwidth]{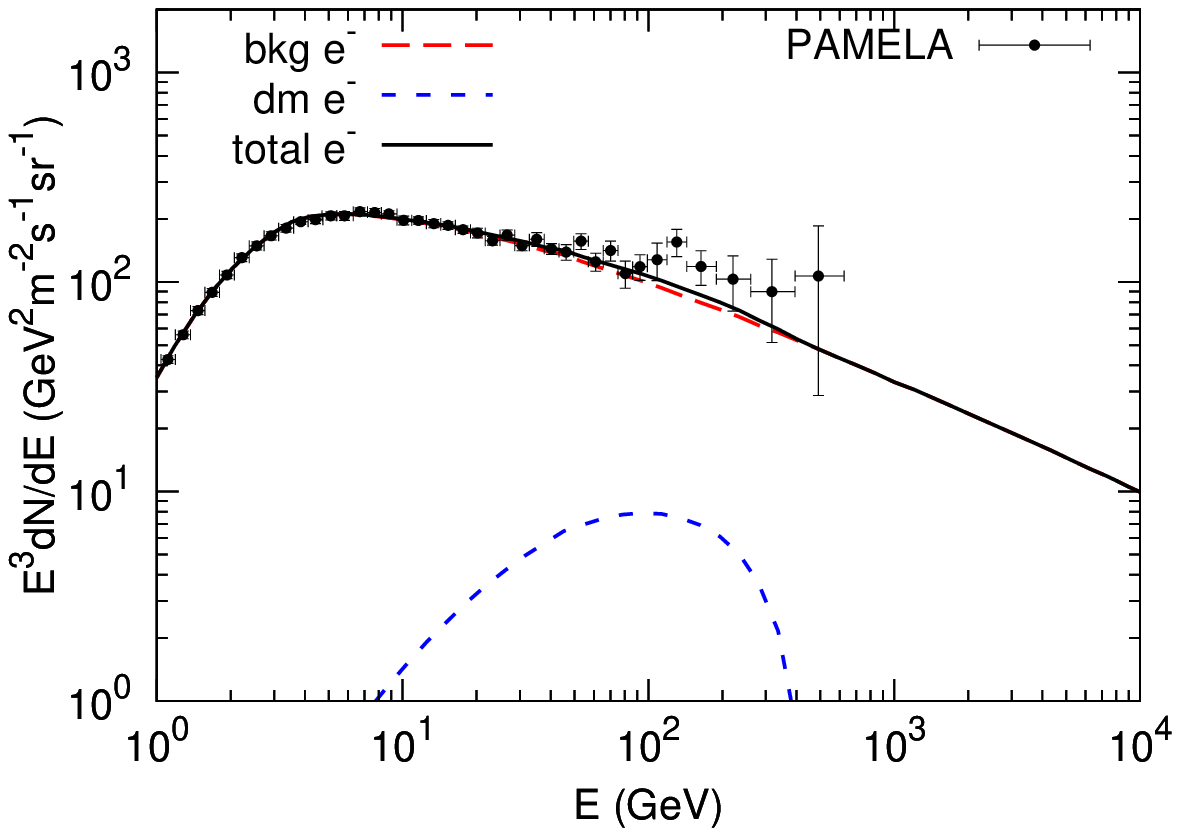}
\caption{Same as Fig. \ref{fig:psr} but the exotic $e^{\pm}$ are
assumed to be from DM annihilation. The annihilation channel is
$\tau^+\tau^-$.
\label{fig:tau}}
\end{figure*}

Since the proton spectrum will affect the secondary positron production, 
and also the determination of the solar modulation parameter, we take the 
proton spectrum into account and redo the fits (labeled as I-b and II-b). 
The fitting results with the best fitting parameters are presented in 
Figs. \ref{fig:psr}-\ref{fig:tau}. The mean values and the $1\sigma$ 
uncertainties of the model parameters are listed in Tables \ref{table:psr}
- \ref{table:tau}, and the best fitting $\chi^2$ over dof are also 
presented in Table \ref{table:chi2}. Qualitatively we find that the 
results are similar with that in the previous subsection. 

\begin{table*}[!htb]
\centering
\caption {Fitting results of pulsar model with proton spectrum relaxed}
\begin{tabular}{c|rr|rr}
\hline \hline
 & \multicolumn{2}{c|}{I-b} & \multicolumn{2}{c}{II-b}  \\ 
\hline
 & best & mean & best & mean \\ 
\hline
$\log(A_p\footnotemark[1])$ & $-8.323$ & $-8.327\pm0.005$ & $-8.328$ & $-8.330\pm0.006$ \\
$\nu_1$  & $1.789$ & $1.797\pm0.019$ & $1.882$ & $1.885\pm0.023$ \\
$\nu_2$  & $2.378$ & $2.388\pm0.011$ & $2.410$ & $2.415\pm0.019$ \\
$\log(p_{\rm br}^p/{\rm MeV})$ & $4.040$ & $4.064\pm0.029$ & $4.099$ & $4.121\pm0.031$ \\
$\log(A_e\footnotemark[2])$ & $-8.977$ & $-8.979\pm0.004$ & $-8.941$ & $-8.948\pm0.008$ \\
$\gamma_1$  & $1.504$ & $1.505\pm0.004$ & $1.535$ & $1.554\pm0.040$ \\
$\gamma_2$  & $2.647$ & $2.645\pm0.011$ & $2.720$ & $2.722\pm0.018$ \\
$\log(p_{\rm br}^e/{\rm MeV})$ & $3.615$ & $3.614\pm0.015$ & $3.614$ & $3.631\pm0.016$ \\
$\log(A_{\rm psr}\footnotemark[3])$ & $-25.104$ & $-25.012\pm0.132$ & $-24.494$ & $-24.411\pm0.335$ \\
$\alpha$  & $1.864$ & $1.881\pm0.026$ &  $1.985$ & $2.006\pm0.068$ \\
$\log(p_c/{\rm MeV})$ & $6.512$ & $6.562\pm0.093$ & $6.205$ & $6.442\pm0.285$ \\
$c_{e^+}$ & $1.306$ & $1.276\pm0.062$ & $1.503$ & $1.468\pm0.127$ \\
$\phi/{\rm MV}$ & $490$ & $489\pm21$ &  $623$ & $614\pm37$ \\
\hline
\hline
\end{tabular}\vspace{3mm}\\
\footnotemark[1]{Normalization at 100 GeV in unit of 
cm$^{-2}$s$^{-1}$sr$^{-1}$MeV$^{-1}$.}\\
\footnotemark[2]{Normalization at 25 GeV in unit of 
cm$^{-2}$s$^{-1}$sr$^{-1}$MeV$^{-1}$.}\\
\footnotemark[3]{Normalization at 1 MeV in unit of 
cm$^{-3}$s$^{-1}$MeV$^{-1}$.}
\label{table:psr}
\end{table*}

For the pulsar scenario, the minimum $\chi^2$ for fit I-b is about 
$288$, which corresponds to a $\sim 3.9\sigma$ deviation from what 
expected for $205$ dof. Compared with the fits of fixed proton 
spectrum, the injection spectrum of positrons from pulsars is a little 
bit softer here. This is because the proton spectrum here is also softer 
than that in the previous subsection.
Therefore a softer pulsar-induced positron spectrum is required to 
give more tens of GeV positrons and to compensate the effect of a 
softer proton spectrum.

\begin{figure*}[!htb]
\centering
\includegraphics[width=0.48\textwidth]{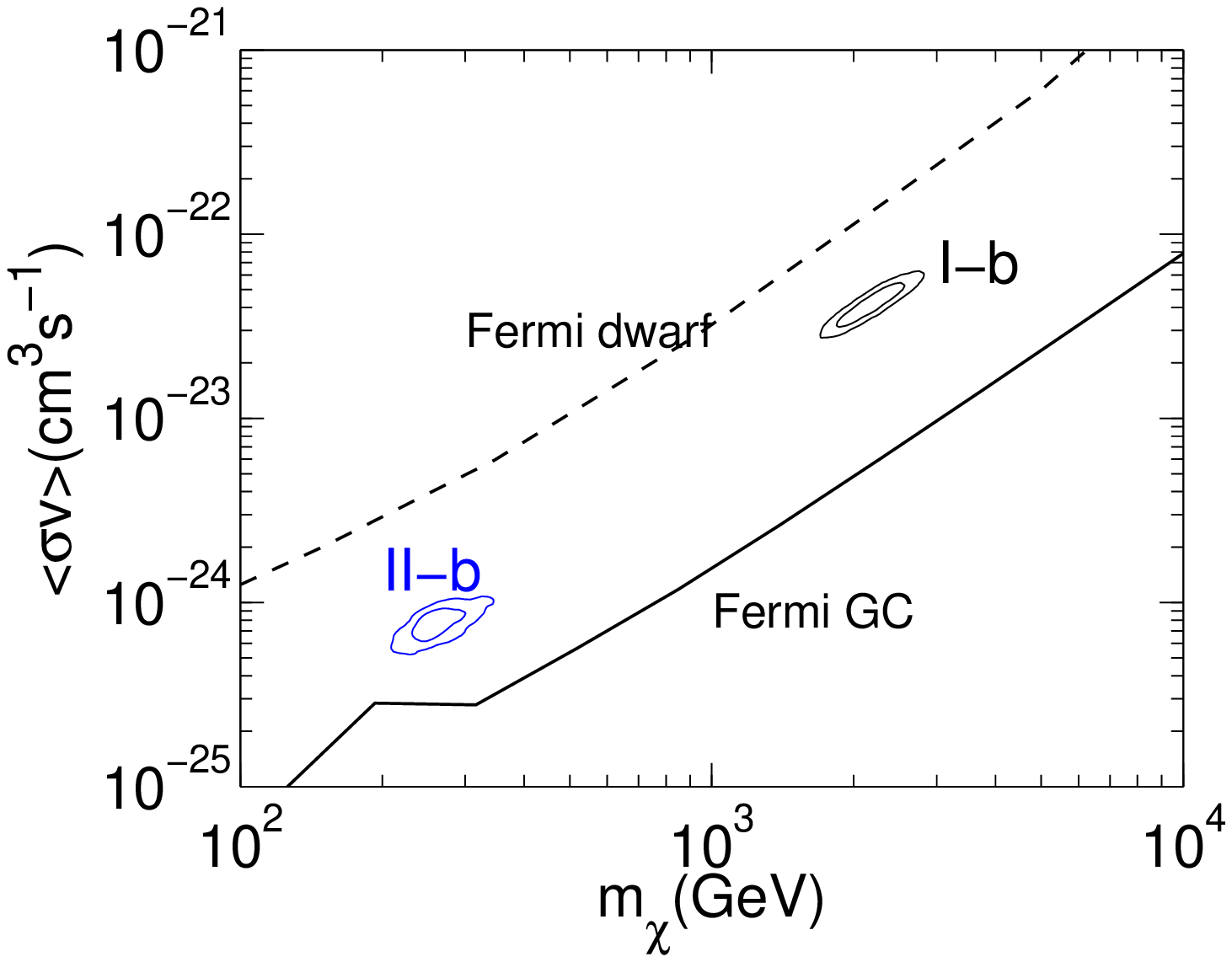}
\includegraphics[width=0.48\textwidth]{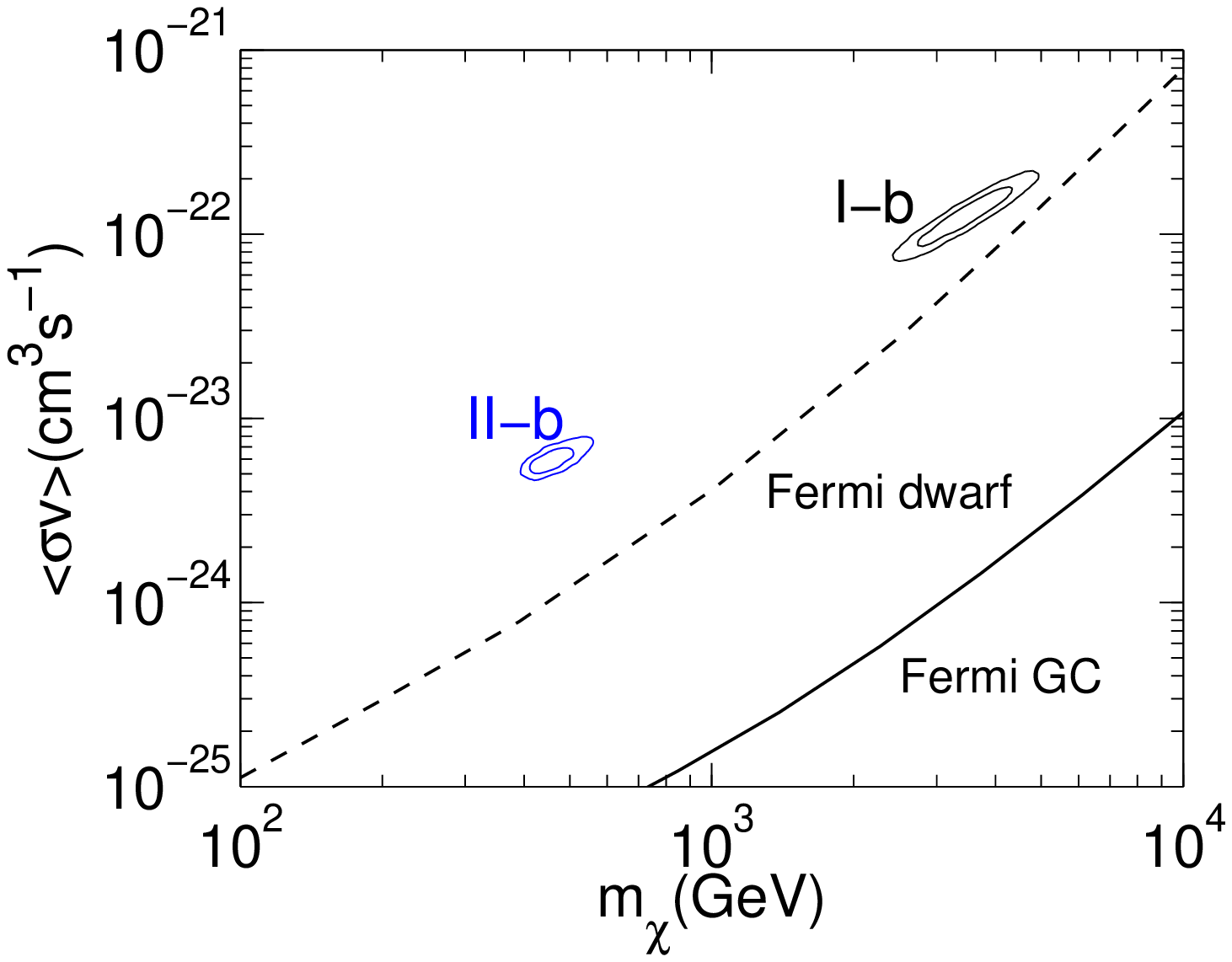}
\caption{$1\sigma$ and $2\sigma$ confidence regions on the DM mass and
cross section plane, for the fits I-b and II-b respectively. The left panel
is for $\mu^+\mu^-$ channel, and the right panel is for $\tau^+\tau^-$ 
channel. The solid lines show the $95\%$ upper limit of Fermi $\gamma$-ray
observations of the Galactic center (with normalization of the local
density corrected) \cite{2012JCAP...11..048H} and dwarf galaxies 
\cite{Drlica-Wagner2012}.
\label{fig:msv_lep}}
\end{figure*}

The DM scenario fits worse than the pulsar scenario. As we have discussed, 
the reason is that the DM-induced positron spectrum is too hard. 
The $1\sigma$ and $2\sigma$ confidence level contours on the 
$m_{\chi}-\sv$ plane for the fits are shown in Fig. 
\ref{fig:msv_lep}. The parameter regions differ only slightly from 
that derived in the previous subsection (Fig. \ref{fig:msv_lep_fixp}). 
The strong constraints on the DM model from $\gamma$-rays are not changed.

From Figs. \ref{fig:mu} and \ref{fig:tau} we note that the proton 
spectrum can not be well fitted for the DM scenario. This is also due to
the hard positron spectrum from DM annihilation into muons and tauons.
A harder proton spectrum will produce more positrons above $\sim10$
GeV, which will compensate the lack of positrons from the hard spectrum
of the DM component. If we reduce the constraints from Fermi/HESS data 
(fits II-b), we see that the proton spectrum fits the data better. 

We also note from Figs. \ref{fig:mu} and \ref{fig:tau} that when not 
including Fermi/HESS data the DM scenario does not give a good description 
to the high energy end of the AMS-02 data, which is different from the 
pulsar case shown in Fig. \ref{fig:psr}. This is because the positron 
spectrum is determined once the mass of DM and its annihilation final 
states are given. To have a minimum $\chi^2$ the mass of DM (i.e. the
shape of the positron spectrum) is usually determined by the data at 
tens of GeV where the errors are very small, instead of the behavior 
of the high energy end data.

\begin{table*}[!htb]
\centering
\caption{Fitting results of DM annihilation into $\mu^+\mu^-$
with proton spectrum relaxed}
\begin{tabular}{c|rr|rr}
\hline \hline
 & \multicolumn{2}{c|}{I-b} & \multicolumn{2}{c}{II-b} \\
\hline
 & best & mean & best & mean \\
\hline
$\log(A_p\footnotemark[1])$ & $-8.275$ & $-8.276\pm0.004$ & $-8.299$ & $-8.298\pm0.004$  \\
$\nu_1$  & $1.936$ & $1.947\pm0.031$ & $2.037$ & $2.026\pm0.022$ \\
$\nu_2$  & $2.271$ & $2.281\pm0.009$ & $2.364$ & $2.365\pm0.010$ \\
$\log(p_{\rm br}^p/{\rm MeV})$ & $3.919$ & $3.929\pm0.045$ & $4.102$ & $4.101\pm0.035$ \\
$\log(A_e\footnotemark[2])$ & $-8.953$ & $-8.950\pm0.006$ & $-8.938$ & $-8.944\pm0.008$ \\
$\gamma_1$  & $1.555$ & $1.588\pm0.054$ & $1.789$ & $1.754\pm0.046$ \\
$\gamma_2$  & $2.768$ & $2.773\pm0.011$ & $2.890$ & $2.895\pm0.016$ \\
$\log(p_{\rm br}^e/{\rm MeV})$ & $3.578$ & $3.573\pm0.027$ & $3.679$ & $3.670\pm0.026$ \\
$\log(m_\chi/{\rm GeV})$  & $3.330$ & $3.338\pm0.045$ &  $2.390$ & $2.417\pm0.038$ \\
$\log(\sv/{\rm cm^3s^{-1}})$ & $-22.397$ & $-22.381\pm0.076$ & $-24.142$ & $-24.117\pm0.060$ \\
$c_{e^+}$ & $1.996$ & $2.037\pm0.064$ & $2.541$ & $2.527\pm0.070$ \\
$\phi/{\rm MV}$ & $702$ & $729\pm39$ &  $879$ & $866\pm36$ \\
\hline
\hline
\end{tabular}\vspace{3mm}\\
\footnotemark[1]{Normalization at 100 GeV in unit of
cm$^{-2}$s$^{-1}$sr$^{-1}$MeV$^{-1}$.}\\
\footnotemark[2]{Normalization at 25 GeV in unit of
cm$^{-2}$s$^{-1}$sr$^{-1}$MeV$^{-1}$.}\\
\label{table:mu}
\end{table*}

\begin{table*}[!htb]
\centering
\caption{Fitting results of DM annihilation into $\tau^+\tau^-$
with proton spectrum relaxed}
\begin{tabular}{c|rr|rr}
\hline \hline
 & \multicolumn{2}{c|}{I-b} & \multicolumn{2}{c}{II-b} \\
\hline
 & best & mean & best & mean  \\
\hline
$\log(A_p\footnotemark[1])$ & $-8.278$ & $-8.282\pm0.004$ &  $-8.310$ & $-8.310\pm0.005$  \\
$\nu_1$  & $1.900$ & $1.892\pm0.023$ &  $1.962$ & $1.951\pm0.030$ \\
$\nu_2$  & $2.281$ & $2.285\pm0.008$ &  $2.387$ & $2.381\pm0.012$ \\
$\log(p_{\rm br}^p/{\rm MeV})$ & $3.936$ & $3.920\pm0.032$ &  $4.136$ & $4.101\pm0.037$ \\
$\log(A_e\footnotemark[2])$ & $-8.962$ & $-8.961\pm0.004$ & $-8.940$ & $-8.944\pm0.008$ \\
$\gamma_1$  & $1.512$ & $1.534\pm0.026$ &  $1.637$ & $1.604\pm0.056$ \\
$\gamma_2$  & $2.730$ & $2.731\pm0.011$ &  $2.794$ & $2.792\pm0.021$ \\
$\log(p_{\rm br}^e/{\rm MeV})$ & $3.558$ & $3.569\pm0.020$ &  $3.633$ & $3.610\pm0.027$ \\
$\log(m_\chi/{\rm GeV})$  & $3.555$ & $3.522\pm0.075$ &  $2.657$ & $2.670\pm0.031$ \\
$\log(\sv/{\rm cm^3s^{-1}})$ & $-21.884$ & $-21.929\pm0.116$ & $-23.222$ & $-23.216\pm0.048$ \\
$c_{e^+}$ & $1.856$ & $1.875\pm0.051$ &  $2.167$ & $2.151\pm0.094$ \\
$\phi/{\rm MV}$ & $650$ & $649\pm29$ & $736$ & $733\pm47$ \\
\hline
\hline
\end{tabular}\vspace{3mm}\\
\footnotemark[1]{Normalization at 100 GeV in unit of
cm$^{-2}$s$^{-1}$sr$^{-1}$MeV$^{-1}$.}\\
\footnotemark[2]{Normalization at 25 GeV in unit of
cm$^{-2}$s$^{-1}$sr$^{-1}$MeV$^{-1}$.}\\
\label{table:tau}
\end{table*}

The solar modulation potential $\phi$ and $c_{e^+}$ at the DM case are 
larger than that in the pulsar case. The reason is that the DM spectrum 
is always harder than the pulsar case. To fit the AMS-02 data at tens 
of GeV, which are very precise, larger $\phi$ and $c_{e^+}$ can 
give relatively higher flux of positrons in this energy range.

\subsection{Further tests}

\begin{figure}[!htb]
\centering
\includegraphics[width=0.7\textwidth]{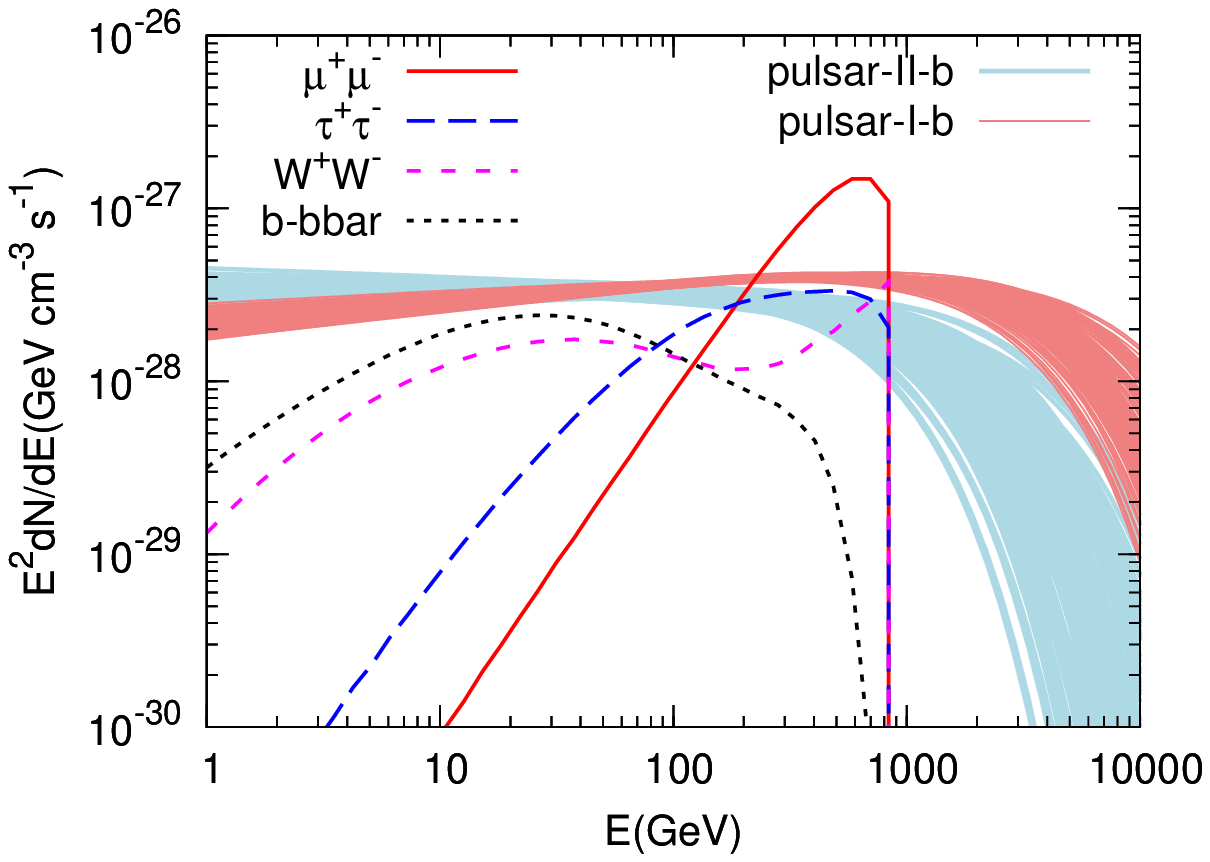}
\caption{Shaded regions are $95\%$ intervals of the exotic positron
source component at the solar location of pulsar models. The 
DM-induced positron spectra for $\mu^+\mu^-$, $\tau^+\tau^-$,
$W^+W^-$ and $b\bar{b}$ channels are shown for comparison.
\label{fig:spec}}
\end{figure}

To better understand how soft a positron spectrum from the extra
sources is needed, we show in Fig. \ref{fig:spec} the $2\sigma$ 
range of source spectra from pulsars at the solar location with
shaded regions. The $2\sigma$ range is defined with $\Delta\chi^2=
\chi^2-\chi^2_{\rm min}=22.7$ for $13$ fitting parameters.
For comparison the DM induced positron spectra for $\mu^+\mu^-$,
$\tau^+\tau^-$, $W^+W^-$ and $b\bar{b}$ channels are also shown.
The mass of DM particle is taken to be $m_{\chi}=1$ TeV and a 
free flux normalization is adopted. It is shown that the 
positron spectra from DM annihilation in the $\mu^+\mu^-$ 
and $\tau^+\tau^-$ channels are much harder than the pulsar component. 
For the $W^+W^-$ and $b\bar{b}$ channels the spectrum is softer and 
we would expect a better fit to the data. 

\begin{figure*}[!htb]
\centering
\includegraphics[width=0.48\textwidth]{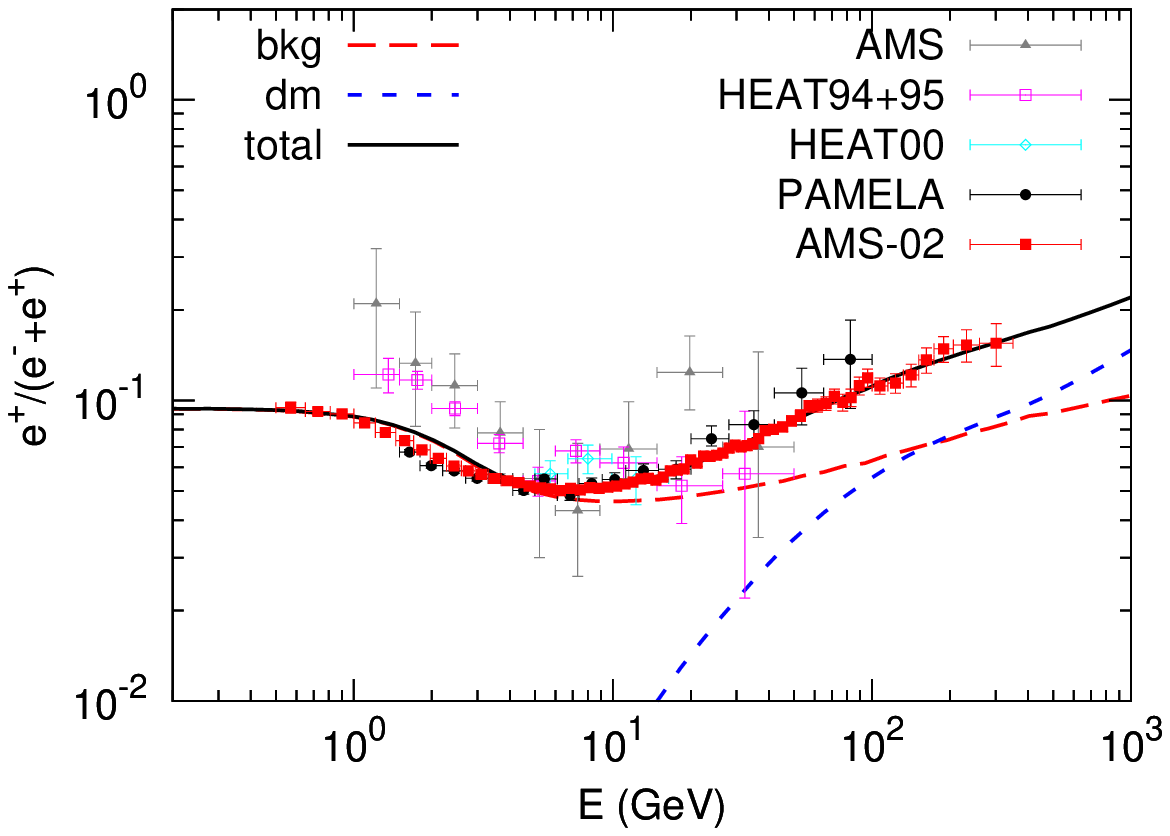}
\includegraphics[width=0.48\textwidth]{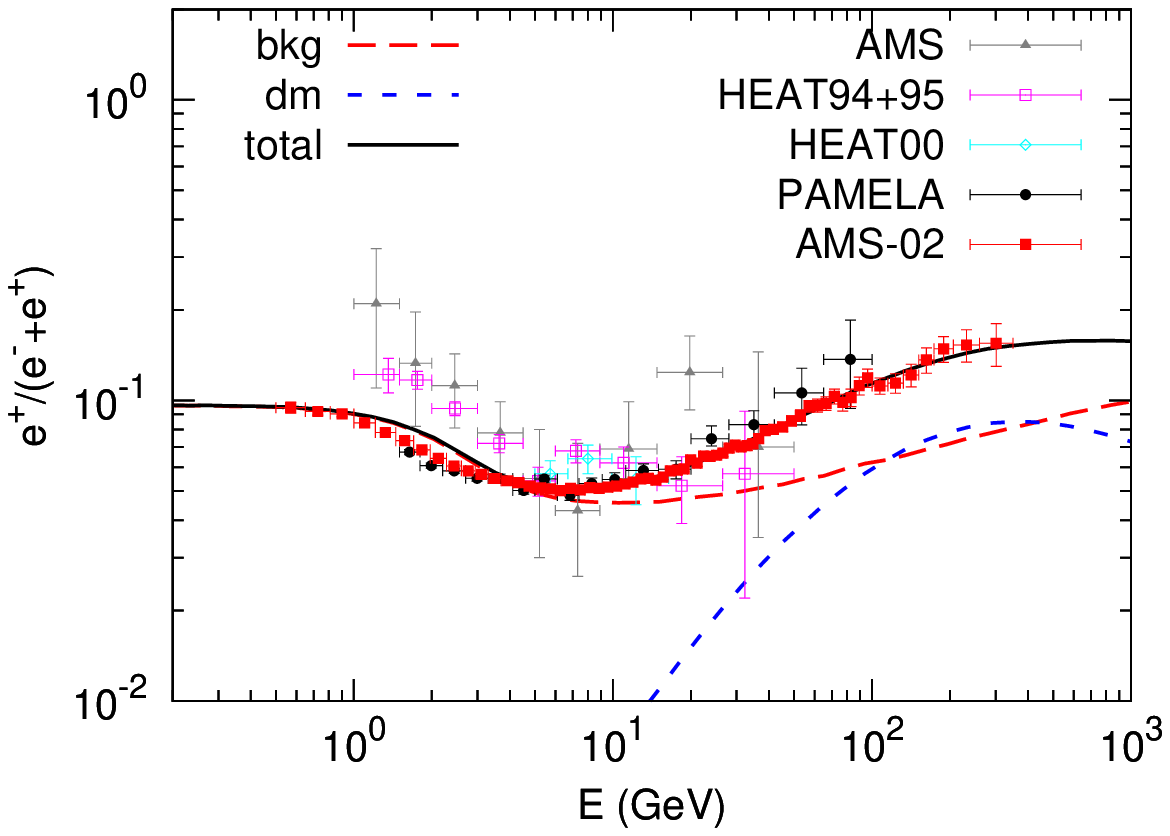}
\includegraphics[width=0.48\textwidth]{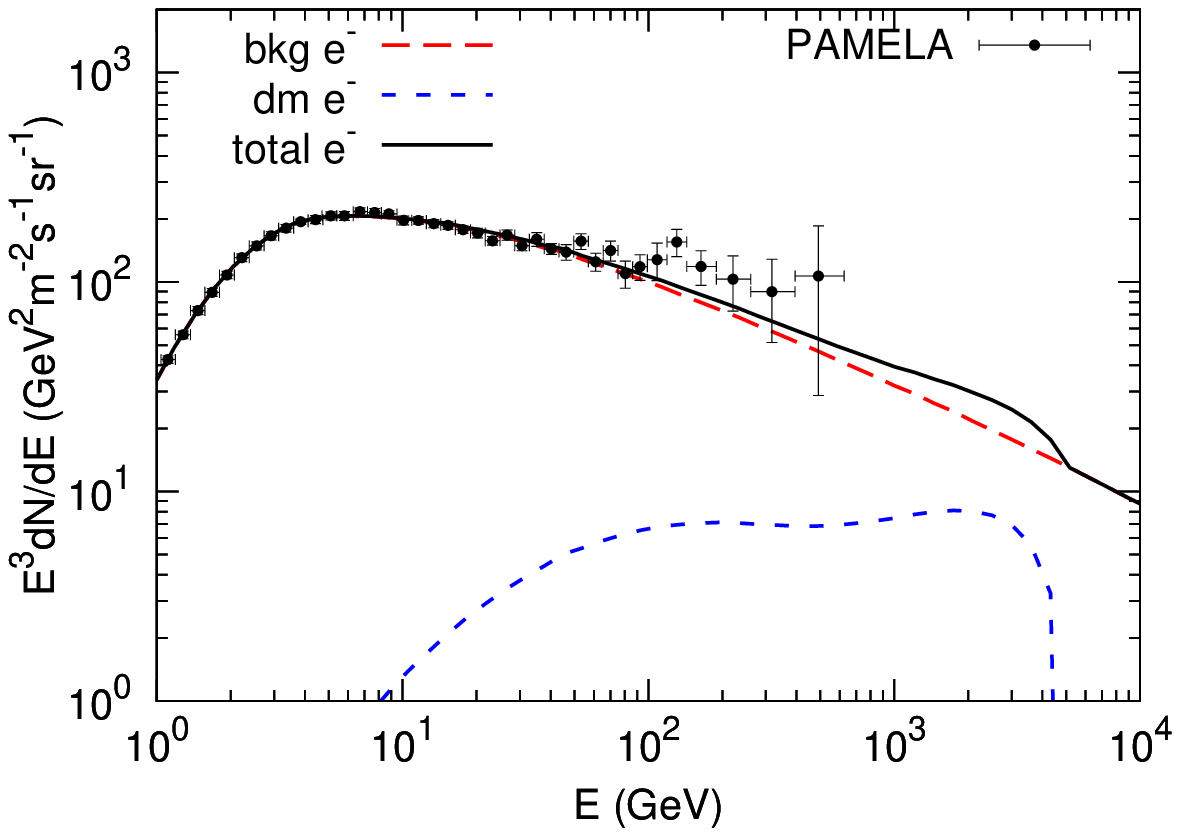}
\includegraphics[width=0.48\textwidth]{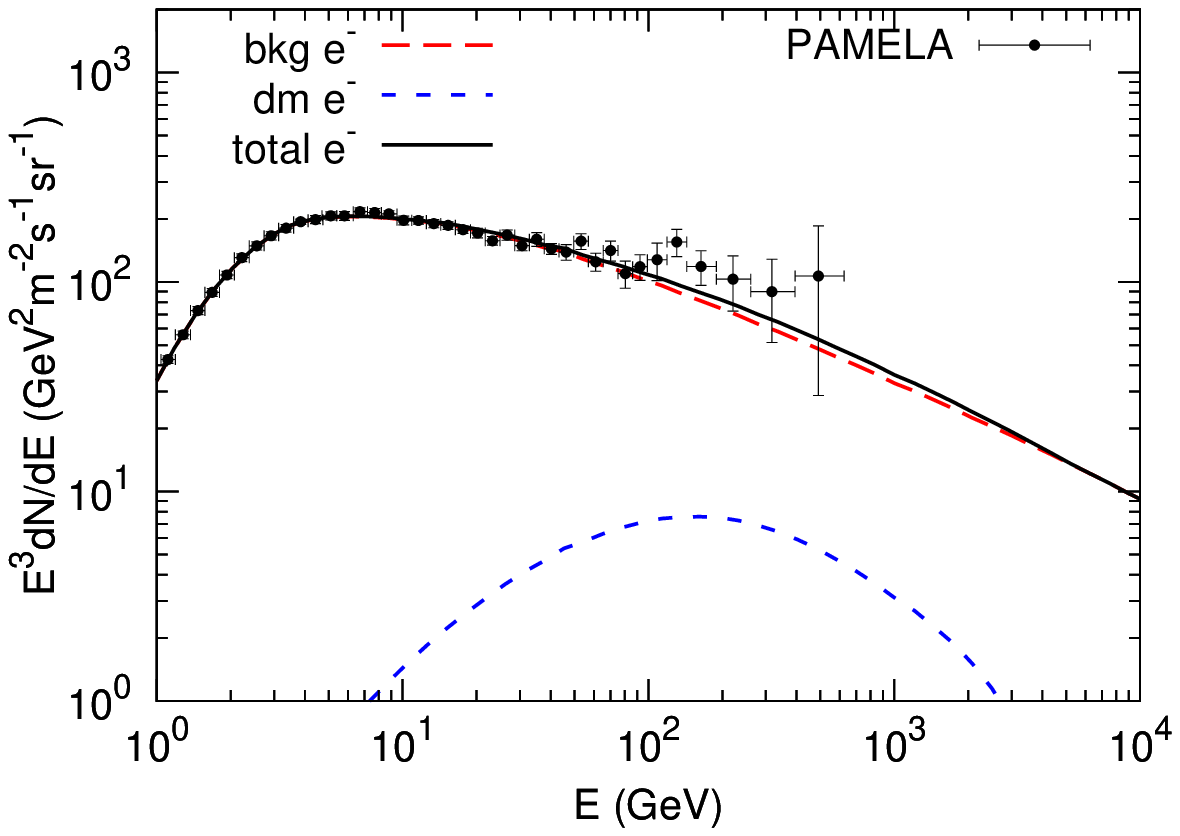}
\caption{The positron fraction (top) and electron spectrum (bottom)
for the best fitting parameters of the fit II-a with DM annihilation
into $W^+W^-$ (left) and $b\bar{b}$ (right) channels.
\label{fig:bw}}
\end{figure*}

As a test we run the fit II-a with DM annihilation to $W^+W^-$ and
$b\bar{b}$ final states. We find the $\chi^2$ values become slightly
smaller ($\sim52.7$ for both $W^+W^-$ and $b\bar{b}$) than that 
of $\tau^+\tau^-$. The positron fraction and electron spectrum for 
the best fitting parameters are shown in Fig. \ref{fig:bw}. 
It is shown that the AMS-02 positron fraction data can be
reproduced in this case.

\begin{figure}[!htb]
\centering
\includegraphics[width=0.7\textwidth]{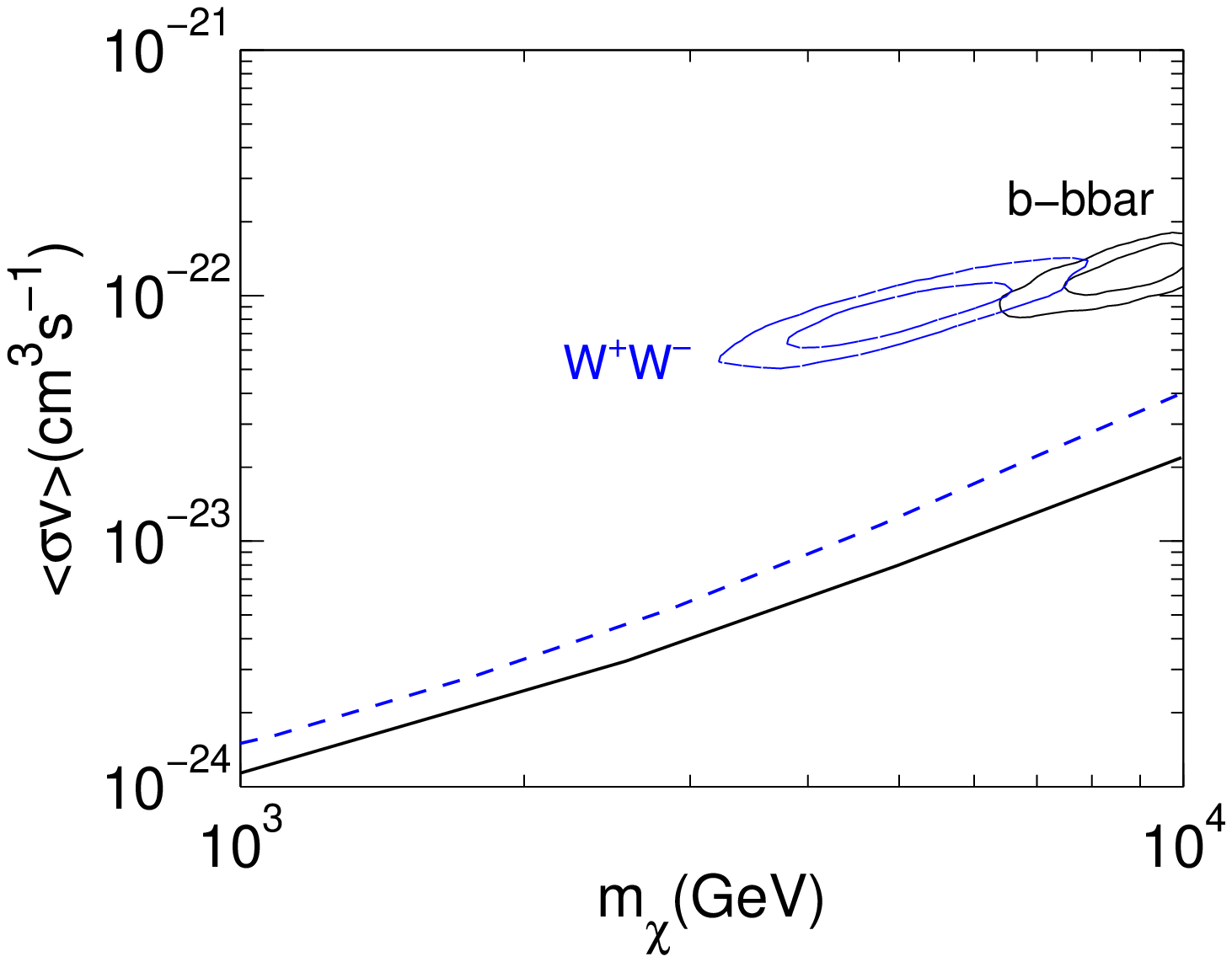}
\caption{$1\sigma$ and $2\sigma$ confidence regions on the DM mass and
cross section plane derived from fit II-a, for $W^+W^-$ and $b\bar{b}$ 
channels. The lines show the upper limits derived using Fermi four year 
observations on the dwarf galaxies for $W^+W^-$ (dashed) and $b\bar{b}$
(solid) channels \cite{Drlica-Wagner2012}.
\label{fig:msv_had}}
\end{figure}

However, it is well known that the PAMELA antiproton data and Fermi 
$\gamma$-ray data set very stringent constraints on the DM annihilation 
into quark and gauge boson final states\cite{2009PhRvL.102g1301D,
2011JCAP...09..007C,2011PhRvL.107x1302A,2011PhRvL.107x1303G}.
Fig. \ref{fig:msv_had} shows the two dimensional contours on the
$m_{\chi}-\sv$ plane for the $W^+W^-$ and $b\bar{b}$ channels
and the $95\%$ exclusion limits on DM annihilation to $b\bar{b}$ 
and $W^+W^-$ channels by Fermi observation of dwarf galaxies 
\cite{Drlica-Wagner2012}. It is shown that the DM scenario is disfavored
to explain the $e^{\pm}$ excesses.

\section{Discussion}
\label{discuss}

\subsection{Uncertainties of the theoretical model}

Since the AMS-02 data are very precise, any uncertainties previously
thought to be not important may affect the fitting result. As we note 
above the positron spectrum around tens of GeV may affect the fitting 
result sensitively. Therefore the change of the shape of background may
affect the fitting results. There are quite a few sources of such 
uncertainties, such as the propagation parameters, the hadronic 
interaction models, and so on.

In this work we use the Kamae et al. (2006) parameterization of 
the $pp$ collision \cite{2006ApJ...647..692K}. As shown in
\cite{2009A&A...501..821D} there were remarkable differences between
different hadronic models. The Kamae et al. (2006) parameterization
included more processes than before and was calibrated with recent
data \cite{2006ApJ...647..692K}. However, it depends strongly on
the Monte Carlo simulations. Therefore we also test the fitting
with old $pp$ collision parameterization \cite{1977PhRvD..15..820B}.
The results show quantitative difference from that shown above.
But, all the conclusions made in the previous section keep unchanged.

We assume single power-law of the high energy ($>10$ GeV) proton
spectrum and neglect the spectral hardening around $200$ GeV.
Through introducing another break of the proton spectrum at $\sim230$
GeV we can get better fit to the data. We have tested that in such 
a case the fitting results in this work change a little in numbers
and the conclusions are not affected.

\subsection{Alternatives of the primary electrons}

As revealed by ATIC, CREAM and PAMELA measurements, the nuclei spectra 
have a hardening above $\sim200$ GV \cite{2007BRASP..71..494P,
2010ApJ...714L..89A,2011Sci...332...69A}. It is possible that the primary 
electron spectrum also has a similar hardening at high energies. By 
including such a modification of the primary electron spectrum may soften 
the {\em tension} and improve the fitting \cite{2014PhLB..728..250F,
2013PhLB..727....1Y}.

It is also possible that the continuous assumption of the primary 
electron sources breaks down at energies above $\sim100$ GeV. The variance 
due to nearby SNRs may result in deviation of the primary electron spectrum 
from power-law assumption and mimic the spectral hardening behavior
\cite{2014arXiv1402.0321D}. In this case all the data may be fitted 
simultaneously.

\subsection{Two or more components of the extra $e^{\pm}$ sources}

It is possible that there are more than one components of the
primary sources of positrons. For example for the pulsar scenario,
the far away pulsar population may contribute to a ``background''
component, and several nearby pulsars may give very distinct 
contributions to the positron spectrum \cite{2009JCAP...01..025H}. 
Such a picture may help to improve the fitting to AMS-02 and Fermi/HESS 
$e^{\pm}$ data. The ``background'' pulsars may give most contributions 
to the AMS-02 positron excesses, while the nearby sources can contribute 
mainly at high energies to reproduce the Fermi/HESS data. However,
the detailed modeling in \cite{2013PhRvD..88b3001Y} seems that it is not 
easy to reconcile different datasets without changing the spectrum of
the primary electrons.

\subsection{To improve the DM scenario}

Finally we discuss the possibilities to improve the DM scenario to give
better explanation of the current data. First we need softer and broader 
spectrum of positrons from DM. Therefore if the annihilation final 
state is not two-body state but four-body, eight-body etc., softer 
positron spectrum may be generated. A mixture of leptonic channels 
and hadronic channels may also give a broader spectrum (see Fig. 
\ref{fig:spec}). Second we have to consider how to avoid the 
constraints from $\gamma$-rays and antiprotons. Decaying DM scenario 
is better since the constraints from $\gamma$-rays are less 
stringent \cite{2010JCAP...03..014P,2010NuPhB.840..284C}. 
As for the antiproton constraints, very massive DM particle may 
avoid the current bounds set by the PAMELA data \cite{2009NuPhB.813....1C}.  

\section{Summary}
\label{summary}

In summary in this work we give a systematical investigation of
the models to explain the cosmic $e^{\pm}$ excesses, based on the
newest AMS-02 data of the positron fraction and other data from
PAMELA, Fermi and HESS. Both the pulsar-like scenario and DM scenario
as the extra primary $e^{\pm}$ sources are studied. Our findings
are as follows.

\begin{itemize}

\item It is found that under the present framework it is difficult
to fit the PAMELA/AMS-02 data and Fermi/HESS data simultaneously. 
The AMS-02 positron fraction data requires less positrons from the extra 
sources than that needed by Fermi/HESS data. It may indicate that either
the model needs to be refined or there is inconsistency between different
data sets. The latter possibility seems to be proved by the preliminary
data of the total $e^+e^-$ spectrum by AMS-02 \cite{2013ICRC-AMS02}.

\item Pulsar-like models can fit the data better than the DM scenario. 
The spectral index of the positrons injected by the pulsars is about $E^{-2}$, 
which is much softer than that derived when using the PAMELA positron 
fraction data.

\item If we fit only the AMS-02 positron fraction and PAMELA electron 
spectrum data both the pulsar-like and DM annihilates/decays into 
$\tau^+\tau^-$, $W^+W^-$ and $b\bar{b}$ can fit the data. However, 
due to the lack of constraints from high energy range, these fits seem
to under-estimate the constribution of the extra sources to the lepton
fluxes.

\item Due to the strong constraints from antiprotons and $\gamma$-rays,
the DM annihilation scenario (into two body final states) seems not 
easy to be consistent with all of the current data. 

\end{itemize}

Our study illustrates the remarkable potential on understanding the
physics of the $e^{\pm}$ excesses from the very precise measurement 
done by AMS-02. According to the PAMELA positron fraction data, the
pulsar-like and DM scenarios are almost identical \cite{2012PhRvD..85d3507L}.
Given the AMS-02 precise measurement, the differences between different 
scenarios become to appear. We are looking forward to more data with
higher precision from AMS-02 to further shed light on the understanding
of the fundamental questions of both astrophysics and particle physics.

\section*{Acknowledgments}
We would like to thank Chen Hesheng, Hu Hongbo, Li Zuhao, L\"u Y\"usheng,
Tang Zhicheng and Xu Weiwei for helpful discussions.
This work is supported by 973 Program under Grant Nos. 2010CB83300 and 
2013CB837000, and by National Natural Science Foundation of China under 
Grant Nos. 11075169, 11135009, 11105155 and 11220101004.

\bibliographystyle{apsrev}
\bibliography{/home/yuanq/work/cygnus/tex/refs}

\begin{thebibliography}{79}
\expandafter\ifx\csname natexlab\endcsname\relax\def\natexlab#1{#1}\fi
\expandafter\ifx\csname bibnamefont\endcsname\relax
  \def\bibnamefont#1{#1}\fi
\expandafter\ifx\csname bibfnamefont\endcsname\relax
  \def\bibfnamefont#1{#1}\fi
\expandafter\ifx\csname citenamefont\endcsname\relax
  \def\citenamefont#1{#1}\fi
\expandafter\ifx\csname url\endcsname\relax
  \def\url#1{\texttt{#1}}\fi
\expandafter\ifx\csname urlprefix\endcsname\relax\def\urlprefix{URL }\fi
\providecommand{\bibinfo}[2]{#2}
\providecommand{\eprint}[2][]{\url{#2}}

\bibitem[{\citenamefont{{Aguilar} et~al.}(2013)\citenamefont{{Aguilar},
  {Alberti}, {Alpat}, {Alvino}, {Ambrosi}, {Andeen}, {Anderhub}, {Arruda},
  {Arrarello}, {Bachlechner} et~al.}}]{2013PhRvL.110n1102A}
\bibinfo{author}{\bibfnamefont{M.}~\bibnamefont{{Aguilar}}},
  \bibnamefont{et~al.}, \bibinfo{journal}{\prl} \textbf{\bibinfo{volume}{110}},
  \bibinfo{eid}{141102} (\bibinfo{year}{2013}).

\bibitem[{\citenamefont{{Adriani}
  et~al.}(2009{\natexlab{a}})\citenamefont{{Adriani}, {Barbarino},
  {Bazilevskaya}, {Bellotti}, {Boezio}, {Bogomolov}, {Bonechi}, {Bongi},
  {Bonvicini}, {Bottai} et~al.}}]{2009Natur.458..607A}
\bibinfo{author}{\bibfnamefont{O.}~\bibnamefont{{Adriani}}},
  \bibnamefont{et~al.}, \bibinfo{journal}{\nat} \textbf{\bibinfo{volume}{458}},
  \bibinfo{pages}{607} (\bibinfo{year}{2009}{\natexlab{a}}),
  \eprint{0810.4995}.

\bibitem[{\citenamefont{{Adriani}
  et~al.}(2010{\natexlab{a}})\citenamefont{{Adriani}, {Barbarino},
  {Bazilevskaya}, {Bellotti}, {Boezio}, {Bogomolov}, {Bonechi}, {Bongi},
  {Bonvicini}, {Borisov} et~al.}}]{2010APh....34....1A}
\bibinfo{author}{\bibfnamefont{O.}~\bibnamefont{{Adriani}}},
  \bibnamefont{et~al.}, \bibinfo{journal}{Astroparticle Physics}
  \textbf{\bibinfo{volume}{34}}, \bibinfo{pages}{1}
  (\bibinfo{year}{2010}{\natexlab{a}}), \eprint{1001.3522}.

\bibitem[{\citenamefont{{Liu} et~al.}(2012)\citenamefont{{Liu}, {Yuan}, {Bi},
  {Li}, and {Zhang}}}]{2012PhRvD..85d3507L}
\bibinfo{author}{\bibfnamefont{J.}~\bibnamefont{{Liu}}},
  \bibinfo{author}{\bibfnamefont{Q.}~\bibnamefont{{Yuan}}},
  \bibinfo{author}{\bibfnamefont{X.-J.} \bibnamefont{{Bi}}},
  \bibinfo{author}{\bibfnamefont{H.}~\bibnamefont{{Li}}}, \bibnamefont{and}
  \bibinfo{author}{\bibfnamefont{X.}~\bibnamefont{{Zhang}}},
  \bibinfo{journal}{\prd} \textbf{\bibinfo{volume}{85}}, \bibinfo{eid}{043507}
  (\bibinfo{year}{2012}), \eprint{1106.3882}.

\bibitem[{\citenamefont{{Y{\"u}ksel} et~al.}(2009)\citenamefont{{Y{\"u}ksel},
  {Kistler}, and {Stanev}}}]{2009PhRvL.103e1101Y}
\bibinfo{author}{\bibfnamefont{H.}~\bibnamefont{{Y{\"u}ksel}}},
  \bibinfo{author}{\bibfnamefont{M.~D.} \bibnamefont{{Kistler}}},
  \bibnamefont{and} \bibinfo{author}{\bibfnamefont{T.}~\bibnamefont{{Stanev}}},
  \bibinfo{journal}{\prl} \textbf{\bibinfo{volume}{103}},
  \bibinfo{pages}{051101} (\bibinfo{year}{2009}), \eprint{0810.2784}.

\bibitem[{\citenamefont{{Hooper} et~al.}(2009)\citenamefont{{Hooper}, {Blasi},
  and {Dario Serpico}}}]{2009JCAP...01..025H}
\bibinfo{author}{\bibfnamefont{D.}~\bibnamefont{{Hooper}}},
  \bibinfo{author}{\bibfnamefont{P.}~\bibnamefont{{Blasi}}}, \bibnamefont{and}
  \bibinfo{author}{\bibfnamefont{P.}~\bibnamefont{{Dario Serpico}}},
  \bibinfo{journal}{\jcap} \textbf{\bibinfo{volume}{1}}, \bibinfo{pages}{25}
  (\bibinfo{year}{2009}), \eprint{0810.1527}.

\bibitem[{\citenamefont{{Profumo}}(2012)}]{2012CEJPh..10....1P}
\bibinfo{author}{\bibfnamefont{S.}~\bibnamefont{{Profumo}}},
  \bibinfo{journal}{Central European Journal of Physics}
  \textbf{\bibinfo{volume}{10}}, \bibinfo{pages}{1} (\bibinfo{year}{2012}),
  \eprint{0812.4457}.

\bibitem[{\citenamefont{{Malyshev} et~al.}(2009)\citenamefont{{Malyshev},
  {Cholis}, and {Gelfand}}}]{2009PhRvD..80f3005M}
\bibinfo{author}{\bibfnamefont{D.}~\bibnamefont{{Malyshev}}},
  \bibinfo{author}{\bibfnamefont{I.}~\bibnamefont{{Cholis}}}, \bibnamefont{and}
  \bibinfo{author}{\bibfnamefont{J.}~\bibnamefont{{Gelfand}}},
  \bibinfo{journal}{\prd} \textbf{\bibinfo{volume}{80}},
  \bibinfo{pages}{063005} (\bibinfo{year}{2009}), \eprint{0903.1310}.

\bibitem[{\citenamefont{{Hu} et~al.}(2009)\citenamefont{{Hu}, {Yuan}, {Wang},
  {Fan}, {Zhang}, and {Bi}}}]{2009ApJ...700L.170H}
\bibinfo{author}{\bibfnamefont{H.-B.} \bibnamefont{{Hu}}},
  \bibinfo{author}{\bibfnamefont{Q.}~\bibnamefont{{Yuan}}},
  \bibinfo{author}{\bibfnamefont{B.}~\bibnamefont{{Wang}}},
  \bibinfo{author}{\bibfnamefont{C.}~\bibnamefont{{Fan}}},
  \bibinfo{author}{\bibfnamefont{J.-L.} \bibnamefont{{Zhang}}},
  \bibnamefont{and} \bibinfo{author}{\bibfnamefont{X.-J.} \bibnamefont{{Bi}}},
  \bibinfo{journal}{\apjl} \textbf{\bibinfo{volume}{700}},
  \bibinfo{pages}{L170} (\bibinfo{year}{2009}), \eprint{0901.1520}.

\bibitem[{\citenamefont{{Blasi}}(2009)}]{2009PhRvL.103e1104B}
\bibinfo{author}{\bibfnamefont{P.}~\bibnamefont{{Blasi}}},
  \bibinfo{journal}{\prl} \textbf{\bibinfo{volume}{103}},
  \bibinfo{pages}{051104} (\bibinfo{year}{2009}), \eprint{0903.2794}.

\bibitem[{\citenamefont{{Bergstr{\"o}m}
  et~al.}(2008)\citenamefont{{Bergstr{\"o}m}, {Bringmann}, and
  {Edsj{\"o}}}}]{2008PhRvD..78j3520B}
\bibinfo{author}{\bibfnamefont{L.}~\bibnamefont{{Bergstr{\"o}m}}},
  \bibinfo{author}{\bibfnamefont{T.}~\bibnamefont{{Bringmann}}},
  \bibnamefont{and}
  \bibinfo{author}{\bibfnamefont{J.}~\bibnamefont{{Edsj{\"o}}}},
  \bibinfo{journal}{\prd} \textbf{\bibinfo{volume}{78}},
  \bibinfo{pages}{103520} (\bibinfo{year}{2008}), \eprint{0808.3725}.

\bibitem[{\citenamefont{{Barger} et~al.}(2009)\citenamefont{{Barger}, {Keung},
  {Marfatia}, and {Shaughnessy}}}]{2009PhLB..672..141B}
\bibinfo{author}{\bibfnamefont{V.}~\bibnamefont{{Barger}}},
  \bibinfo{author}{\bibfnamefont{W.-Y.} \bibnamefont{{Keung}}},
  \bibinfo{author}{\bibfnamefont{D.}~\bibnamefont{{Marfatia}}},
  \bibnamefont{and}
  \bibinfo{author}{\bibfnamefont{G.}~\bibnamefont{{Shaughnessy}}},
  \bibinfo{journal}{Phys. Lett. B} \textbf{\bibinfo{volume}{672}},
  \bibinfo{pages}{141} (\bibinfo{year}{2009}), \eprint{0809.0162}.

\bibitem[{\citenamefont{{Cirelli} et~al.}(2009)\citenamefont{{Cirelli},
  {Kadastik}, {Raidal}, and {Strumia}}}]{2009NuPhB.813....1C}
\bibinfo{author}{\bibfnamefont{M.}~\bibnamefont{{Cirelli}}},
  \bibinfo{author}{\bibfnamefont{M.}~\bibnamefont{{Kadastik}}},
  \bibinfo{author}{\bibfnamefont{M.}~\bibnamefont{{Raidal}}}, \bibnamefont{and}
  \bibinfo{author}{\bibfnamefont{A.}~\bibnamefont{{Strumia}}},
  \bibinfo{journal}{Nuclear Physics B} \textbf{\bibinfo{volume}{813}},
  \bibinfo{pages}{1} (\bibinfo{year}{2009}), [Addendum-ibid. B {\bf 873},
  530 (2013)], \eprint{0809.2409}.

\bibitem[{\citenamefont{{Yin} et~al.}(2009)\citenamefont{{Yin}, {Yuan}, {Liu},
  {Zhang}, {Bi}, {Zhu}, and {Zhang}}}]{2009PhRvD..79b3512Y}
\bibinfo{author}{\bibfnamefont{P.~F.} \bibnamefont{{Yin}}},
  \bibinfo{author}{\bibfnamefont{Q.}~\bibnamefont{{Yuan}}},
  \bibinfo{author}{\bibfnamefont{J.}~\bibnamefont{{Liu}}},
  \bibinfo{author}{\bibfnamefont{J.}~\bibnamefont{{Zhang}}},
  \bibinfo{author}{\bibfnamefont{X.~J.} \bibnamefont{{Bi}}},
  \bibinfo{author}{\bibfnamefont{S.~H.} \bibnamefont{{Zhu}}}, \bibnamefont{and}
  \bibinfo{author}{\bibfnamefont{X.~M.} \bibnamefont{{Zhang}}},
  \bibinfo{journal}{\prd} \textbf{\bibinfo{volume}{79}},
  \bibinfo{pages}{023512} (\bibinfo{year}{2009}), \eprint{0811.0176}.

\bibitem[{\citenamefont{{Zhang} et~al.}(2009)\citenamefont{{Zhang}, {Bi},
  {Liu}, {Liu}, {Yin}, {Yuan}, and {Zhu}}}]{2009PhRvD..80b3007Z}
\bibinfo{author}{\bibfnamefont{J.}~\bibnamefont{{Zhang}}},
  \bibinfo{author}{\bibfnamefont{X.~J.} \bibnamefont{{Bi}}},
  \bibinfo{author}{\bibfnamefont{J.}~\bibnamefont{{Liu}}},
  \bibinfo{author}{\bibfnamefont{S.~M.} \bibnamefont{{Liu}}},
  \bibinfo{author}{\bibfnamefont{P.~F.} \bibnamefont{{Yin}}},
  \bibinfo{author}{\bibfnamefont{Q.}~\bibnamefont{{Yuan}}}, \bibnamefont{and}
  \bibinfo{author}{\bibfnamefont{S.~H.} \bibnamefont{{Zhu}}},
  \bibinfo{journal}{\prd} \textbf{\bibinfo{volume}{80}},
  \bibinfo{pages}{023007} (\bibinfo{year}{2009}), \eprint{0812.0522}.

\bibitem[{\citenamefont{{Adriani}
  et~al.}(2011{\natexlab{a}})\citenamefont{{Adriani}, {Barbarino},
  {Bazilevskaya}, {Bellotti}, {Boezio}, {Bogomolov}, {Bongi}, {Bonvicini},
  {Borisov}, {Bottai} et~al.}}]{2011PhRvL.106t1101A}
\bibinfo{author}{\bibfnamefont{O.}~\bibnamefont{{Adriani}}},
  \bibnamefont{et~al.}, \bibinfo{journal}{\prl} \textbf{\bibinfo{volume}{106}},
  \bibinfo{pages}{201101} (\bibinfo{year}{2011}{\natexlab{a}}).

\bibitem[{\citenamefont{{Adriani}
  et~al.}(2009{\natexlab{b}})\citenamefont{{Adriani}, {Barbarino},
  {Bazilevskaya}, {Bellotti}, {Boezio}, {Bogomolov}, {Bonechi}, {Bongi},
  {Bonvicini}, {Bottai} et~al.}}]{2009PhRvL.102e1101A}
\bibinfo{author}{\bibfnamefont{O.}~\bibnamefont{{Adriani}}},
  \bibnamefont{et~al.}, \bibinfo{journal}{\prl} \textbf{\bibinfo{volume}{102}},
  \bibinfo{pages}{051101} (\bibinfo{year}{2009}{\natexlab{b}}),
  \eprint{0810.4994}.

\bibitem[{\citenamefont{{Adriani}
  et~al.}(2010{\natexlab{b}})\citenamefont{{Adriani}, {Barbarino},
  {Bazilevskaya}, {Bellotti}, {Boezio}, {Bogomolov}, {Bonechi}, {Bongi},
  {Bonvicini}, {Borisov} et~al.}}]{2010PhRvL.105l1101A}
\bibinfo{author}{\bibfnamefont{O.}~\bibnamefont{{Adriani}}},
  \bibnamefont{et~al.}, \bibinfo{journal}{\prl} \textbf{\bibinfo{volume}{105}},
  \bibinfo{pages}{121101} (\bibinfo{year}{2010}{\natexlab{b}}),
  \eprint{1007.0821}.

\bibitem[{\citenamefont{{Adriani}
  et~al.}(2011{\natexlab{b}})\citenamefont{{Adriani}, {Barbarino},
  {Bazilevskaya}, {Bellotti}, {Boezio}, {Bogomolov}, {Bonechi}, {Bongi},
  {Bonvicini}, {Borisov} et~al.}}]{2011Sci...332...69A}
\bibinfo{author}{\bibfnamefont{O.}~\bibnamefont{{Adriani}}},
  \bibnamefont{et~al.}, \bibinfo{journal}{Science}
  \textbf{\bibinfo{volume}{332}}, \bibinfo{pages}{69}
  (\bibinfo{year}{2011}{\natexlab{b}}), \eprint{1103.4055}.

\bibitem[{\citenamefont{{Abdo} et~al.}(2009)\citenamefont{{Abdo}, {Ackermann},
  {Ajello}, {Atwood}, {Axelsson}, {Baldini}, {Ballet}, {Barbiellini},
  {Bastieri}, {Battelino} et~al.}}]{2009PhRvL.102r1101A}
\bibinfo{author}{\bibfnamefont{A.~A.} \bibnamefont{{Abdo}}},
  \bibnamefont{et~al.}, \bibinfo{journal}{\prl} \textbf{\bibinfo{volume}{102}},
  \bibinfo{pages}{181101} (\bibinfo{year}{2009}), \eprint{0905.0025}.

\bibitem[{\citenamefont{{Ackermann} et~al.}(2010)\citenamefont{{Ackermann},
  {Ajello}, {Atwood}, {Baldini}, {Ballet}, {Barbiellini}, {Bastieri},
  {Baughman}, {Bechtol}, {Bellardi} et~al.}}]{2010PhRvD..82i2004A}
\bibinfo{author}{\bibfnamefont{M.}~\bibnamefont{{Ackermann}}},
  \bibnamefont{et~al.}, \bibinfo{journal}{\prd} \textbf{\bibinfo{volume}{82}},
  \bibinfo{pages}{092004} (\bibinfo{year}{2010}).

\bibitem[{\citenamefont{{Chang} et~al.}(2008)\citenamefont{{Chang}, {Adams},
  {Ahn}, {Bashindzhagyan}, {Christl}, {Ganel}, {Guzik}, {Isbert}, {Kim},
  {Kuznetsov} et~al.}}]{2008Natur.456..362C}
\bibinfo{author}{\bibfnamefont{J.}~\bibnamefont{{Chang}}},
  \bibnamefont{et~al.}, \bibinfo{journal}{\nat} \textbf{\bibinfo{volume}{456}},
  \bibinfo{pages}{362} (\bibinfo{year}{2008}).

\bibitem[{\citenamefont{{Aharonian} et~al.}(2008)\citenamefont{{Aharonian},
  {Akhperjanian}, {Barres de Almeida}, {Bazer-Bachi}, {Becherini}, {Behera},
  {Benbow}, {Bernl{\"o}hr}, {Boisson}, {Bochow} et~al.}}]{2008PhRvL.101z1104A}
\bibinfo{author}{\bibfnamefont{F.}~\bibnamefont{{Aharonian}}},
  \bibnamefont{et~al.}, \bibinfo{journal}{\prl} \textbf{\bibinfo{volume}{101}},
  \bibinfo{pages}{261104} (\bibinfo{year}{2008}), \eprint{0811.3894}.

\bibitem[{\citenamefont{{Aharonian} et~al.}(2009)\citenamefont{{Aharonian},
  {Akhperjanian}, {Anton}, {Barres de Almeida}, {Bazer-Bachi}, {Becherini},
  {Behera}, {Bernl{\"o}hr}, {Bochow}, {Boisson} et~al.}}]{2009A&A...508..561A}
\bibinfo{author}{\bibfnamefont{F.}~\bibnamefont{{Aharonian}}},
  \bibnamefont{et~al.}, \bibinfo{journal}{\aap} \textbf{\bibinfo{volume}{508}},
  \bibinfo{pages}{561} (\bibinfo{year}{2009}), \eprint{0905.0105}.

\bibitem[{\citenamefont{{Liu} et~al.}(2010)\citenamefont{{Liu}, {Yuan}, {Bi},
  {Li}, and {Zhang}}}]{2010PhRvD..81b3516L}
\bibinfo{author}{\bibfnamefont{J.}~\bibnamefont{{Liu}}},
  \bibinfo{author}{\bibfnamefont{Q.}~\bibnamefont{{Yuan}}},
  \bibinfo{author}{\bibfnamefont{X.~J.} \bibnamefont{{Bi}}},
  \bibinfo{author}{\bibfnamefont{H.}~\bibnamefont{{Li}}}, \bibnamefont{and}
  \bibinfo{author}{\bibfnamefont{X.~M.} \bibnamefont{{Zhang}}},
  \bibinfo{journal}{\prd} \textbf{\bibinfo{volume}{81}},
  \bibinfo{pages}{023516} (\bibinfo{year}{2010}), \eprint{0906.3858}.

\bibitem[{\citenamefont{{Strong} and {Moskalenko}}(1998)}]{1998ApJ...509..212S}
\bibinfo{author}{\bibfnamefont{A.~W.} \bibnamefont{{Strong}}} \bibnamefont{and}
  \bibinfo{author}{\bibfnamefont{I.~V.} \bibnamefont{{Moskalenko}}},
  \bibinfo{journal}{\apj} \textbf{\bibinfo{volume}{509}}, \bibinfo{pages}{212}
  (\bibinfo{year}{1998}), \eprint{astro-ph/9807150}.

\bibitem[{\citenamefont{{Gaisser}}(1990)}]{1990cup..book.....G}
\bibinfo{author}{\bibfnamefont{T.~K.} \bibnamefont{{Gaisser}}},
  \emph{\bibinfo{title}{{Cosmic rays and particle physics}}}
  (\bibinfo{publisher}{Cambridge and New York, Cambridge University Press,
  1990, 292 p.}, \bibinfo{year}{1990}).

\bibitem[{\citenamefont{{Maurin} et~al.}(2001)\citenamefont{{Maurin}, {Donato},
  {Taillet}, and {Salati}}}]{2001ApJ...555..585M}
\bibinfo{author}{\bibfnamefont{D.}~\bibnamefont{{Maurin}}},
  \bibinfo{author}{\bibfnamefont{F.}~\bibnamefont{{Donato}}},
  \bibinfo{author}{\bibfnamefont{R.}~\bibnamefont{{Taillet}}},
  \bibnamefont{and} \bibinfo{author}{\bibfnamefont{P.}~\bibnamefont{{Salati}}},
  \bibinfo{journal}{\apj} \textbf{\bibinfo{volume}{555}}, \bibinfo{pages}{585}
  (\bibinfo{year}{2001}).

\bibitem[{\citenamefont{{di Bernardo} et~al.}(2010)\citenamefont{{di Bernardo},
  {Evoli}, {Gaggero}, {Grasso}, and {Maccione}}}]{2010APh....34..274D}
\bibinfo{author}{\bibfnamefont{G.}~\bibnamefont{{di Bernardo}}},
  \bibinfo{author}{\bibfnamefont{C.}~\bibnamefont{{Evoli}}},
  \bibinfo{author}{\bibfnamefont{D.}~\bibnamefont{{Gaggero}}},
  \bibinfo{author}{\bibfnamefont{D.}~\bibnamefont{{Grasso}}}, \bibnamefont{and}
  \bibinfo{author}{\bibfnamefont{L.}~\bibnamefont{{Maccione}}},
  \bibinfo{journal}{Astroparticle Physics} \textbf{\bibinfo{volume}{34}},
  \bibinfo{pages}{274} (\bibinfo{year}{2010}), \eprint{0909.4548}.

\bibitem[{\citenamefont{{Lin} et~al.}(2013)\citenamefont{{Lin}, {Cai}, and {et
  al.}}}]{mcmc:prop}
\bibinfo{author}{\bibfnamefont{S.~J.} \bibnamefont{{Lin}}},
  \bibinfo{author}{\bibfnamefont{C.~F.} \bibnamefont{{Cai}}}, \bibnamefont{and}
  \bibinfo{author}{\bibnamefont{{et al.}}}, \bibinfo{journal}{in preparation}
  (\bibinfo{year}{2013}).

\bibitem[{\citenamefont{{Strong} et~al.}(2007)\citenamefont{{Strong},
  {Moskalenko}, and {Ptuskin}}}]{2007ARNPS..57..285S}
\bibinfo{author}{\bibfnamefont{A.~W.} \bibnamefont{{Strong}}},
  \bibinfo{author}{\bibfnamefont{I.~V.} \bibnamefont{{Moskalenko}}},
  \bibnamefont{and} \bibinfo{author}{\bibfnamefont{V.~S.}
  \bibnamefont{{Ptuskin}}}, \bibinfo{journal}{Annual Review of Nuclear and
  Particle Science} \textbf{\bibinfo{volume}{57}}, \bibinfo{pages}{285}
  (\bibinfo{year}{2007}), \eprint{astro-ph/0701517}.

\bibitem[{\citenamefont{{Seo} and {Ptuskin}}(1994)}]{1994ApJ...431..705S}
\bibinfo{author}{\bibfnamefont{E.~S.} \bibnamefont{{Seo}}} \bibnamefont{and}
  \bibinfo{author}{\bibfnamefont{V.~S.} \bibnamefont{{Ptuskin}}},
  \bibinfo{journal}{\apj} \textbf{\bibinfo{volume}{431}}, \bibinfo{pages}{705}
  (\bibinfo{year}{1994}).

\bibitem[{\citenamefont{{Evoli} et~al.}(2008)\citenamefont{{Evoli}, {Gaggero},
  {Grasso}, and {Maccione}}}]{2008JCAP...10..018E}
\bibinfo{author}{\bibfnamefont{C.}~\bibnamefont{{Evoli}}},
  \bibinfo{author}{\bibfnamefont{D.}~\bibnamefont{{Gaggero}}},
  \bibinfo{author}{\bibfnamefont{D.}~\bibnamefont{{Grasso}}}, \bibnamefont{and}
  \bibinfo{author}{\bibfnamefont{L.}~\bibnamefont{{Maccione}}},
  \bibinfo{journal}{\jcap} \textbf{\bibinfo{volume}{10}}, \bibinfo{pages}{18}
  (\bibinfo{year}{2008}), \eprint{0807.4730}.

\bibitem[{\citenamefont{{Putze} et~al.}(2009)\citenamefont{{Putze}, {Derome},
  {Maurin}, {Perotto}, and {Taillet}}}]{2009A&A...497..991P}
\bibinfo{author}{\bibfnamefont{A.}~\bibnamefont{{Putze}}},
  \bibinfo{author}{\bibfnamefont{L.}~\bibnamefont{{Derome}}},
  \bibinfo{author}{\bibfnamefont{D.}~\bibnamefont{{Maurin}}},
  \bibinfo{author}{\bibfnamefont{L.}~\bibnamefont{{Perotto}}},
  \bibnamefont{and}
  \bibinfo{author}{\bibfnamefont{R.}~\bibnamefont{{Taillet}}},
  \bibinfo{journal}{\aap} \textbf{\bibinfo{volume}{497}}, \bibinfo{pages}{991}
  (\bibinfo{year}{2009}), \eprint{0808.2437}.

\bibitem[{\citenamefont{{Putze} et~al.}(2010)\citenamefont{{Putze}, {Derome},
  and {Maurin}}}]{2010A&A...516A..66P}
\bibinfo{author}{\bibfnamefont{A.}~\bibnamefont{{Putze}}},
  \bibinfo{author}{\bibfnamefont{L.}~\bibnamefont{{Derome}}}, \bibnamefont{and}
  \bibinfo{author}{\bibfnamefont{D.}~\bibnamefont{{Maurin}}},
  \bibinfo{journal}{\aap} \textbf{\bibinfo{volume}{516}}, \bibinfo{pages}{A66}
  (\bibinfo{year}{2010}), \eprint{1001.0551}.

\bibitem[{\citenamefont{{Trotta} et~al.}(2011)\citenamefont{{Trotta},
  {J{\'o}hannesson}, {Moskalenko}, {Porter}, {Ruiz de Austri}, and
  {Strong}}}]{2011ApJ...729..106T}
\bibinfo{author}{\bibfnamefont{R.}~\bibnamefont{{Trotta}}},
  \bibinfo{author}{\bibfnamefont{G.}~\bibnamefont{{J{\'o}hannesson}}},
  \bibinfo{author}{\bibfnamefont{I.~V.} \bibnamefont{{Moskalenko}}},
  \bibinfo{author}{\bibfnamefont{T.~A.} \bibnamefont{{Porter}}},
  \bibinfo{author}{\bibfnamefont{R.}~\bibnamefont{{Ruiz de Austri}}},
  \bibnamefont{and} \bibinfo{author}{\bibfnamefont{A.~W.}
  \bibnamefont{{Strong}}}, \bibinfo{journal}{\apj}
  \textbf{\bibinfo{volume}{729}}, \bibinfo{pages}{106} (\bibinfo{year}{2011}),
  \eprint{1011.0037}.

\bibitem[{\citenamefont{{Ackermann} et~al.}(2012)\citenamefont{{Ackermann},
  {Ajello}, {Atwood}, {Baldini}, {Barbiellini}, {Bastieri}, {Bechtol},
  {Bellazzini}, {Blandford}, {Bloom} et~al.}}]{2012ApJ...761...91A}
\bibinfo{author}{\bibfnamefont{M.}~\bibnamefont{{Ackermann}}},
  \bibnamefont{et~al.}, \bibinfo{journal}{\apj} \textbf{\bibinfo{volume}{761}},
  \bibinfo{eid}{91} (\bibinfo{year}{2012}), \eprint{1205.6474}.

\bibitem[{\citenamefont{{Kamae} et~al.}(2006)\citenamefont{{Kamae}, {Karlsson},
  {Mizuno}, {Abe}, and {Koi}}}]{2006ApJ...647..692K}
\bibinfo{author}{\bibfnamefont{T.}~\bibnamefont{{Kamae}}},
  \bibinfo{author}{\bibfnamefont{N.}~\bibnamefont{{Karlsson}}},
  \bibinfo{author}{\bibfnamefont{T.}~\bibnamefont{{Mizuno}}},
  \bibinfo{author}{\bibfnamefont{T.}~\bibnamefont{{Abe}}}, \bibnamefont{and}
  \bibinfo{author}{\bibfnamefont{T.}~\bibnamefont{{Koi}}},
  \bibinfo{journal}{\apj} \textbf{\bibinfo{volume}{647}}, \bibinfo{pages}{692}
  (\bibinfo{year}{2006}), \eprint{astro-ph/0605581}.

\bibitem[{\citenamefont{{Mori}}(2009)}]{2009APh....31..341M}
\bibinfo{author}{\bibfnamefont{M.}~\bibnamefont{{Mori}}},
  \bibinfo{journal}{Astroparticle Physics} \textbf{\bibinfo{volume}{31}},
  \bibinfo{pages}{341} (\bibinfo{year}{2009}), \eprint{0903.3260}.

\bibitem[{\citenamefont{{Lorimer}}(2004)}]{2004IAUS..218..105L}
\bibinfo{author}{\bibfnamefont{D.~R.} \bibnamefont{{Lorimer}}}, in
  \emph{\bibinfo{booktitle}{Young Neutron Stars and Their Environments}},
  edited by \bibinfo{editor}{\bibnamefont{{F.~Camilo \& B.~M.~Gaensler}}}
  (\bibinfo{year}{2004}), vol. \bibinfo{volume}{218} of
  \emph{\bibinfo{series}{IAU Symposium}}, p. \bibinfo{pages}{105}.

\bibitem[{\citenamefont{{Fujita} et~al.}(2009)\citenamefont{{Fujita}, {Kohri},
  {Yamazaki}, and {Ioka}}}]{2009PhRvD..80f3003F}
\bibinfo{author}{\bibfnamefont{Y.}~\bibnamefont{{Fujita}}},
  \bibinfo{author}{\bibfnamefont{K.}~\bibnamefont{{Kohri}}},
  \bibinfo{author}{\bibfnamefont{R.}~\bibnamefont{{Yamazaki}}},
  \bibnamefont{and} \bibinfo{author}{\bibfnamefont{K.}~\bibnamefont{{Ioka}}},
  \bibinfo{journal}{\prd} \textbf{\bibinfo{volume}{80}}, \bibinfo{eid}{063003}
  (\bibinfo{year}{2009}), \eprint{0903.5298}.

\bibitem[{\citenamefont{{Blasi} and {Serpico}}(2009)}]{2009PhRvL.103h1103B}
\bibinfo{author}{\bibfnamefont{P.}~\bibnamefont{{Blasi}}} \bibnamefont{and}
  \bibinfo{author}{\bibfnamefont{P.~D.} \bibnamefont{{Serpico}}},
  \bibinfo{journal}{\prl} \textbf{\bibinfo{volume}{103}}, \bibinfo{eid}{081103}
  (\bibinfo{year}{2009}), \eprint{0904.0871}.

\bibitem[{\citenamefont{{Mertsch} and {Sarkar}}(2009)}]{2009PhRvL.103h1104M}
\bibinfo{author}{\bibfnamefont{P.}~\bibnamefont{{Mertsch}}} \bibnamefont{and}
  \bibinfo{author}{\bibfnamefont{S.}~\bibnamefont{{Sarkar}}},
  \bibinfo{journal}{\prl} \textbf{\bibinfo{volume}{103}}, \bibinfo{eid}{081104}
  (\bibinfo{year}{2009}), \eprint{0905.3152}.

\bibitem[{\citenamefont{{Ahlers} et~al.}(2009)\citenamefont{{Ahlers},
  {Mertsch}, and {Sarkar}}}]{2009PhRvD..80l3017A}
\bibinfo{author}{\bibfnamefont{M.}~\bibnamefont{{Ahlers}}},
  \bibinfo{author}{\bibfnamefont{P.}~\bibnamefont{{Mertsch}}},
  \bibnamefont{and} \bibinfo{author}{\bibfnamefont{S.}~\bibnamefont{{Sarkar}}},
  \bibinfo{journal}{\prd} \textbf{\bibinfo{volume}{80}}, \bibinfo{eid}{123017}
  (\bibinfo{year}{2009}), \eprint{0909.4060}.

\bibitem[{\citenamefont{{Navarro} et~al.}(1997)\citenamefont{{Navarro},
  {Frenk}, and {White}}}]{1997ApJ...490..493N}
\bibinfo{author}{\bibfnamefont{J.~F.} \bibnamefont{{Navarro}}},
  \bibinfo{author}{\bibfnamefont{C.~S.} \bibnamefont{{Frenk}}},
  \bibnamefont{and} \bibinfo{author}{\bibfnamefont{S.~D.~M.}
  \bibnamefont{{White}}}, \bibinfo{journal}{\apj}
  \textbf{\bibinfo{volume}{490}}, \bibinfo{pages}{493} (\bibinfo{year}{1997}),
  \eprint{astro-ph/9611107}.

\bibitem[{\citenamefont{{Catena} and {Ullio}}(2010)}]{2010JCAP...08..004C}
\bibinfo{author}{\bibfnamefont{R.}~\bibnamefont{{Catena}}} \bibnamefont{and}
  \bibinfo{author}{\bibfnamefont{P.}~\bibnamefont{{Ullio}}},
  \bibinfo{journal}{\jcap} \textbf{\bibinfo{volume}{8}}, \bibinfo{pages}{4}
  (\bibinfo{year}{2010}), \eprint{0907.0018}.

\bibitem[{\citenamefont{{Salucci} et~al.}(2010)\citenamefont{{Salucci},
  {Nesti}, {Gentile}, and {Frigerio Martins}}}]{2010A&A...523A..83S}
\bibinfo{author}{\bibfnamefont{P.}~\bibnamefont{{Salucci}}},
  \bibinfo{author}{\bibfnamefont{F.}~\bibnamefont{{Nesti}}},
  \bibinfo{author}{\bibfnamefont{G.}~\bibnamefont{{Gentile}}},
  \bibnamefont{and} \bibinfo{author}{\bibfnamefont{C.}~\bibnamefont{{Frigerio
  Martins}}}, \bibinfo{journal}{\aap} \textbf{\bibinfo{volume}{523}},
  \bibinfo{pages}{A83} (\bibinfo{year}{2010}), \eprint{1003.3101}.

\bibitem[{\citenamefont{{Pato} et~al.}(2010)\citenamefont{{Pato}, {Agertz},
  {Bertone}, {Moore}, and {Teyssier}}}]{2010PhRvD..82b3531P}
\bibinfo{author}{\bibfnamefont{M.}~\bibnamefont{{Pato}}},
  \bibinfo{author}{\bibfnamefont{O.}~\bibnamefont{{Agertz}}},
  \bibinfo{author}{\bibfnamefont{G.}~\bibnamefont{{Bertone}}},
  \bibinfo{author}{\bibfnamefont{B.}~\bibnamefont{{Moore}}}, \bibnamefont{and}
  \bibinfo{author}{\bibfnamefont{R.}~\bibnamefont{{Teyssier}}},
  \bibinfo{journal}{\prd} \textbf{\bibinfo{volume}{82}},
  \bibinfo{pages}{023531} (\bibinfo{year}{2010}), \eprint{1006.1322}.

\bibitem[{\citenamefont{{Donato} et~al.}(2009)\citenamefont{{Donato}, {Maurin},
  {Brun}, {Delahaye}, and {Salati}}}]{2009PhRvL.102g1301D}
\bibinfo{author}{\bibfnamefont{F.}~\bibnamefont{{Donato}}},
  \bibinfo{author}{\bibfnamefont{D.}~\bibnamefont{{Maurin}}},
  \bibinfo{author}{\bibfnamefont{P.}~\bibnamefont{{Brun}}},
  \bibinfo{author}{\bibfnamefont{T.}~\bibnamefont{{Delahaye}}},
  \bibnamefont{and} \bibinfo{author}{\bibfnamefont{P.}~\bibnamefont{{Salati}}},
  \bibinfo{journal}{\prl} \textbf{\bibinfo{volume}{102}},
  \bibinfo{pages}{071301} (\bibinfo{year}{2009}), \eprint{0810.5292}.

\bibitem[{\citenamefont{{Cholis}}(2011)}]{2011JCAP...09..007C}
\bibinfo{author}{\bibfnamefont{I.}~\bibnamefont{{Cholis}}},
  \bibinfo{journal}{\jcap} \textbf{\bibinfo{volume}{9}}, \bibinfo{eid}{007}
  (\bibinfo{year}{2011}), \eprint{1007.1160}.

\bibitem[{\citenamefont{{Sj{\"o}strand}
  et~al.}(2006)\citenamefont{{Sj{\"o}strand}, {Mrenna}, and
  {Skands}}}]{2006JHEP...05..026S}
\bibinfo{author}{\bibfnamefont{T.}~\bibnamefont{{Sj{\"o}strand}}},
  \bibinfo{author}{\bibfnamefont{S.}~\bibnamefont{{Mrenna}}}, \bibnamefont{and}
  \bibinfo{author}{\bibfnamefont{P.}~\bibnamefont{{Skands}}},
  \bibinfo{journal}{Journal of High Energy Physics}
  \textbf{\bibinfo{volume}{5}}, \bibinfo{pages}{26} (\bibinfo{year}{2006}),
  \eprint{hep-ph/0603175}.

\bibitem[{\citenamefont{{Gleeson} and {Axford}}(1968)}]{1968ApJ...154.1011G}
\bibinfo{author}{\bibfnamefont{L.~J.} \bibnamefont{{Gleeson}}}
  \bibnamefont{and} \bibinfo{author}{\bibfnamefont{W.~I.}
  \bibnamefont{{Axford}}}, \bibinfo{journal}{\apj}
  \textbf{\bibinfo{volume}{154}}, \bibinfo{pages}{1011} (\bibinfo{year}{1968}).

\bibitem[{\citenamefont{{Clem} et~al.}(1996)\citenamefont{{Clem}, {Clements},
  {Esposito}, {Evenson}, {Huber}, {L'Heureux}, {Meyer}, and
  {Constantin}}}]{1996ApJ...464..507C}
\bibinfo{author}{\bibfnamefont{J.~M.} \bibnamefont{{Clem}}},
  \bibinfo{author}{\bibfnamefont{D.~P.} \bibnamefont{{Clements}}},
  \bibinfo{author}{\bibfnamefont{J.}~\bibnamefont{{Esposito}}},
  \bibinfo{author}{\bibfnamefont{P.}~\bibnamefont{{Evenson}}},
  \bibinfo{author}{\bibfnamefont{D.}~\bibnamefont{{Huber}}},
  \bibinfo{author}{\bibfnamefont{J.}~\bibnamefont{{L'Heureux}}},
  \bibinfo{author}{\bibfnamefont{P.}~\bibnamefont{{Meyer}}}, \bibnamefont{and}
  \bibinfo{author}{\bibfnamefont{C.}~\bibnamefont{{Constantin}}},
  \bibinfo{journal}{\apj} \textbf{\bibinfo{volume}{464}}, \bibinfo{pages}{507}
  (\bibinfo{year}{1996}).

\bibitem[{\citenamefont{{Beischer} et~al.}(2009)\citenamefont{{Beischer}, {von
  Doetinchem}, {Gast}, {Kirn}, and {Schael}}}]{2009NJPh...11j5021B}
\bibinfo{author}{\bibfnamefont{B.}~\bibnamefont{{Beischer}}},
  \bibinfo{author}{\bibfnamefont{P.}~\bibnamefont{{von Doetinchem}}},
  \bibinfo{author}{\bibfnamefont{H.}~\bibnamefont{{Gast}}},
  \bibinfo{author}{\bibfnamefont{T.}~\bibnamefont{{Kirn}}}, \bibnamefont{and}
  \bibinfo{author}{\bibfnamefont{S.}~\bibnamefont{{Schael}}},
  \bibinfo{journal}{New Journal of Physics} \textbf{\bibinfo{volume}{11}},
  \bibinfo{pages}{105021} (\bibinfo{year}{2009}).

\bibitem[{\citenamefont{{Della Torre} et~al.}(2012)\citenamefont{{Della Torre},
  {Bobik}, {Boschini}, {Consolandi}, {Gervasi}, {Grandi}, {Kudela}, {Pensotti},
  {Rancoita}, {Rozza} et~al.}}]{2012AdSpR..49.1587D}
\bibinfo{author}{\bibfnamefont{S.}~\bibnamefont{{Della Torre}}},
  \bibnamefont{et~al.}, \bibinfo{journal}{Advances in Space Research}
  \textbf{\bibinfo{volume}{49}}, \bibinfo{pages}{1587} (\bibinfo{year}{2012}).

\bibitem[{\citenamefont{{Maccione}}(2013)}]{2013PhRvL.110h1101M}
\bibinfo{author}{\bibfnamefont{L.}~\bibnamefont{{Maccione}}},
  \bibinfo{journal}{\prl} \textbf{\bibinfo{volume}{110}}, \bibinfo{eid}{081101}
  (\bibinfo{year}{2013}), \eprint{1211.6905}.

\bibitem[{\citenamefont{{Panov} et~al.}(2007)\citenamefont{{Panov}, {Adams},
  {Ahn}, {Batkov}, {Bashindzhagyan}, {Watts}, {Wefel}, {Wu}, {Ganel}, {Guzik}
  et~al.}}]{2007BRASP..71..494P}
\bibinfo{author}{\bibfnamefont{A.~D.} \bibnamefont{{Panov}}},
  \bibnamefont{et~al.}, \bibinfo{journal}{Bulletin of the Russian Academy of
  Science, Phys.} \textbf{\bibinfo{volume}{71}}, \bibinfo{pages}{494}
  (\bibinfo{year}{2007}), \eprint{astro-ph/0612377}.

\bibitem[{\citenamefont{{Ahn} et~al.}(2010)\citenamefont{{Ahn}, {Allison},
  {Bagliesi}, {Beatty}, {Bigongiari}, {Childers}, {Conklin}, {Coutu},
  {DuVernois}, {Ganel} et~al.}}]{2010ApJ...714L..89A}
\bibinfo{author}{\bibfnamefont{H.~S.} \bibnamefont{{Ahn}}},
  \bibnamefont{et~al.}, \bibinfo{journal}{\apjl}
  \textbf{\bibinfo{volume}{714}}, \bibinfo{pages}{L89} (\bibinfo{year}{2010}),
  \eprint{1004.1123}.

\bibitem[{\citenamefont{{AMS-02 collaboration}}(2013)}]{2013ICRC-AMS02}
\bibinfo{author}{\bibnamefont{{AMS-02 collaboration}}}, in
  \emph{\bibinfo{booktitle}{International Cosmic Ray Conference}}
  (\bibinfo{publisher}{http://www.ams02.org/2013/07/new-results-from-ams-presented-at-icrc-2013/}, \bibinfo{year}{2013}).

\bibitem[{\citenamefont{{Yuan} et~al.}(2011)\citenamefont{{Yuan}, {Zhang}, and
  {Bi}}}]{2011PhRvD..84d3002Y}
\bibinfo{author}{\bibfnamefont{Q.}~\bibnamefont{{Yuan}}},
  \bibinfo{author}{\bibfnamefont{B.}~\bibnamefont{{Zhang}}}, \bibnamefont{and}
  \bibinfo{author}{\bibfnamefont{X.-J.} \bibnamefont{{Bi}}},
  \bibinfo{journal}{\prd} \textbf{\bibinfo{volume}{84}},
  \bibinfo{pages}{043002} (\bibinfo{year}{2011}), \eprint{1104.3357}.

\bibitem[{\citenamefont{{Alcaraz} et~al.}(2000)\citenamefont{{Alcaraz},
  {Alpat}, {Ambrosi}, {Anderhub}, {Ao}, {Arefiev}, {Azzarello}, {Babucci},
  {Baldini}, {Basile} et~al.}}]{2000PhLB..490...27A}
\bibinfo{author}{\bibfnamefont{J.}~\bibnamefont{{Alcaraz}}},
  \bibnamefont{et~al.}, \bibinfo{journal}{Phys. Lett. B}
  \textbf{\bibinfo{volume}{490}}, \bibinfo{pages}{27} (\bibinfo{year}{2000}).

\bibitem[{\citenamefont{{Sanuki} et~al.}(2000)\citenamefont{{Sanuki}, {Motoki},
  {Matsumoto}, {Seo}, {Wang}, {Abe}, {Anraku}, {Asaoka}, {Fujikawa}, {Imori}
  et~al.}}]{2000ApJ...545.1135S}
\bibinfo{author}{\bibfnamefont{T.}~\bibnamefont{{Sanuki}}},
  \bibnamefont{et~al.}, \bibinfo{journal}{\apj} \textbf{\bibinfo{volume}{545}},
  \bibinfo{pages}{1135} (\bibinfo{year}{2000}), \eprint{astro-ph/0002481}.

\bibitem[{\citenamefont{{Aguilar} et~al.}(2007)\citenamefont{{Aguilar},
  {Alcaraz}, {Allaby}, {Alpat}, {Ambrosi}, {Anderhub}, {Ao}, {Arefiev},
  {Azzarello}, {Baldini} et~al.}}]{2007PhLB..646..145A}
\bibinfo{author}{\bibfnamefont{M.}~\bibnamefont{{Aguilar}}},
  \bibnamefont{et~al.}, \bibinfo{journal}{Phys. Lett. B}
  \textbf{\bibinfo{volume}{646}}, \bibinfo{pages}{145} (\bibinfo{year}{2007}),
  \eprint{astro-ph/0703154}.

\bibitem[{\citenamefont{{Barwick} et~al.}(1997)\citenamefont{{Barwick},
  {Beatty}, {Bhattacharyya}, {Bower}, {Chaput}, {Coutu}, {de Nolfo}, {Knapp},
  {Lowder}, {McKee} et~al.}}]{1997ApJ...482L.191B}
\bibinfo{author}{\bibfnamefont{S.~W.} \bibnamefont{{Barwick}}},
  \bibnamefont{et~al.}, \bibinfo{journal}{\apjl}
  \textbf{\bibinfo{volume}{482}}, \bibinfo{pages}{L191} (\bibinfo{year}{1997}),
  \eprint{astro-ph/9703192}.

\bibitem[{\citenamefont{{Coutu} et~al.}(2001)\citenamefont{{Coutu}, {Beach},
  {Beatty}, {Bhattacharyya}, {Bower}, {Duvernois}, {Labrador}, {McKee},
  {Minnick}, {Muller} et~al.}}]{2001ICRC....5.1687C}
\bibinfo{author}{\bibfnamefont{S.}~\bibnamefont{{Coutu}}},
  \bibnamefont{et~al.}, in \emph{\bibinfo{booktitle}{International Cosmic Ray
  Conference}} (\bibinfo{year}{2001}), vol.~\bibinfo{volume}{5} of
  \emph{\bibinfo{series}{International Cosmic Ray Conference}}, p.
  \bibinfo{pages}{1687}.

\bibitem[{\citenamefont{{Jin} et~al.}(2013)\citenamefont{{Jin}, {Wu}, and
  {Zhou}}}]{2013JCAP...11..026J}
\bibinfo{author}{\bibfnamefont{H.-B.} \bibnamefont{{Jin}}},
  \bibinfo{author}{\bibfnamefont{Y.-L.} \bibnamefont{{Wu}}}, \bibnamefont{and}
  \bibinfo{author}{\bibfnamefont{Y.-F.} \bibnamefont{{Zhou}}},
  \bibinfo{journal}{\jcap} \textbf{\bibinfo{volume}{11}}, \bibinfo{eid}{026}
  (\bibinfo{year}{2013}), \eprint{1304.1997}.

\bibitem[{\citenamefont{{Cholis} and {Hooper}}(2013)}]{2013PhRvD..88b3013C}
\bibinfo{author}{\bibfnamefont{I.}~\bibnamefont{{Cholis}}} \bibnamefont{and}
  \bibinfo{author}{\bibfnamefont{D.}~\bibnamefont{{Hooper}}},
  \bibinfo{journal}{\prd} \textbf{\bibinfo{volume}{88}}, \bibinfo{eid}{023013}
  (\bibinfo{year}{2013}), \eprint{1304.1840}.

\bibitem[{\citenamefont{{Huang} et~al.}(2012)\citenamefont{{Huang}, {Yuan},
  {Yin}, {Bi}, and {Chen}}}]{2012JCAP...11..048H}
\bibinfo{author}{\bibfnamefont{X.}~\bibnamefont{{Huang}}},
  \bibinfo{author}{\bibfnamefont{Q.}~\bibnamefont{{Yuan}}},
  \bibinfo{author}{\bibfnamefont{P.-F.} \bibnamefont{{Yin}}},
  \bibinfo{author}{\bibfnamefont{X.-J.} \bibnamefont{{Bi}}}, \bibnamefont{and}
  \bibinfo{author}{\bibfnamefont{X.}~\bibnamefont{{Chen}}},
  \bibinfo{journal}{\jcap} \textbf{\bibinfo{volume}{11}}, \bibinfo{eid}{048}
  (\bibinfo{year}{2012}), \eprint{1208.0267}.

\bibitem[{\citenamefont{{Drlica-Wagner}}(2012)}]{Drlica-Wagner2012}
\bibinfo{author}{\bibfnamefont{A.}~\bibnamefont{{Drlica-Wagner}}}, in
  \emph{\bibinfo{booktitle}{Fermi Symposium 2012}} (\bibinfo{year}{2012}).

\bibitem[{\citenamefont{{Ackermann} et~al.}(2011)\citenamefont{{Ackermann},
  {Ajello}, {Albert}, {Atwood}, {Baldini}, {Ballet}, {Barbiellini}, {Bastieri},
  {Bechtol}, {Bellazzini} et~al.}}]{2011PhRvL.107x1302A}
\bibinfo{author}{\bibfnamefont{M.}~\bibnamefont{{Ackermann}}},
  \bibnamefont{et~al.}, \bibinfo{journal}{\prl} \textbf{\bibinfo{volume}{107}},
  \bibinfo{eid}{241302} (\bibinfo{year}{2011}), \eprint{1108.3546}.

\bibitem[{\citenamefont{{Geringer-Sameth} and
  {Koushiappas}}(2011)}]{2011PhRvL.107x1303G}
\bibinfo{author}{\bibfnamefont{A.}~\bibnamefont{{Geringer-Sameth}}}
  \bibnamefont{and} \bibinfo{author}{\bibfnamefont{S.~M.}
  \bibnamefont{{Koushiappas}}}, \bibinfo{journal}{\prl}
  \textbf{\bibinfo{volume}{107}}, \bibinfo{eid}{241303} (\bibinfo{year}{2011}),
  \eprint{1108.2914}.

\bibitem[{\citenamefont{{Delahaye} et~al.}(2009)\citenamefont{{Delahaye},
  {Lineros}, {Donato}, {Fornengo}, {Lavalle}, {Salati}, and
  {Taillet}}}]{2009A&A...501..821D}
\bibinfo{author}{\bibfnamefont{T.}~\bibnamefont{{Delahaye}}},
  \bibinfo{author}{\bibfnamefont{R.}~\bibnamefont{{Lineros}}},
  \bibinfo{author}{\bibfnamefont{F.}~\bibnamefont{{Donato}}},
  \bibinfo{author}{\bibfnamefont{N.}~\bibnamefont{{Fornengo}}},
  \bibinfo{author}{\bibfnamefont{J.}~\bibnamefont{{Lavalle}}},
  \bibinfo{author}{\bibfnamefont{P.}~\bibnamefont{{Salati}}}, \bibnamefont{and}
  \bibinfo{author}{\bibfnamefont{R.}~\bibnamefont{{Taillet}}},
  \bibinfo{journal}{\aap} \textbf{\bibinfo{volume}{501}}, \bibinfo{pages}{821}
  (\bibinfo{year}{2009}), \eprint{0809.5268}.

\bibitem[{\citenamefont{{Badhwar} et~al.}(1977)\citenamefont{{Badhwar},
  {Golden}, and {Stephens}}}]{1977PhRvD..15..820B}
\bibinfo{author}{\bibfnamefont{G.~D.} \bibnamefont{{Badhwar}}},
  \bibinfo{author}{\bibfnamefont{R.~L.} \bibnamefont{{Golden}}},
  \bibnamefont{and} \bibinfo{author}{\bibfnamefont{S.~A.}
  \bibnamefont{{Stephens}}}, \bibinfo{journal}{\prd}
  \textbf{\bibinfo{volume}{15}}, \bibinfo{pages}{820} (\bibinfo{year}{1977}).

\bibitem[{\citenamefont{{Feng} et~al.}(2014)\citenamefont{{Feng}, {Yang}, {He},
  {Dong}, {Fan}, and {Chang}}}]{2014PhLB..728..250F}
\bibinfo{author}{\bibfnamefont{L.}~\bibnamefont{{Feng}}},
  \bibinfo{author}{\bibfnamefont{R.-Z.} \bibnamefont{{Yang}}},
  \bibinfo{author}{\bibfnamefont{H.-N.} \bibnamefont{{He}}},
  \bibinfo{author}{\bibfnamefont{T.-K.} \bibnamefont{{Dong}}},
  \bibinfo{author}{\bibfnamefont{Y.-Z.} \bibnamefont{{Fan}}}, \bibnamefont{and}
  \bibinfo{author}{\bibfnamefont{J.}~\bibnamefont{{Chang}}},
  \bibinfo{journal}{Physics Letters B} \textbf{\bibinfo{volume}{728}},
  \bibinfo{pages}{250} (\bibinfo{year}{2014}), \eprint{1303.0530}.

\bibitem[{\citenamefont{{Yuan} and {Bi}}(2013)}]{2013PhLB..727....1Y}
\bibinfo{author}{\bibfnamefont{Q.}~\bibnamefont{{Yuan}}} \bibnamefont{and}
  \bibinfo{author}{\bibfnamefont{X.-J.} \bibnamefont{{Bi}}},
  \bibinfo{journal}{Physics Letters B} \textbf{\bibinfo{volume}{727}},
  \bibinfo{pages}{1} (\bibinfo{year}{2013}), \eprint{1304.2687}.

\bibitem[{\citenamefont{{Di Mauro} et~al.}(2014)\citenamefont{{Di Mauro},
  {Donato}, {Fornengo}, {Lineros}, and {Vittino}}}]{2014arXiv1402.0321D}
\bibinfo{author}{\bibfnamefont{M.}~\bibnamefont{{Di Mauro}}},
  \bibinfo{author}{\bibfnamefont{F.}~\bibnamefont{{Donato}}},
  \bibinfo{author}{\bibfnamefont{N.}~\bibnamefont{{Fornengo}}},
  \bibinfo{author}{\bibfnamefont{R.}~\bibnamefont{{Lineros}}},
  \bibnamefont{and}
  \bibinfo{author}{\bibfnamefont{A.}~\bibnamefont{{Vittino}}},
  \bibinfo{journal}{\jcap} \textbf{\bibinfo{volume}{04}}, \bibinfo{pages}{006}
  (\bibinfo{year}{2014}), \eprint{1402.0321}.

\bibitem[{\citenamefont{{Yin} et~al.}(2013)\citenamefont{{Yin}, {Yu}, {Yuan},
  and {Bi}}}]{2013PhRvD..88b3001Y}
\bibinfo{author}{\bibfnamefont{P.-F.} \bibnamefont{{Yin}}},
  \bibinfo{author}{\bibfnamefont{Z.-H.} \bibnamefont{{Yu}}},
  \bibinfo{author}{\bibfnamefont{Q.}~\bibnamefont{{Yuan}}}, \bibnamefont{and}
  \bibinfo{author}{\bibfnamefont{X.-J.} \bibnamefont{{Bi}}},
  \bibinfo{journal}{\prd} \textbf{\bibinfo{volume}{88}}, \bibinfo{eid}{023001}
  (\bibinfo{year}{2013}), \eprint{1304.4128}.

\bibitem[{\citenamefont{{Papucci} and {Strumia}}(2010)}]{2010JCAP...03..014P}
\bibinfo{author}{\bibfnamefont{M.}~\bibnamefont{{Papucci}}} \bibnamefont{and}
  \bibinfo{author}{\bibfnamefont{A.}~\bibnamefont{{Strumia}}},
  \bibinfo{journal}{\jcap} \textbf{\bibinfo{volume}{3}}, \bibinfo{pages}{14}
  (\bibinfo{year}{2010}), \eprint{0912.0742}.

\bibitem[{\citenamefont{{Cirelli} et~al.}(2010)\citenamefont{{Cirelli},
  {Panci}, and {Serpico}}}]{2010NuPhB.840..284C}
\bibinfo{author}{\bibfnamefont{M.}~\bibnamefont{{Cirelli}}},
  \bibinfo{author}{\bibfnamefont{P.}~\bibnamefont{{Panci}}}, \bibnamefont{and}
  \bibinfo{author}{\bibfnamefont{P.~D.} \bibnamefont{{Serpico}}},
  \bibinfo{journal}{Nuclear Physics B} \textbf{\bibinfo{volume}{840}},
  \bibinfo{pages}{284} (\bibinfo{year}{2010}), \eprint{0912.0663}.

\end{thebibliography}

\end{document}